\global\def\draftcontrol{0}

   \def\versionno{real vertex}

\catcode`\@=11

\expandafter\ifx\csname draftcontrol\endcsname\relax\global\def\draftcontrol{0} 
\fi 

{\count255=\time\divide\count255 by 60 
\xdef\hourmin{\number\count255} 
\multiply\count255 by-60\advance\count255 by\time 
\xdef\hourmin{\hourmin:\ifnum\count255<10 0\fi\the\count255}} 
\def\draftdate{\number\month/\number\day/\number\year\ \ \ \hourmin } 

\newcommand\makepapertitle{\par

  \begingroup 
    \renewcommand\thefootnote{\@fnsymbol\c@footnote}%
    \def\@makefnmark{\rlap{\@textsuperscript{\normalfont\@thefnmark}}}%
    \long\def\@makefntext##1{\parindent 1em\noindent 
            \hb@xt@1.8em{%
                \hss\@textsuperscript{\normalfont\@thefnmark}}##1}%
     \newpage 
     \global\@topnum\z@   
     \@makepapertitle 
     \thispagestyle{empty}\@thanks 
  \endgroup 
  \setcounter{footnote}{0}%
  \global\let\thanks\relax 
  \global\let\makepapertitle\relax 
  \global\let\@makepapertitle\relax 
  \global\let\@thanks\@empty 
  \global\let\@author\@empty 
  \global\let\@date\@empty 
  \global\let\@title\@empty 
  \global\let\title\relax 
  \global\let\author\relax 
  \global\let\date\relax 
  \global\let\and\relax 
  \def\version{\let\version\@version\@gobble} 
} 
\def\@makepapertitle{%
  \newpage 
   \ifnum\draftcontrol=1 {} 
   \version\versionno 
   \vskip 5.5em%
   \else 
   \hfill\hbox to 3.5cm {\parbox{4.5cm}{\@pubnum}\hss}%
   \vskip 6.5em%
   \fi 
   \begin{center}%
   \let \footnote \thanks 
      {\hskip -0\textwidth \hbox to 1\textwidth%
        {\centerline{\Large\bf{\noindent\@title}}}}%
     \vskip 2em%
     {\normalsize
       \lineskip .5em%
       \begin{tabular}[t]{c}%
         \@author 
       \end{tabular}\par}%
     \vskip 1.5em%
     {\@bstract}%
     \end{center}%
     \vfill
     \@date%
     \vskip 1.5em%
   \par 
} 

\gdef\@pubnum{} 
\def\pubnum#1{%
  \gdef\@pubnum{#1}} 

\gdef\@bstract{} 
\def\Abstract#1{%
  \gdef\@bstract{%
   \parbox{\textwidth-0pc}{%
   \centerline{\bf Abstract}\penalty1000 
   \noindent
   \renewcommand\baselinestretch{1.0} 
   {#1}}} 
} 

\gdef\@email{}
\def\email#1{%
   \gdef\@email{%
   Email: {\tt #1}}
}

\def\ps@paper{\let\@mkboth\@gobbletwo%
     \ifnum\draftcontrol=1 
        \def\@oddfoot{\hbox to \textwidth{\tiny \versionno \hfil\tiny\draftdate}%
        \hskip -\textwidth \hbox to \textwidth{\hfil\rm\thepage\hfil}}%
     \else\def\@oddfoot{\hbox to \textwidth{\hfil\rm\thepage\hfil}} 
     \fi 
     \let\@evenfoot\@oddfoot 
} 

\def\body{\clearpage 
          \pagestyle{paper} 
        } 
\newenvironment{acknowledgments}{%
\vskip 3.25ex 
\addcontentsline{toc}{section}{Acknowledgments}
\noindent {\bf Acknowledgments} 
} 

\def\@version#1{\ifnum\draftcontrol=1 
\typeout{}\typeout{#1}\typeout{} 
\vskip3mm\centerline{\hbox{\fbox{\normalsize{\tt DRAFT -- #1 -- } 
                   {\draftdate}}}}\vskip3mm 
\fi} 
\let\version\@version 
\long\def\eqlabel#1{\ifnum\draftcontrol=1 
                    \tag@false  
                    \tag*{(\theequation) \hbox to -0.2cm{\hspace{0cm}\small{#1}\hss}} 
                    \refstepcounter{equation}  
                    \edef\@currentlabel{\theequation} 
                    \ltx@label{#1}          
                    \else 
                    \label{#1} 
                    \fi 
                    } 
\let\st@bibitem\@bibitem 
\let\st@lbibitem\@lbibitem 
\ifnum\draftcontrol=1 
  \def\@bibitem#1{%
    \st@bibitem{#1}\a@@label{#1}\ignorespaces} 
  \def\@lbibitem[#1]#2{%
    \st@lbibitem[#1]{#2}\a@@label{#2}\ignorespaces} 
  \def\a@@label#1{%
    \gdef\a@lab{\smash{\normalfont\small#1}} 
    \ifvmode 
      \if@inlabel 
        \global\setbox\@labels\hbox{%
          \llap{\a@lab\let\a@lab\relax 
                \kern\@totalleftmargin\kern\marginparsep}%
          \box\@labels}%
      \fi 
    \fi} 
\fi 

\documentclass[12pt,letterpaper]{article} 

\usepackage{amsmath,bm,amsfonts,amssymb,array,calc,amsthm,rotating}
\usepackage{epsfig,psfrag} 
\usepackage{rotating}
\usepackage{amscd}
\usepackage{graphicx}
\usepackage{color}
\usepackage[colorlinks=false]{hyperref}

\renewcommand\baselinestretch{1.25} 
\renewcommand\section{\@startsection {section}{1}{\z@}%
                                   {-3.5ex \@plus -1ex \@minus -.2ex}%
                                   {2.3ex \@plus.2ex}%
                                   {\normalfont\large\bfseries}} 
\renewcommand\subsection{\@startsection{subsection}{2}{\z@}%
                                   {-3.25ex\@plus -1ex \@minus -.2ex}%
                                   {1.5ex \@plus .2ex}%
                                   {\normalfont\normalsize\bfseries}} 
\renewcommand\subsubsection{\@startsection{subsubsection}{3}{\z@}%
                                   {-3.25ex\@plus -1ex \@minus -.2ex}%
                                   {1.5ex \@plus .2ex}%
                                   {\normalfont\normalsize\it}} 
\renewcommand\paragraph{\@startsection{paragraph}{4}{\z@}%
                                   {-3.25ex\@plus -1ex \@minus -.2ex}%
                                   {1.5ex \@plus .2ex}%
                                   {\normalfont\normalsize\bf}} 
\renewcommand\subparagraph{\@startsection{subparagraph}{5}{\z@}%
                                   {-1.25ex\@plus -1ex \@minus -.2ex}%
                                   {0ex \@plus .2ex}%
                                   {\normalfont\normalsize\it}}

\numberwithin{equation}{section}

\long\def\@makecaption#1#2{%
  \vskip\abovecaptionskip
  \sbox\@tempboxa{{\bf #1:} #2}%
  \ifdim \wd\@tempboxa >\hsize
    {\small\bf #1:} {\small #2}\par
  \else
    \global \@minipagefalse
    \hb@xt@\hsize{\hfil\box\@tempboxa\hfil}%
  \fi
  \vskip\belowcaptionskip}

\renewcommand*\l@section[2]{%
  \ifnum \c@tocdepth >\z@
    \addpenalty\@secpenalty
    \addvspace{.5em \@plus\p@}%
    \setlength\@tempdima{1.5em}%
    \begingroup
      \parindent \z@ \rightskip \@pnumwidth
      \parfillskip -\@pnumwidth
      \leavevmode \bfseries
      \advance\leftskip\@tempdima
      \hskip -\leftskip
      #1\nobreak\hfil \nobreak\hb@xt@\@pnumwidth{\hss #2}\par
    \endgroup
  \fi}
\renewcommand*\l@subsection{\addvspace{.0em \@plus\p@}\@dottedtocline{2}{1.5em}{2.3em}}
\renewcommand*\l@subsubsection{\addvspace{-.2em \@plus\p@}\@dottedtocline{3}{3.8em}{3.2em}}

\def\hepth#1{\href{http://xxx.arxiv.org/abs/hep-th/#1}{{arXiv:hep-th/#1}}}

\def\math#1{\href{http://xxx.arxiv.org/abs/math/#1}{{arXiv:math/#1}}}

\def\mathag#1{\href{http://xxx.arxiv.org/abs/math.AG/#1}{{arXiv:math.ag/#1}}}

\def\arxiv#1#2{\href{http://xxx.arxiv.org/abs/#1}{{arXiv:#1 [#2]}}}

\definecolor{refcol}{rgb}{0.2,0.2,0.8}
\definecolor{eqcol}{rgb}{.6,0,0}
\definecolor{purple}{cmyk}{0,1,0,0}

\gdef\@citecolor{refcol}
\gdef\@linkcolor{eqcol}
\def\colorlinkspurple{\gdef\@urlcolor{purple}}
\def\colorlinksblue{\gdef\@urlcolor{blue}}
\def\colorlinksred{\gdef\@urlcolor{red}}

\def\ie{{\it i.e.}} 
\def\eg{{\it e.g.}} 

\def\cf{{\it cf.}}

\def\revise#1       {\raisebox{-0em}{\rule{3pt}{1em}}%
                     \marginpar{\raisebox{.5em}{\vrule width3pt\ 
                     \vrule width0pt height 0pt depth0.5em 
                     \hbox to 0cm{\hspace{0cm}{%
                     \parbox[t]{4em}{\raggedright\footnotesize{#1}}}\hss}}}}

\def\calc         {{\cal C}} 
 
\def\cale         {{\cal E}} 
\def\calf         {{\cal F}} 
\def\calg         {{\cal G}}

\def\calm         {{\cal M}} 
 
\def\calo         {{\cal O}}

\def\calw         {{\cal W}}

\def\complex      {{\mathbb C}} 
 
\def\projective   {{\mathbb P}} 
 
\def\reals        {{\mathbb R}} 
\def\zet          {{\mathbb Z}} 
\def\CP{\complex\projective}
\def\RP{\reals\projective}

\def\ee           {{\it e}} 
\def\ii           {{\it i}} 
 
\def\tr           {{\rm Tr}} 
 
\def\Im           {{\rm Im\hskip0.1em}}

\newcommand\topa[2]{\genfrac{}{}{0pt}{2}{\scriptstyle #1}{\scriptstyle #2}}

\def\sqr#1#2{{\vcenter{\vbox{\hrule height.#2pt   
 \hbox{\vrule width.#2pt height#1pt \kern#1pt 
 \vrule width.#2pt}\hrule height.#2pt}}}}

\newcommand{\beq}{\begin{equation}}
\newcommand{\eq}{\end{equation}}

\renewcommand{\k}{\kappa}

\newcommand{\ket}[1]{\left|#1\right>}

\newcommand{\Ga}{\Gamma}

\renewcommand{\P}{\mathbb P}
\newcommand{\Z}{\mathbb Z}

\renewcommand{\O}{\mathcal O}

\newcommand{\N}{\mathcal N}

\renewcommand{\d}{\partial}

\newcommand{\req}[1]{(\ref{#1})}

\catcode`\@=12 

\begin{document} 


\title{The Real Topological Vertex at Work}

\pubnum{
arXiv:0909.1324\\
CERN-PH-TH/2009-161 \\
IPMU09-0110
}
\date{September 2009}

\author{
Daniel Krefl$^{a}$, Sara Pasquetti$^{b}$ and Johannes Walcher$^{b}$ \\[0.2cm]
\it ${}^{a}$ IPMU, The University of Tokyo, Kashiwa, Japan \\
\it $^{b}$ PH-TH Division, CERN, Geneva, Switzerland
}

\Abstract{
We develop the real vertex formalism for the computation of the topological string partition
function with D-branes and O-planes at the fixed point locus of an anti-holomorphic involution
acting non-trivially on the toric diagram of any local toric Calabi-Yau manifold. Our results
cover in particular the real vertex with non-trivial fixed leg. We give a careful derivation 
of the relevant ingredients using duality with Chern-Simons theory on orbifolds.
We show that the real vertex can also be interpreted in terms of a statistical 
model of symmetric crystal melting. Using this latter connection, we also assess the constant 
map contribution in Calabi-Yau orientifold models. We find that there are no perturbative 
contributions beyond one-loop, but a non-trivial sum over non-perturbative sectors, which 
we compare with the non-perturbative contribution to the closed string expansion.
}

\makepapertitle

\body

\version\versionno

\vskip 1em

\tableofcontents
\newpage

\section{Introduction}

The topological vertex \cite{akmv} provides a complete solution of the topological string
on local toric Calabi-Yau manifolds. By conception, the vertex is an {\it open 
string} amplitude. However, the relevant D-branes (external, non-compact, toric branes)
live in the realm of large-$N$ duality, and can therefore be absorbed in the geometry by
appropriate geometric transitions. This is indeed how the topological vertex was 
originally derived, and for this reason, it is most naturally used for the computation 
of {\it closed string} amplitudes. 

It is a natural question to ask for extensions of the vertex formalism to situations
involving D-branes that are not necessarily amenable to large-$N$ duality. Two of 
the phenomena not covered by the standard 
formalism are boundary condition changing open strings (which are non toric) and the
generation of a topological tadpole (toric branes only have a large-$N$ limit because
they do not generate such tadpoles). Concerning the second point, it was recently 
found in \cite{krwa} that the partition function of the real topological string 
(namely, with D-branes and O-planes at the fixed point locus of an anti-holomorphic involution
\cite{tadpole}) on local $\projective^2$ indeed does admit a representation in the topological 
vertex formalism. (For previous studies of toric orientifolds after the topological
vertex, see \cite{bouchard1,bouchard2}. For recent discussions of tadpole cancellation in topological strings and their orientifolds, see \cite{cook,bonelli}.) The key new ingredient is a real version (as a
kind of squareroot) of the topological vertex. Several questions were however not
addressed in \cite{krwa}, and the purpose of the present work is to close these gaps.

We begin in section \ref{begin} with reviewing some general aspects of toric orientifold 
geometries. Considerations of the conifold will then be enough to obtain most of the real 
vertex formalism in section \ref{formalism}. The subtler aspects are several new 
sign rules, which we state in section \ref{formalism}, and justify in the remaining sections 
by comparison with Chern-Simons 
theory, examples, as well as global consistency conditions of the physical interpretation.

In section \ref{chernsimons}, we turn to the derivation of the real topological vertex 
from Chern-Simons theory. The relevant orientifold of the deformed conifold acts on the 
base $S^3$ with a fixed point locus in codimension 2. For the evaluation of these
amplitudes, we will rely on some old work of Ho\v rava  \cite{horava}. 

In section \ref{melting}, we study the real vertex from the point of view of the statistical
model of melting crystal introduced in \cite{crystal}. We show that the real vertex
amplitude (at least in case of trivial fixed leg) precisely computes the partition function
of melting crystal configurations that are invariant under the exchange of two of the 
axes. (More precisely, as we will see in section \ref{formalism}, the real vertex actually
comes in two versions. Both of them have a melting crystal interpretation.) In particular, 
in analogy with the MacMahon function, which is the counting function for plane partitions, 
capturing the constant map contribution of the topological string, we propose that the 
counting function for symmetric plane partitions (the ``real'' MacMahon function),
encodes the constant map contribution of general Calabi-Yau orientifolds. A perhaps 
soothing consequence of this conjecture is that the constant map sector receives 
perturbative contributions from open and non-orientable worldsheets only at tree
and one-loop level. The remainder of the real MacMahon function is non-perturbative
in the string coupling. 

We end in section \ref{examples} with some explicit examples, before concluding in section 
\ref{conclusions}.

\section{Classification of toric orientifolds}
\label{begin}

We are interested in orientifold models of A-type topological strings on a local
toric Calabi-Yau manifold $X$. We intend to wrap our D-branes, which are necessary 
for tadpole cancellation, on top of the orientifold plane. Thus, to specify our 
model, all we need to do is identify a suitable anti-holomorphic involution
$\sigma:X\to X$. Here we have in mind that the involution should be compatible
with the toric symmetries of $X$ so that we could in principle use localization
techniques to compute A-model invariants, and ultimately will be able to compute
using the (real) topological vertex formalism. The complete specification of the 
background includes of course also the closed and open string moduli, which is a 
point to which we will return in section \ref{formalism}.

\subsection{Anti-holomorphic involutions from symmetries of the toric diagram}

One way to identify suitable anti-holomorphic involutions is to use the gauged 
linear sigma model description of toric manifolds. Let's have $N+3$ chiral fields 
$z_i$, $i=1,\ldots,N+3$, with charges $Q_i^A$ (of $\sum_i Q_i^A=0$) under the 
$U(1)^N$ gauge group with Fayet-Iliopoulos (FI) parameters $t^A$, $A=1,\ldots, N$. 
In the absence of a GLSM superpotential (toric Calabi-Yaus are always rigid), one 
anti-holomorphic involution that always exists is simply conjugating all fields 
$\sigma_0:z_i\to \bar z_i$. Other involutions can be obtained by dressing this basic 
involution with global (holomorphic) symmetries $g$ of the model, such that the 
dressed involution $\sigma_g=g\circ \sigma_0$ is still involutive. This means that
$(\sigma_g)^2$ is equivalent to the identity modulo a $U(1)^N$ gauge 
transformation. Typical symmetries $g$ one can consider include permutations
of variables (possibly combined with permutations of the gauge group
factors), as well as phase rotations on the variables. Note that dressings 
that can be related to each other by conjugating $\sigma_g$ with a global 
symmetry should be considered equivalent. (This is not the same as
conjugating $g$!) We will see these options at work in examples below, but 
otherwise shall not develop the GLSM point of view any further here.

For purposes of the topological vertex, as explained in \cite{akmv}, we think
of our toric manifold $X$ in terms of a certain $T^2\times \reals$-fibration 
over a three-dimensional base. This three-dimensional base is the image of $X$ 
under the ``moment maps''
\begin{equation}
\eqlabel{moment}
r_\alpha= |z_1|^2-|z_3|^2 \,,\qquad
r_\beta = |z_2|^2-|z_3|^2 \,,\qquad
r_\gamma = {\rm Im}({\textstyle\prod_i} z_i)
\end{equation}
(where we have arbitrarily distinguished the first three fields), and the $T^2\times
\reals$ fibers are the orbits of the Hamiltonian flows generated by $r_\alpha,r_\beta,
r_\gamma$ with respect to the symplectic form $\omega=\sum_i dz_i\wedge 
d\bar z_i$, modulo symplectic reduction to $\sum Q_i^A |z_i|^2=t^A$. The $T^2$ 
fibers degenerate over a certain piecewise linear trivalent graph $\Gamma$ in the 
$r_\alpha$-$r_\beta$ plane, known as the toric, or $(p,q)$-web, diagram. (For full 
details, see \cite{akmv}.) 

The anti-holomorphic involutions of our interest can be described in terms of
their action on this toric diagram. (As we will see below, the topological vertex
formalism does not distinguish between different dressing phases, which lead to 
different actions on the $T^2$ fibers.) Essentially, $\sigma$ acts as a linear 
involution on the $r_\alpha$-$r_\beta$ plane and induces a symmetry of $X$ if it maps
the toric diagram to itself. Now note that since $\sigma(\omega)=-\omega$, the
isometries of $T^2$ compatible with the involution are generated by eigenvectors of 
$\sigma$ with eigenvalue $-1$. In particular, the trivial action on $\Gamma$ does
not preserve any symmetries\footnote{There are cases, such as the conifold,
where the trivial involution preserves additional accidental $U(1)$ symmetries.
There is then a different presentation of the manifold which makes this manifest.}, 
which is not enough for our purposes. This leaves 
two possibilities. When $\sigma$ acts with two negative eigenvalues, we are
dealing with a ``point reflection''. By an appropriate choice of phases, one
can arrange that $\sigma$ acts without fixed points on the total space of the 
fibration, which is our Calabi-Yau $X$. Thus, the orientifold plane is empty, and 
we should not wrap any D-brane for tadpole cancellation. The implementation of this 
situation at the level of the topological vertex was studied in detail in 
\cite{bouchard1}, and will be subsummed in our formalism below. The last possibility 
is that $\sigma$ acts in the $r_\alpha$-$r_\beta$ plane with eigenvalues $(+1,-1)$. 
This corresponds to ``reflection at a line''. In that case, the action of $\sigma$ 
on $X$ will typically have fixed points, and we should wrap D-branes on top of these 
orientifold planes.

\begin{figure}
\psfrag{0}[cc][][0.7]{$R_0$}
\psfrag{1}[cc][][0.7]{$R_2$}
\psfrag{2}[cc][][0.7]{$R_1$}
\begin{center}
\includegraphics[scale=0.5]{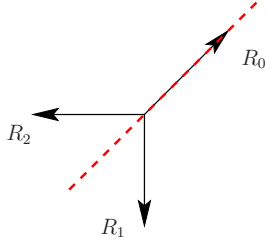}
\caption{The local geometry around a fixed vertex is $\complex^3$ with involution 
acting as $(z_0,z_1,z_2)\to (\bar z_0,\bar z_2,\bar z_1)$. The $R_i$ denote the 
$U(\infty)$ representations propagating on the legs in the topological vertex 
formalism. Invariance under the involution requires $R_0=R_0^t$, and $R_2=R_1^t$. }
\label{c3fig}
\end{center}
\end{figure}

The line reflection is the richest of these possibilities. Consider in particular
the intersection of the fixed line in the $r_\alpha$-$r_\beta$ plane with the
toric diagram $\Gamma$. The finite legs of $\Gamma$ that are fixed under the
involution fall in two classes. They can either intersect the fixed line 
perpendicularly (which we will call fixed legs of type 2) or be fixed point-wise 
(type 3). (The fixed legs that can occur in the point reflection we refer to as 
of type 1.) Fixed legs of type 3 end in two fixed vertices, around which
the local geometry is as depicted in figure \ref{c3fig}. This is the geometry
that necessitates the real topological vertex.

\subsection{Geometric transitions}
\label{geotrans}

It is instructive and useful for the subsequent Chern-Simons discussion to review 
here the possible A-type orientifolds of the conifold, their toric realization,
and their relation under the conifold transition. 

Recall first (see, \eg, \cite{raul}) the classification of anti-holomorphic
involutions of the resolved and deformed conifold as K\"ahler manifolds (forgetting
the torus action). On the resolved side, the possible involutions are distinguished by 
the action on the base $\projective^1$ of $\calo(-1)\oplus\calo(-1)$. Namely, this action 
can be fixed point free, $z\to -1/\bar z$, where $z$ is an inhomogeneous coordinate
on $\projective^1$, or the fixed point locus is an $S^1$, $z\to 1/\bar z$.
The deformed conifold, $T^*S^3$ can be thought of as the hypersurface
$\sum x_i^2=\mu$ in $\complex^4$. Fixing an anti-holomorphic involution
imposes a reality condition on the deformation parameter $\mu$. The 
topological type of the fixed point locus depends on the signs in
$\sum\pm x_i^2=\mu$, with all variables real. When all signs are $+1$, the fixed
point locus is $S^3$ for $\mu>0$, and empty for $\mu<0$. Under conifold
transition, this is related to the fixed point free action on the resolved
conifold. For an odd number of $-1$'s, the fixed point locus is
$S^2\times\reals$ or $\reals^3\cup\reals^3$ (depending on the sign of $\mu$). 
This involution is not compatible with the small resolution as the K\"ahler 
deformation at $\mu=0$ is projected out. Finally, for two $+1$'s and two $-1$'s,
the fixed point locus is $S^1\times\reals^2$, and this is related to
the involution of the resolved conifold whose fixed locus is also 
$S^1\times\reals^2$.

\begin{figure}[t]
\psfrag{1}[][0.7]{$z_3$}
\psfrag{2}[][0.7]{$z_4$}
\psfrag{3}[][0.7]{$z_1$}
\psfrag{4}[][0.7]{$z_2$}
\psfrag{I}[cc][][0.7]{$1$}
\psfrag{II}[cc][][0.7]{$2$}
\psfrag{III}[cc][][0.7]{$3$}
\begin{center}
\includegraphics[scale=0.5]{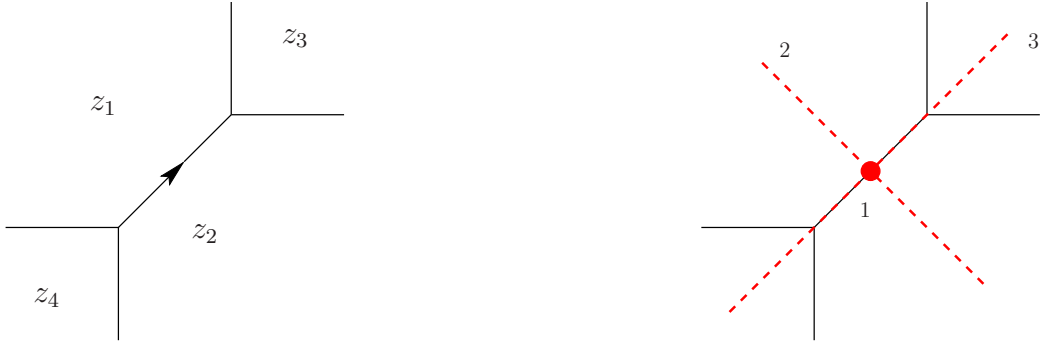}
\caption{Left: $(p,q)$-web of the resolved conifold. The $z_i$ indicate
which coordinate vanishes on the corresponding toric divisor.
Right: Illustration of the three involutive symmetries of the web diagram.}
\label{conifig1}
\end{center}
\end{figure}

We now describe how these involutions are realized at the level of the toric geometry. 
The web diagram of the resolved conifold is shown in figure \ref{conifig1}. It 
has three apparent non-trivial symmetries: A point reflection at the center of the 
compact edge (we refer to this as involution 1), a reflection at a line perpendicular to
the compact leg (involution 2), and a reflection at the line containing the compact leg
(involution 3). To understand the corresponding fixed point loci, we lift the action to
the fields of the GLSM: $(z_1,z_2,z_3,z_4)$, of charge $(1,1,-1,-1)$, with FI $t>0$. 
We deduce that involution 2 acts by $(z_1,z_2,z_3,z_4)\to (\bar z_2,\bar z_1,\bar 
z_3,\bar z_4)$, while involution 3 sends $(z_1,z_2,z_3,z_4)$ to $(\bar z_1,\bar 
z_2,\bar z_4,\bar z_3)$ (up to conjugation by global symmetries). Those two involutions 
are related by a flop, see figure \ref{coniflopfig} on page \pageref{coniflopfig} and the 
fixed point locus on $\projective^1$ 
is $S^1$. (The full O-plane has topology $S^1\times \reals^2$.) Turning to involution 1, 
complex conjugation must be dressed by exchanging both, $z_1,z_2$ and $z_3,z_4$. In addition, 
we can allow for a non-trivial phase dressing. Namely, we have the two possible lifts: 
$(z_1,z_2,z_3,z_4)\to (\bar z_2,\bar z_1,\bar z_4,\bar z_3)$ and $(z_1,z_2,z_3,z_4)\to 
(\bar z_2,-\bar z_1,\bar z_4,-\bar z_3)$. (The point being that the second involution squares to
$(-1,-1,-1,-1)$, which is gauge equivalent to the identity.) The fixed point locus
for the first choice is $S^1$, in the second the action is fixed point free.

\section{The real topological vertex}
\label{formalism}

The topological vertex formalism computes the A-model topological string
partition function $Z_X$ of a toric Calabi-Yau manifold $X$ from the toric diagram 
$\Gamma$. Orient each compact leg of the toric diagram, and attach a Young diagram 
$R_i$, $i=1,\ldots,l$. Then $Z_X$ is obtained by summing over the $R_i$ a certain 
rational function $P_{(R_i)}(q,Q_A)$ of $N+1$ variables (one, $Q_A=
\ee^{-t_A}$, for each K\"ahler class, and one, $q=\ee^{g_s}$, for the string 
coupling) constructed from contributions at each vertex and each compact leg.
\begin{equation}
\eqlabel{contributions}
Z_X(q,Q_A) = \sum_{(R_i)} P_{(R_i)}(q,Q_A)\,.
\end{equation}
As explained in \cite{krwa}, the basic idea behind the real topological vertex
is that the action of the involution on $\Gamma$ induces a symmetry of the sum 
over $R_i$'s. The terms fixed under the involution are, appropriately interpreted,
perfect squares when the parameters $Q_A$ are also appropriately restricted, 
$\sigma(Q_A)=Q_A$, and the real topological string partition function of
$X$ with involution $\sigma$, is given 
by
\begin{equation}
Z^\sigma_X = \sum_{(R_i)=\sigma(R_i)} \pm \sqrt{P_{(R_i)}} \,.
\end{equation}
Really, the real topological vertex is the theory of signs in taking this
squareroot. To isolate the genuine open+unoriented contribution, we reduce with 
respect to the closed string partition function,
\begin{equation}
Z'^\sigma_X = \frac{Z^\sigma_X}{\sqrt{Z_X}}\,,
\end{equation}
such that the (reduced) real topological string free energy reads
\beq\eqlabel{realfenergy}
\mathcal G'^\sigma_{X}=\log\left(Z'^\sigma_X\right) \,.
\end{equation}
Note that the restriction on the $Q_A$ is the usual orientifold projection of 
K\"ahler parameters. Taking the squareroot of $P_{(R_i)}$ often also involves a
squareroot of some of the $Q_A$. The resulting sign degree of freedom can usually
be interpreted as a discrete Wilson line on the D-brane wrapped around the
corresponding component of the fixed point locus, \ie, as a discrete open 
string modulus.

\subsection{Conifold}
\label{coninv}

Let us see how this procedure works for the involutions of the conifold described in the
previous section. The closed topological string partition function is in the topological 
vertex formalism computed as
\begin{equation}
\eqlabel{summation}
Z_{{\rm con.}}(q,Q)=\sum_R (-1)^{\ell(R)}Q^{\ell(R)}C_{\cdot\cdot R}(q)C_{\cdot\cdot R^t}(q)\,.
\end{equation}
Here, $Q=\ee^{-t}$ is the single K\"ahler parameter of the geometry, $q=\ee^{g_s}$
with $g_s$ the topological string coupling. The sum runs over all partitions
$R=(\lambda_1,\lambda_2,\ldots)$ with number of boxes computed by 
$\ell(R)=\sum \lambda_i$. $C_{\cdot\cdot R}(q)$ is the 1-legged topological vertex,
which can be expressed in terms of the standard Schur function
\begin{equation}
C_{\cdot\cdot R}(q) = s_R(q^\rho)\,.
\end{equation}
(The notation means evaluating $s_R$ at $x_i=q^{-i+1/2}$, for $i=1,2,
\ldots$) Using the elementary Schur function identity
\begin{equation} 
\sum_R s_R(x)s_{R^t}(y)=\prod_{i,j=1}^\infty(1+x_i y_j)\,,
\end{equation}
we obtain the standard expression \cite{gova2} for $Z$,
\begin{equation}
Z_{{\rm con.}}(q,Q)=\prod_{n=0}^\infty(1- Q\, q^{-n})^n=
\exp\biggl(-\sum_{k=1}^\infty\frac{Q^k}{k\bigl(q^{k/2}-q^{-k/2 }\bigr)^2}\biggr)\,.
\end{equation}

\paragraph{Involution 1}

The point reflection acts on the summation variable in \eqref{summation} by $R\to R^t$. The sign 
of the squareroot of the fixed configurations has been found in \cite{bouchard1} 
to be determined by the {\it rank} of the corresponding self-conjugate representation, 
\ie, the number of boxes on the diagonal
\begin{equation}
\eqlabel{rdef}
r(R) = \#\{ \lambda_i\ge i\}\,.
\end{equation}
Occasionally, we will refer to this sign as an ``$r$-type sign''.
We obtain
\begin{equation}
\eqlabel{weobtain}
Z_{{\rm con.}}^1=\sum_{R=R^t}(-1)^{(\ell(R)\mp r(R))/2}Q^{\ell(R)/2}C_{\cdot\cdot R}(q)\,.
\end{equation}
Note that the sign of the squareroot $Q^{1/2}$ is equivalent to the sign
$\mp 1$ in front of $r(R)$. We can now use the Schur function identity 
\cite{macdonald}
\begin{equation}
\sum_{R=R^t}(-1)^{(\ell(R)\mp r(R))/2}s_R(x)=\prod_{i=1}^\infty(1\pm x_i)
\prod_{1\leq i<j}^\infty(1-x_ix_j)\,,
\end{equation}
to obtain the expression
\begin{equation}
\eqlabel{Z1}
Z_{{\rm con.}}^1=\exp\biggl(-\frac{1}{2}\sum_{k=1}^\infty 
\frac{Q^k}{k\bigl(q^{k/2})-q^{-k/2}\bigr)^2}
-\sum_{k=1}^\infty \frac{(\pm Q)^{k/2}}{k(q^{k/2}-q^{-k/2}\bigr)}\biggr)\,.
\end{equation}
This is the expected result. It can be interpreted as the partition function of 
Chern-Simons theory on $S^3$ with orthogonal or symplectic gauge group, depending on
the sign $\pm 1$, which is the sign of the crosscap amplitude in the topological string
interpretation. This is consistent with the geometric transition to orientifold of 
deformed conifold with $S^3$ fixed locus, as was first observed in \cite{sinha}. 

\paragraph{Involution 2}

This involution leaves $R$ in \eqref{summation} invariant, so the summation variable is 
not restricted. To see that we nevertheless have a perfect square, we need to use the 
identity
\begin{equation}
s_{R^t}(q) =  q^{-\kappa(R)/2} s_{R}(q)\,,
\end{equation}
where $\kappa(R)= \sum \lambda_i(\lambda_i-2i+1)$. The sign of the squareroot we 
can borrow from \cite{krwa}. We denote the {\it number of boxes in even rows} by 
$p(R)$,
\begin{equation}
\eqlabel{pdef}
p(R) = \sum_i \lambda_{2i}\,.
\end{equation}
Note that one may write
\begin{equation}
\eqlabel{cdef}
p(R) = (\ell(R)-c(R))/2\,,
\end{equation}
where $c(R)$ is the {\it number of columns of odd height}. This is a quantity that
features in section I.5 of \cite{macdonald}, thus exhibiting the representation 
theoretic relevance of $p(R)$. Occasionally, we will refer to this sign as a 
``$c$-type sign''. Note that as for the $r$-type sign, the $c$-type sign is itself
only determined up to a sign, $(-1)^{\ell(R)}$. Namely, a replacement of $r(R)$ with
$-r(R)$ can be absorbed in a corresponding open string degree of freedom (the sign of
$Q^{1/2}$). We obtain
\begin{equation}
\eqlabel{unable}
Z_{{\rm con.}}^2 =\sum_{R}(-1)^{(\ell(R)\mp c(R))/2}Q^{\ell(R)/2}q^{-\kappa(R)/4}
C_{\cdot\cdot R}(q)\,.
\end{equation}
We have not been able to identify any Schur identity to sum this expression. But low
degree expansion in $Q$ reveals that in fact
\begin{equation}
Z_{{\rm con.}}^2 = Z_{{\rm con.}}^1\,.
\end{equation}
This is also the expected result: For involution 2 we need to wrap a D-brane on the 
fixed point locus, and the sign of the squareroot $Q^{1/2}$ is the value of the
discrete Wilson line at the critical locus of the superpotential. (Going to this
critical locus also eliminates any potential framing dependence.) The difference 
between involution 1 and involution 2 is merely whether we give a BPS/enumerative 
interpretation of the second term in the exponential of \eqref{Z1} as a crosscap or 
disk respectively.

Note that as for the $r$-type sign, the sign of $c(R)$ in the $c$-type sign is equivalent 
to the sign of the squareroot of $Q^{1/2}$. However, in contrast to the $r$-type sign, 
the $c$-type sign comes with an additional degree of freedom. Namely, we may as well 
have used $c(R^t)$, which is generally not equal to $c(R)$. For the conifold, the two 
possibilities yield the same result. However, for more complicated models taking account
of this additional degree of freedom turns out to be crucial. We will come back to this 
point in the more general discussion below.

\paragraph{Involution 3}

\begin{figure}
\psfrag{a}[cc][][0.7]{type 2}
\psfrag{b}[cc][][0.7]{type 3}
\psfrag{F}[cc][][1]{flop}
\begin{center}
\includegraphics[scale=0.5]{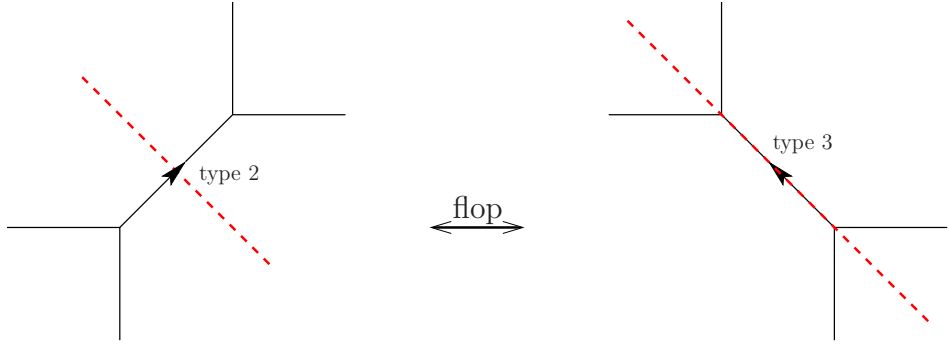}
\caption{A flop of the resolved conifold relates the involutions with fixed leg of 
type 2 and type 3.}
\label{coniflopfig}
\end{center}
\end{figure}

We expect this involution to also give the same result as involution 1 and 2 (after all, 
as a string background, this is identical to involution 2). Stated more generally, we 
expect that the flop invariance of the closed topological vertex carries over to 
orientifolds, as illustrated in figure \ref{coniflopfig}. Especially, the flop 
transition transforms a type 2 fixed leg to a type 3 fixed leg and vice versa.

On the other hand, the naive application of the squareroot philosophy requires
the real vertex with non-trivial fixed leg, in other words, the squareroot
of $C_{\cdot\cdot R}(q)= s_R(q^\rho)$. However, recalling the formula
\begin{equation}
s_R(q^\rho)=q^{n(R)-l(R)/2}\prod_{(i,j)\in R}\left(1-q^{h(i,j)}\right)^{-1}\,,
\end{equation}
with $n(R)=\sum_i(i-1)\lambda_i$ and $h(i,j)$ the Hook length $h(i,j)=1+\lambda_i+
\lambda_j^t-i-j$, we see that $s_R(q)$ is {\it not} a perfect square even for 
$R=R^t$ (where only the terms on the diagonal, $i=j$, remain unpaired). This 
was pointed out in \cite{krwa}.

Thinking somewhat more generally about fixed vertices with non-trivial
fixed leg in arbitrary toric diagrams, we notice that we can save the situation 
by exploiting the fact that such a fixed leg will always end on another fixed
vertex. (Otherwise, the leg is external, and carries a trivial representation.) 
The two squareroots can be combined to a rational expression, and
this is what we propose for fixed legs of type 3 in the general case. (More
precisely, we will have to amend one additional sign in the real vertex, as 
compared with \cite{krwa}.)

Returning to involution 3 of the conifold, we use the same formulas and signs
as for involution 1, and obtain trivially
\begin{equation}
Z_{{\rm con.}}^3=Z_{{\rm con.}}^1\,,
\end{equation}
as required.

\subsection{The general case}
\label{generalcase}

We don't need to recall the full rules of the ordinary topological vertex formalism 
for computing the rational function $P_{(R_i)}(Q^A,q)$ in \eqref{contributions}. It 
suffices to observe that the contributions from legs and vertices that are not
fixed by the involution $\sigma$ (and instead, are paired by it) automatically give 
rise to a perfect square when representations are restricted to the fixed configurations. 
In other words, this part of the amplitude is given by the ordinary vertex formalism, 
evaluated on the smooth part of the quotient of the toric diagram by $\sigma$. It 
remains to describe the contribution from legs and vertices that are fixed under $\sigma$. 
This is most easily done by considering the gluing of two vertices adjacent to a
fixed leg, and then taking a squareroot. A special case of this is obtained by adding
external representations to the involutions of the conifold described above. 

In the usual vertex formalism, the gluing of two vertices along a common edge is given by
\begin{equation}
\sum_{R_i} C_{R_j R_k R_i} \ee^{-t_i {\ell(R_i)}} (-1)^{(n_i+1)\ell(R_i)}
q^{-n_i \kappa_{R_i}/2} C_{R_i^t R_j' R_k'} \,.
\end{equation}
Here, one is assuming that the edges decorated with $R_j$, $R_k$ are outgoing 
in direction $v_j$, $v_k$ at the first vertex, and the edges decorated with $R_j'$, $R_k'$ are 
outgoing in direction $v_j'$, $v_k'$ at the second vertex, respectively. The cycling ordering 
(with respect to some fixed orientation of the $r_\alpha$-$r_\beta$ plane) is as 
indicated on the $C$'s, which represent the topological vertex in canonical framing. Then 
$n_i$ is the integer $n_i= v_j'\wedge v_j$ accounting for the adjustement of framing between 
the two vertices. Let us now describe the squareroot of this amplitude for the 
three types of fixed legs discussed at the end of section \ref{begin}.

Fixed legs of type 1 (point reflection at the center of the line) were treated in 
\cite{bouchard1}. Inspection shows that in this case, $v_j'=-v_j$, $v_k'=-v_k$, and 
hence $n_i=0$. The restriction to invariant configurations of Young diagrams imposes 
$R_i=R_i^t$, $R_j'=R_j$, $R_k'=R_k$. The squareroot is
\begin{equation}
\sum_{R_i=R_i^t} \ee^{-t_i\ell(R_i)/2} (-1)^{(\ell(R_i) \pm r(R_i))/2} C_{R_j R_k R_i}\,,
\end{equation}
where we inserted an $r$-type sign, as defined in \eqref{rdef}.

For fixed legs of type 2, the $j$ leg is mapped to the $k'$ leg, and the $k$ leg
to the $j'$ leg. Invariant configurations
have $R_k'=R_j^t$ and $R_j'=R_k^t$, with no restriction on $R_i$. (Compared with type 1,
there is an additional transposition because of the orientation reversal of the 
$r_\alpha$-$r_\beta$ plane.) The amplitude is then a sum of perfect squares because
of the symmetry
\begin{equation}
C_{R_i^t R_k^t R_j^t} = q^{-\kappa_{R_i}/2-\kappa_{R_j}/2-\kappa_{R_k}/2} C_{R_i R_j R_k}\,,
\end{equation}
of the topological vertex. The sign of the squareroot is in principle determined by the 
$c$-type sign defined around \eqref{pdef}. 
\begin{equation}
\sum_{R_i} \ee^{-t_i \ell(R_i)/2} (-1)^{(n_i+1)(\ell(R_i)\pm c(R_i))/2} q^{-(n_i+1)\kappa_{R_i}/4
-\kappa_{R_j}/4-\kappa_{R_k}/4} C_{R_j R_k R_i}\,.
\end{equation}
However, as alluded to above, it is necessary to allow for the replacement of $c(R_i)$ with 
$c(R_i^t)$ in this expression. The choice between the two options depends on a global
consistency condition that we explain in subsection \ref{elusive} below. We will corroborate
this rule at hand of examples in section \ref{examples}.

Finally, we discuss fixed legs of type 3. In that case, the invariant configurations
have $R_j=R_k^t$, and $R_j'={R_k'}^t$. To find the squareroot of the resulting 
expression, we have to delve deeper into the structure of the topological vertex
\begin{equation}
\eqlabel{delve}
C_{R_jR_kR_i} = q^{(\kappa_{R_i}+\kappa_{R_j})/2} \sum_{Q_j,Q_k,Q}
N_{Q Q_k^t}^{R_k} N_{Q Q_j}^{R_j^t} \frac{W_{R_i^t Q_k^t}W_{R_i Q_j}}
{W_{R_i}}\,.
\end{equation}
We also recall that the real vertex was introduced in \cite{krwa} as the squareroot of this 
expression for $R_k=R_j^t$, when the summand is almost a perfect square. As anticipated in the 
previous subsection, we deal with the non-trivial fixed leg, $R_i=R_i^t$, by simply taking the 
squareroot of the lone $W_{R_i}$ in the denominator,
\begin{equation}
\eqlabel{straight}
C_{R_j R_i}^{\rm real} = q^{\kappa_{R_j}/4} \sum_{Q_j, Q} N_{Q Q_j}^{R_j^t} 
\frac{W_{R_i Q_j}}{\sqrt{W_{R_i}}}\,,
\end{equation}
and noting that the squareroots of the two real vertices recombine to give a rational function of $q$ 
for the full contribution of a fixed leg of type 3.

Somewhat similarly to fixed legs of type 2, we have found that it is necessary to include an additional 
sign degree of freedom in the real vertex. Together with \eqref{straight}, which we'll 
call the ``straight'' real vertex, we define the ``twisted'' real vertex as
\begin{equation}
\eqlabel{twisted}
\tilde C_{R_j R_i}^{\rm real} = q^{\kappa_{R_j}/4} \sum_{Q_j, Q} (-1)^{\ell(Q)} N_{Q Q_j}^{R_j^t} 
\frac{W_{R_i Q_j}}{\sqrt{W_{R_i}}}\,.
\end{equation}
According to the global sign rule which we explain below, the full contribution of a fixed leg of type
3 is then given by
\begin{equation}
\eqlabel{fullcont}
\sum_{R_i=R_i^t} C_{R_j R_i}^{\rm real} \ee^{-t_i\ell(R_i)/2}
(-1)^{(\ell(R_i)\mp r(R_i))/2} \tilde C_{R_j' R_i}^{\rm real} \,.
\end{equation}

For later reference, note that by using the expression of $W$ in terms of Schur functions 
\begin{equation}
W_{R_iQ_j}=s_{R_i}(q^\rho)s_{Q_j}(q^{R_i+\rho})\,,
\end{equation}
with $q^{\rho+R_j}=(q^{R_j^1-1/2},q^{R_j^2-3/2},\dots)$, where $R^i_j$ denotes the $i$-th part of 
the partition $R_j$, and the expression $s_{\lambda/\mu}(x)=\sum_\nu 
N^\lambda_{\mu\nu}s_\nu(x)$ for skew Schur functions, the untwisted \req{straight} and twisted 
\req{twisted} real vertex can be expressed in terms of Schur functions as
\begin{equation}
\eqlabel{rvertexschur}
 C^{\rm real}_{R_j R_i }=q^{\kappa_{R_j}/4}\sqrt{s_{R_i}(q^\rho)}\sum_{Q} s_{R_j^t/Q}(q^{\rho+R_i})\,,
\end{equation}
and
\beq
\eqlabel{Trvertexschur}
\tilde C^{\rm real}_{R_j R_i }=q^{\kappa_{R_j}/4}\sqrt{s_{R_i}(q^\rho)}
\sum_{Q} (-1)^{\ell(Q)} s_{R_j^t/Q}(q^{\rho+R_i})\,.
\eq
respectively. Note that \req{rvertexschur} is, as expected, just a termwise squareroot of the full 
topological vertex expressed in Schur functions (\cf, \cite{akmv}).

\subsection{A global sign rule}
\label{elusive}

To complete the real vertex formalism, we need to specify how to correlate the choice between 
$c(R)$ and $c(R^t)$ on fixed legs of type 2 with the choice between straight and twisted real 
vertex at the ends of fixed legs of type 3. As is not uncommon for orientifolds, this is a non-local 
issue, \ie, not every combination is globally consistent. Note that the question only arises when 
the involution changes the orientation of the plane (otherwise all fixed legs are of type 1). 
The fixed line of this involution then divides the $r_\alpha$-$r_\beta$-plane of the toric diagram
in two parts, which we choose to call ``above the fixed line'' and ``below the fixed line'', 
respectively. This then also determines a left-right orientation of the fixed line itself.

In the topological vertex, one has to take account of the cyclic ordering of the outgoing legs. We 
work in conventions in which $C_{R_0 R_1 R_2}$ is the amplitude when the representations 
$R_0$, $R_1$, and $R_2$ appear in clockwise order. (See figure \ref{c3fig}.) The real vertex is
the squareroot of this amplitude when $R_0=R_0^t$ and $R_2=R_1^t$, and we label it by 
$R_0$ and $R_1$, which is the representation following the fixed leg in clockwise direction. 
The choice between straight \eqref{straight} and twisted \eqref{twisted} real vertex now
depends on the orientation of the fixed vertex relative to the chosen orientation of the
$r_\alpha$-$r_\beta$-plane: When the fixed leg points to the left, we use $C^{\rm real}_{R_1 R_0}$,
and when it points to the right, we use $\tilde C^{\rm real}_{R_1 R_0}$. Note that this implies
that in the former case, the leg carrying $R_1$ lies above the fixed line, while in the latter
case, it lies below the fixed line.

For fixed legs of type 2, we use $c(R)$ when the leg crosses the fixed line from above, and
we use $c(R^t)$ when it crosses from below.

We emphasize that in this prescription, every word counts. Once the orientations are fixed, we 
could make one global choice:\footnote{There are geometries for which
accidentatlly more choices are possible, but the rule as we have stated it is the only globally
consistent one.} Whether to use $C^{\rm real}$ or $\tilde C^{\rm real}$ for fixed legs pointing 
left. The opposite choice (\ie, $\tilde C^{\rm real}$ for fixed legs pointing left, $C^{\rm real}$
for those pointing right, $c(R^t)$ for legs crossing from above, and $c(R)$ for those crossing
from below) also gives a consistent theory, although the sign of some (real Gopakumar-Vafa) 
invariants might change. (The choice of sign in front of $r(R)$ and $c(R)$ also changes the sign 
of these invariants, but as emphasized before, this can be absorbed in the discrete Wilson lines.)

\begin{figure}
\begin{center}
\psfrag{R1}[cc][][0.75]{$R_1$}
\psfrag{R2}[cc][][0.75]{$R_2$}
\psfrag{R3}[cc][][0.75]{$R_3=R_1$}
\epsfig{scale=0.5,file=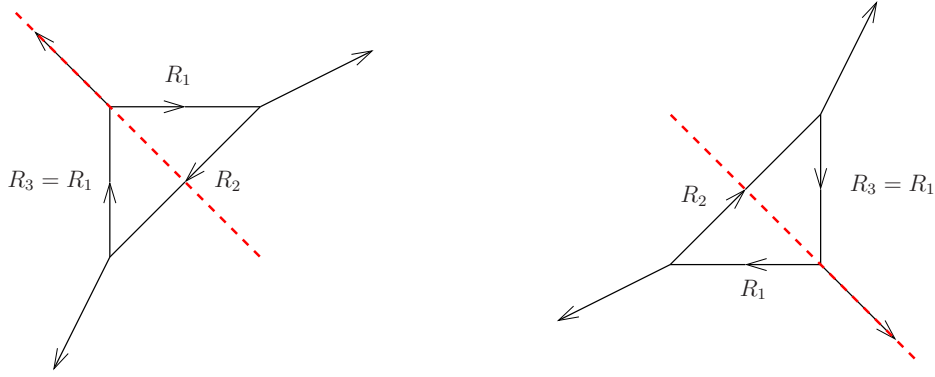}
\caption{Two web diagrams of local $\P^2$ with involution with different orientation for
purposes of the real vertex.}
\label{P2fig}
\end{center}
\end{figure}

To illustrate the rule, we consider local $\P^2$ in the real vertex formalism, see figure \ref{P2fig}.
Evaluation of the rule above gives for the diagram on the left,
\begin{equation}
\eqlabel{left}
\sum_{R_1 R_2} 
(-1)^{\ell(R_1)+(\ell(R_2)-c(R_2))/2}
q^{5\kappa_{R_1}/4+\kappa_{R_2}/4}
\ee^{-t(\ell(R_1)+\ell(R_2)/2)} 
C^{\rm real}_{R_1\cdot} C_{R_1^t \cdot R_2} \,,
\end{equation}
and for that on the right
\begin{equation}
\eqlabel{right}
\sum_{R_1 R_2} 
(-1)^{\ell(R_1)+(\ell(R_2)-c(R_2^t))/2}
q^{5\kappa_{R_1}/4+\kappa_{R_2}/4}
\ee^{-t(\ell(R_1)+\ell(R_2)/2)} 
\tilde C^{\rm real}_{R_1\cdot} C_{R_1^t \cdot R_2} \,.
\end{equation}
One may check that \eqref{left} and \eqref{right} give the same result for the real topological
string partition function of local $\P^2$.

\section{Chern-Simons}
\label{chernsimons}

The ordinary topologically vertex was originally derived in \cite{akmv} by exploiting the
duality between topological strings on resolved conifold with external branes and Chern-Simons 
theory with Wilson loop insertions. The purpose of the present section is to give a
similar derivation of the real vertex, which we have obtained in the previous section as the 
squareroot of the ordinary vertex. Let us first give a brief review of the derivation of
\cite{akmv}. 

\subsection{Review of ordinary vertex}

\begin{figure}
\psfrag{a}[cc][][0.7]{$O^+_V(U_2,V_2)$}
\psfrag{b}[cc][][0.7]{$O^-_V(U_1,\hat V_1)$}
\psfrag{c}[cc][][0.7]{$O^-_V(V_1, V_3)$}
\psfrag{d}[cc][][0.7]{$O^-_V(U_3, V_3)$}
\psfrag{t}[cc][][0.7]{$t\to \infty$}
\psfrag{l1}[cc][][0.7]{$L_1$}
\psfrag{l2}[cc][][0.7]{$L_2$}
\psfrag{l3}[cc][][0.7]{$L_3$}
\psfrag{la1}[cc][][0.7]{$L_1$}
\psfrag{la2}[cc][][0.7]{$L_2$}
\psfrag{la3}[cc][][0.7]{$L_3$}
\psfrag{lb1}[cc][][0.7]{$L_1$}
\psfrag{lb2}[cc][][0.7]{$L_2$}
\psfrag{lb3}[cc][][0.7]{$L_3$}
\begin{center}
\includegraphics[scale=0.8]{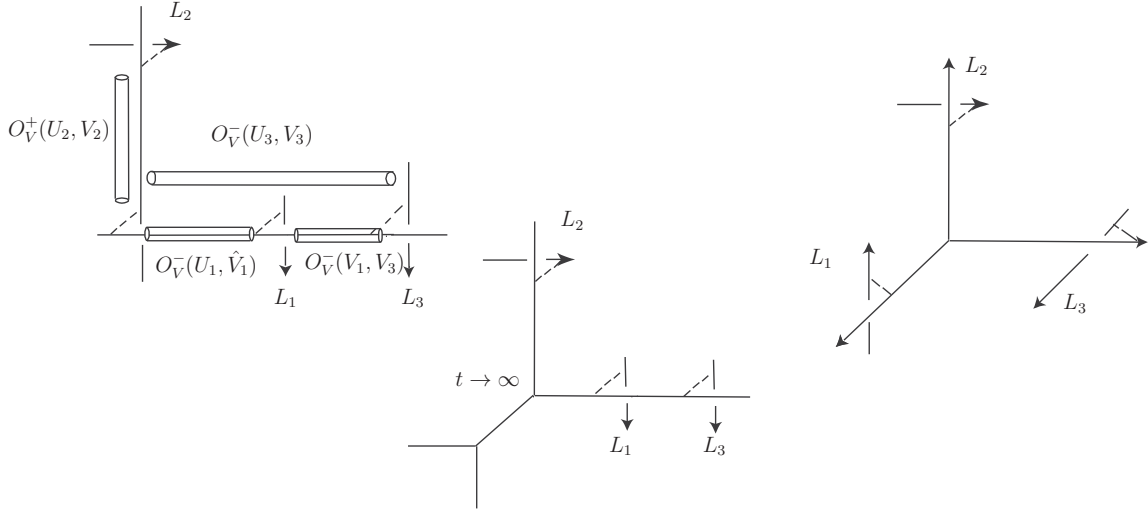}
\caption{The D-brane configuration used by \cite{akmv} for the derivation of the topological vertex.}
\label{relevant}
\end{center}
\end{figure}

The derivation of the topological vertex begins with the conifold geometry with three toric 
Lagrangian branes inserted as probes, in a configuration sketched in figure \ref{relevant}.

This amplitude is evaluated in the Chern-Simons description, as we will review below. The
vertex itself, which corresponds to the brane configuration in $\complex^3$ depicted in figure 
\ref{relevant} on the right, is obtained by first taking the K\"ahler parameter of the conifold
$t\to\infty$, to obtain the configuration in the middle of figure \ref{relevant}, and then
moving the Lagrangian brane $L_1$ to the third outgoing edge and adjusting framing.

The role of the probe branes in the topological vertex formalism is to ``cut'' the toric diagram 
of a toric Calabi-Yau threefold into $\complex^3$-patches. This cutting procedure requires for 
each brane the choice of a framing, depicted in figure \ref{relevant} by arrows
(as well as an additional subtlety, recorded below).

The probe branes couple to the Chern-Simons theory via annulus worldsheets, also depicted in 
figure \ref{relevant}. The net effect of these annuli is the insertion of a certain Wilson line,
known as the Ooguri-Vafa operator, into the Chern-Simons path integral. More precisely, the Ooguri-Vafa 
operator comes in two guises, depending on whether the branes connected by the annulus lie on 
opposite or the same sides of the corresponding lines of the toric diagram:
\def\OV{O_V}
\begin{equation}
\eqlabel{ov}
\begin{split}
\OV^+(U,V) &= \sum_{Q} \tr_Q U \ee^{-r \ell(Q)} \tr_Q V\,, \\
\OV^-(U,V) &=\sum_Q \tr_{Q^t} U \ee^{-r\ell(Q)} (-1)^{\ell(Q)} \tr_Q V \,.
\end{split}
\end{equation}
Here, $U$, and $V$ are the holonomies of the D-brane gauge field around the two ends of the 
cylinder, $r$ is its length, and the sums are over all $U(\infty)$ representations. In the 
above configuration, we have the four Ooguri-Vafa operators,
\begin{equation}
\eqlabel{ovs}
\OV^-(U_1,\widehat{V_1})\,,\qquad
\OV^+(U_2,V_2)\,,\qquad
\OV^-(U_3,V_3)\,,\qquad
\OV^-(V_1,V_3)\,.
\end{equation}
On the dynamical branes the Wilson lines   $U_1$ and $U_3$ are parallel and
  form with   $U_2$  a double Hopf link whose normalized Chern-Simons 
expectation value is given by
\begin{equation}
\eqlabel{gloss}
\frac{\langle \tr_{Q_1^t} U_1 \tr_{Q_3^t} U_1 \tr_{Q_2} U_2\rangle}{\langle \cdot\rangle} 
= \frac{\calw_{Q_1^t Q_2}\calw_{Q_3^t Q_2}}{\calw_{Q_2}}\,,
\end{equation}
where $\calw_{R_1 R_2}= S_{R_1 R_2}/S_{00}$ is the Hopf link invariant  and $S_{R_1R_2}$ is
the modular S-matrix of the relevant WZW-model. \footnote{We are here glossing over the
orientations of the relevant link components.}
We also  recall the relations
\begin{equation}
\eqlabel{justso}
g_s = \frac{2\pi\ii}{k+N} \,,\qquad
t = N g_s\,,
\end{equation}
between topological string coupling $g_s$, K\"ahler parameter $t$, and Chern-Simons parameters 
rank, $N$, and level, $k$. 
In the limit of interest, $t\to \infty$, one defines
\begin{equation}
W_{R_1R_2} = \lim_{t\to\infty} \ee^{-\frac{\ell(R_1)+\ell(R_2)}2 t}
\calw_{R_1R_2}\,,
\end{equation}
and this yields for the configuration in the middle of figure \ref{relevant}: \footnote{The 
requisite renormalization of the Ooguri-Vafa operators can again be absorbed  in the $V_i$.
We have also  suppressed the lengths of the cylinders, since they can be absorbed in the
complexified Wilson line $V_1,V_2,V_3$ of the non-dynamical branes.}
\begin{equation}
\sum_{Q_1, Q_2,Q_3,Q} (-1)^{\ell(Q_1)+\ell(Q_2)+\ell(Q_3)}
\frac{W_{Q_1^t Q_2} W_{Q_3^t Q_2}}{W_{Q_2}} \tr_{Q_1}\widehat {V_1}
\tr_{Q^t} V_1\tr_{Q_2}V_2\tr_{Q}V_3\tr_{Q_3}V_3\,.
\end{equation}
By considering the simplified situation with only branes $L_1$ and $L_2$ present, and exploiting
symmetries of the vertex, one finds that moving $L_1$ to the empty leg amounts
to replacing $(-1)^{\ell(Q_1)} W_{Q_1^tQ_2}\tr_{Q_1}\widehat{V_1} \tr_{Q_2}V_2$
with $(-1)^{\ell(Q_2)} q^{\kappa_{Q_2}/2} W_{Q_2^t Q_1} \tr_{Q_1}V_1\tr_{Q_2}V_2$ in this expression. 
Upon fusing the representations with the same $V_i$, (\eg, $\tr_{Q}V_3\tr_{Q_3}V_3 = \sum_{R_3} 
N_{Q Q_3}^{R_3} \tr_{R_3}V_3$), and absorbing some innocuous signs into the $V_i$, we transform to
the representation basis by extracting the coefficient of $\tr_{R_i}V_i$ for three abritrarily
chosen representations $R_1,R_2,R_3$. Adjusting the framing, this yields \eqref{delve}:
\begin{equation}
C_{R_1R_2R_3} = q^{\frac{\kappa_{R_2}+\kappa_{R_3}}2}
\sum_{Q_1,Q_3,Q} N_{QQ_1}^{R_1} N_{Q Q_3}^{R_3^t}
\frac{W_{R_2^t Q_1}W_{R_2 Q_3}}{W_{R_2}}\,.
\end{equation}

Before proceeding, we wish to make an important observation about one arbitrariness in the 
above derivation. In principle, we could have derived a similar amplitude with some of the 
probe branes {\it above} instead of below the plane of the toric diagram. This would change 
some of the Ooguri-Vafa operators in \eqref{ovs} and result in a slightly different final expression.
(Most significantly, a sign $(-1)^{\ell(Q)}$.) Of course, one would then have to match these 
choices when gluing vertices back together. In the usual formalism, this is clearly best 
accomplished by putting all probe branes below the plane, as is usually done. The context of 
the real vertex, however, imposes a different constraint on the probe branes, as we shall see 
below.

\subsection{Ho\texorpdfstring{\v r}{r}ava operators}

In the following, we will employ the same strategy we used in the previous subsection, 
to derive  the real topological vertex. Note first that involution 1 of the conifold requires 
on the gauge theory side the computation of ${\it SO}(N)/{\it Sp}(N)$ Chern-Simons amplitudes 
on $S^3$, or ${\it SU}(N)$ amplitudes on $\RP^3$, depending on whether we choose to deform with 
$\mu$ positive or negative, cmp.\ subsection \ref{geotrans}. The ${\it SO/Sp}$ picture was 
used extensively in \cite{bouchard1}, and we have little reason to reconsider it here. The 
${\it SU}(N)$ on $\RP^3$ picture has not so far been developed. But note that since the 
topological amplitudes do not depend on the deformation parameter, ${\it SU}(N)$ on $\RP^3$ 
will give exactly the same answer as ${\it SO/Sp}(N)$ on $S^3$. In particular, the discrete 
Wilson line around the non-trivial one-cycle of $\RP^3$ is identified wih the distinction 
between orthogonal and symplectic gauge group.

For involutions 2 and 3, we require the computation of Chern-Simons amplitudes on the three-orbifold
$S^3/\zet_2$, where the $\zet_2$ acts with an $S^1$ fixed locus. \footnote{If $S^3\cong
\{\sum x_i^2=1\}\subset\reals^4$, the relevant involution is $(x_1,x_2,x_3,x_4)\to
(x_1,x_2,-x_3,-x_4)$ .}
Chern-Simons theory in the presence of precisely such $\zet_2$ orbifold singularities was 
studied a good while ago by Ho\v rava \cite{horava}. The main result of \cite{horava} that we 
can use is the equivalence of Chern-Simons theory on orbifolds with
Chern-Simons theory on smooth manifolds with additional Wilson line insertions. Let's recall 
the basic idea.

In general, of course, field theories on orbifolds are ill defined because of the violation
of unitarity at the orbifold singularities. As argued in \cite{horava}, however, the absence 
of local excitations eliminates this dificulty in topological field theories such as 
Chern-Simons theory. In fact, this must be so because Chern-Simons theory is equivalent to an 
open string theory \cite{wittencs}, and we know that string theory on orbifolds is perfectly 
well defined.

For Chern-Simons theory, the most interesting orbifold singularities are those in codimension 
2. Topologically, we can construct a smooth three-manifold $X$ by cutting out a tubular 
neighborhood of the singular locus of the orbifold and replacing it with a collection of solid 
tori. (Locally, a connected component of the tubular neighborhood is $S^1\times D/\Gamma$, 
where $\Gamma$ is a 
discrete group acting on the unit disk $D$, fixing the origin. This can be smoothed in an obvious 
way.) It was argued in \cite{horava} that Chern-Simons theory on such an orbifold $O$ is equivalent
to Chern-Simons theory on $X$, obtained by replacing each component $C_\alpha$ of the singular locus 
with a collection of Wilson lines,
\begin{equation}
\eqlabel{hor}
H(C_\alpha) = \sum_{R} c_{R,\alpha} W_R(C_\alpha)\,,
\end{equation}
where $c_{R,\alpha}$ are some complex coefficients whose calculation is described in \cite{horava}. 
Namely,
\begin{equation}
\eqlabel{add}
\langle \Phi\rangle_{O} = \langle\Phi \prod_{\alpha} H(C_\alpha) \rangle_X\,,
\end{equation}
where $\Phi$ is any additional observable. There are several subtleties in the formulas \eqref{hor} 
and \eqref{add}. We mention just a few. For one, the relation between the set of observables on the
theory on $X$ to the observables of the orbifold theory is not obvious. This is related to 
the statement that the gauge group of the theory on $X$ might differ from the gauge group on $O$ by 
some discrete factor. Moreover, the Wilson lines appearing in the expansion \eqref{hor} need to 
be appropriately framed. We shall refer to $H(C_\alpha)$ as the Ho\v rava operator.

In the problem of our interest, $O=S^3/\zet_2$ with singular locus $S^1$, it is not hard to 
see that $X$ is again a copy of $S^3$ (albeit of half the size of the original one). We will not
compute the Ho\v rava operator \eqref{hor} from first principles, but instead use the results
for the vacuum amplitudes from subsection \ref{coninv}.

We begin with involution 3 of the conifold.\footnote{At first sight, it appears surprising
that we would need two different Ho\v rava operators, given that involutions 2 and 3 are
related by a flop of the resolved conifold, and indistinguishable on the deformed conifold.
A similar puzzle arises, without orientifold, on the conifold with probe branes on the external 
legs. The toric picture shows that the distinction between the two situation is the channel in
which we choose to factorize.} Referring back to \eqref{weobtain}, and using that $C_{\cdot \cdot R}=
W_R=\ee^{-t\ell(R)/2}\calw_R$, $\calw_R=\langle \tr_R U\rangle_{S^3}/\langle\cdot \rangle_{S^3}$, 
we see that we can write
\begin{equation}
\begin{split}
Z^3_{\rm con.} &= \langle \cdot\rangle_{S^3/\zet_2} \\
&= \frac 1{S_{00}} \sum_{R=R^t} (-1)^{(\ell(R)-r(R))/2} \ee^{-t\ell(R)} 
\langle \tr_R U\rangle_{S^3}\,.
\end{split}
\end{equation}
From this we deduce that the Ho\v rava operator is given by,
\begin{equation}
\eqlabel{H3}
H_3 = \frac 1{S_{00}} \sum_{R=R^t} (-1)^{(\ell(R)-r(R))/2} \ee^{-t\ell(R)}\tr_R U\,,
\end{equation}
with canonically framed Wilson lines $U=P\exp \oint A$ along the fixed $S^1$.

For involution 2, we consider \eqref{unable}, and identify the factor $q^{-\kappa(R)/4}$ as a 
fractional framing of the Wilson line. Thus, we can write
\begin{equation}
Z^2_{\rm con.} = \frac{1}{S_{00}} \sum_R (-1)^{(\ell(R)-c(R))/2} \ee^{-t\ell(R)} 
\langle \tr_R U_{1/2}\rangle_{S^3}\, ,
\end{equation}
where the subscript indicates the fractional framing of the Wilson line. This gives the 
Ho\v rava operator
\begin{equation}
\eqlabel{H2}
H_2 = \frac{1}{S_{00}}\sum_R (-1)^{(\ell(R)-c(R))/2}\ee^{-t\ell(R)} 
\tr_R U_{1/2}\,.
\end{equation}

\subsection{Derivation of the real vertex}

\begin{figure}
\psfrag{a}[cc][][0.7]{$O^+_V(U_2,V_2)$}
\psfrag{b}[cc][][0.7]{$O^+_V(U_1,V_1)$}
\psfrag{c}[cc][][0.7]{$O^+_V(V_1,V_2)$}
\psfrag{l1}[cc][][0.7]{$L_1$}
\psfrag{l2}[cc][][0.7]{$L_2$}
\psfrag{t}[cc][][0.7]{$t\to\infty$}
\begin{center}
\includegraphics[scale=0.7]{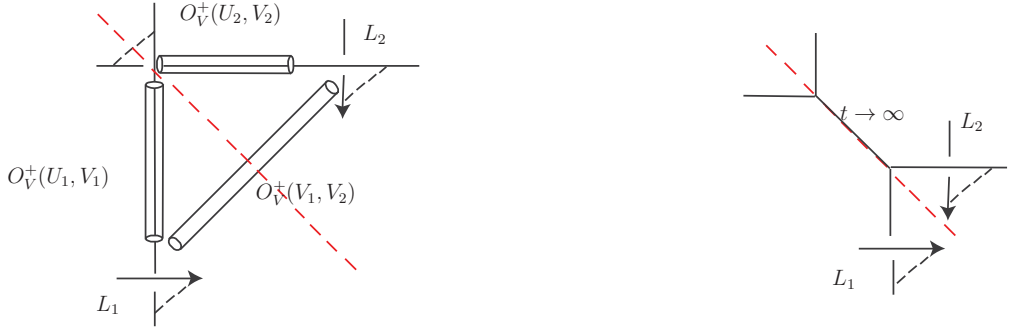}
\caption{The D-brane configuration relevant for the derivation of the real topological vertex.}
\label{notrelevant}
\end{center}
\end{figure}

With Ho\v rava operators in hand, we now proceed with the derivation of the full real vertex.
We consider the D-brane configuration depicted in figure \ref{notrelevant}.

Note that since any A-type involution acts by complex conjugation, and the direction perpendicular 
to the $r_\alpha$-$r_\beta$ plane of the toric diagram is given by $r_\gamma=\Im(\prod z_i)$ (see 
\eqref{moment}), probe branes switch the side under the involution. One way to obtain an invariant 
brane configuration is to insert $L_1$ above the plane, and $L_2$ below the plane. The other option 
will be considered below. Before taking account of the involution, we have the following Ooguri-Vafa 
operators:
\begin{equation}
\eqlabel{change}
\OV^+ (U_1,V_1) \,, \qquad
\OV^+ (U_2,V_2) \,,\qquad
\OV^+ (V_1,V_2) \,,
\end{equation}
where the third operator accounts for the annulus stretching between $L_1$ and $L_2$.

We now wish to quotient this configuration by the A-type involution exchanging $L_1$ and $L_2$. 
This first of all leads to the restriction $V_1=V_2$. Moreover, we have to replace the third
Ooguri-Vafa operator above, which came from integrating out the annulus, with the corresponding 
operator for a M\"obius strip. This operator is given by
\begin{equation}
\sum_{Q} \tr_{Q} V_1 \,,
\end{equation}
so we obtain the amplitude
\begin{equation}
Z^3_{\rm con.}(V_1)=\frac{1}{S_{00}} \sum_{Q, Q_1} \langle \tr_{Q_1} U_1 \rangle_{S^3/\zet_2} 
\ee^{-t \ell(Q_1)} \tr_{Q_1} V_1 \tr_{Q}V_1\,.
\end{equation}
Applying the rule \eqref{add} with Ho\v rava operator given by \eqref{H3}, this becomes
\begin{equation}
\sum_{\topa{R=R^t}{Q,Q_1}} \calw_{Q_1 R} \ee^{-t(\ell(R)+\ell(Q_1))} (-1)^{(\ell(R)-r(R))/2}
\tr_{Q_1} V_1\tr_Q V_1\,.
\end{equation}
In the limit $t\to\infty$,  transforming to the representation basis
and fusing the representations $Q$ and $Q_1$, this gives the real vertex 
with trivial fixed leg:
\begin{equation}
C^{\rm real}_{R_1 \cdot} = q^{\kappa_{R_1}/4} \sum_{Q Q_1} N_{QQ_1}^{R_1} W_{Q_1}\,,
\end{equation}
where the prefactor $q^{\kappa_{R_1}/4}$ adjusts to the canonical framing.

\begin{figure}
\psfrag{a}[cc][][0.7]{$O^-_V(U_2,V_2)$}
\psfrag{b}[cc][][0.7]{$O^-_V(U_1, V_1)$}
\psfrag{c}[cc][][0.7]{$O^+_V(V_1, V_2)$}
\psfrag{l1}[cc][][0.7]{$L_1$}
\psfrag{l2}[cc][][0.7]{$L_2$}
\psfrag{la1}[cc][][0.7]{$L_1$}
\psfrag{lb2}[cc][][0.7]{$L_2$}
\psfrag{t}[cc][][0.7]{$t\to\infty$}
\begin{center}
\includegraphics[scale=0.7]{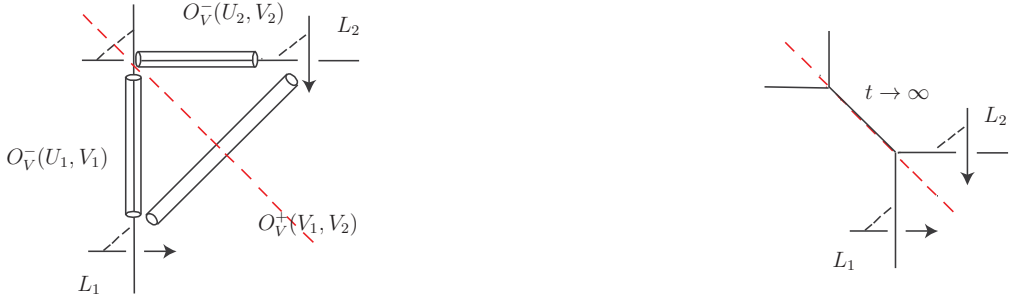}
\caption{Derivation of the twisted real vertex.}
\label{twrelevant}
\end{center}
\end{figure}

Now let us consider the situation in which $L_1$ is inserted below the plane, and $L_2$ above  the
plane, as depicted in figure \ref{twrelevant},
This changes the Ooguri-Vafa operators \eqref{change} into
\begin{equation}
\OV^- (U_1,V_1) \,, \qquad
\OV^- (U_2,V_2) \,,\qquad
\OV^+ (V_1,V_2) \,.
\end{equation}
Repeating the same steps as above, we obtain \footnote{
We use  that $(-1)^{(\ell(Q_1)+\ell(R_1))} N_{QQ_1}^{R_1}=(-1)^{\ell(Q)} N_{QQ_1}^{R_1}$.}
\begin{equation}
\tilde C^{\rm real}_{R_1 \cdot} = q^{\kappa_{R_1}/4}
\sum_{Q Q_1} (-1)^{\ell(Q)} N_{QQ_1}^{R_1} W_{Q_1}\,,
\end{equation}
which, because of the extra sign, we named  the ``twisted'' real vertex in equation  
\eqref{twisted}.

What about the real vertex with non-trivial representation on the fixed leg? We have
not been able to identify a brane configuration that would allow for its direct computation
from Chern-Simons theory. However, as we have remarked in subsection \ref{generalcase},
we never need this amplitude in practice as long as we exclude external branes on fixed 
legs. Instead, we derive the amplitude \eqref{fullcont} for the (orientifold of) conifold 
with branes on all external legs. All amplitudes can be built on that together with the 
ordinary topological vertex.

\begin{figure}
\psfrag{a}[cc][][0.7]{$O^+_V(U_2,V_2)$}
\psfrag{b}[cc][][0.7]{$O^+_V(U_1, V_1)$}
\psfrag{c}[cc][][0.7]{$O^+_V(V_1, V_2)$}
\psfrag{d}[cc][][0.7]{$O^-_V(U_3,V_3)$}
\psfrag{e}[cc][][0.7]{$O^-_V(U_4, V_4)$}
\psfrag{f}[cc][][0.7]{$O^+_V(V_3, V_4)$}
\psfrag{l1}[cc][][0.7]{$L_1$}
\psfrag{l2}[cc][][0.7]{$L_2$}
\psfrag{l3}[cc][][0.7]{$L_3$}
\psfrag{l4}[cc][][0.7]{$L_4$}
\begin{center}
\includegraphics[scale=0.8]{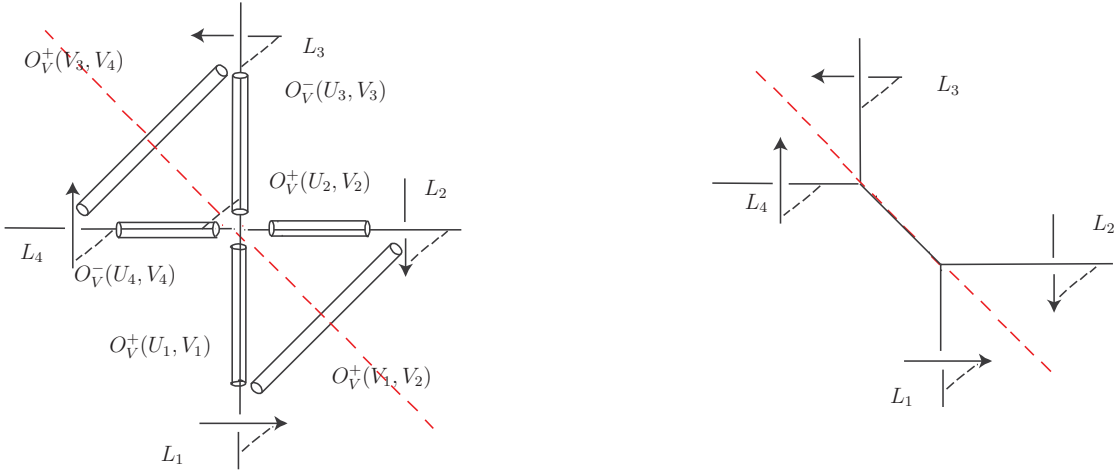}
\caption{Derivation of the real conifold vertex.}
\label{fullytwisted}
\end{center}
\end{figure}

The relevant brane configuration is shown in figure \ref{fullytwisted}. Note that we have
inserted branes $L_1$ and $L_4$ above the plane and $L_2$ and $L_3$ below the plane.
This is the correct choice compatible with cutting the toric diagram into pieces ``from
above'' on one side of the fixed line, and ``from below'' on the other side of the
fixed line, as required by orientifold invariance.

Applying the by now familiar procedure, we obtain the following expression for this
amplitude in the representation basis
\begin{equation}
\begin{split}
&Z^3_{\rm con.}(R_1, R_3)= \\
&=\sum \ee^{-(\ell(R)+\ell(Q_1)+\ell(Q_3)) t}  N^{R_1}_{Q Q_1}  N^{R_3^t}_{\tilde Q Q_3}
(-1)^{\ell( \tilde Q)} 
(-1)^{(\ell(R)-r(R))/2}      \langle \tr_R U \tr_{Q_3}U_3 \tr_{Q_1}U_1\rangle_{S_3}\\
&=
\sum  \ee^{-\ell(R) t/2}  N^{R_1}_{Q Q_1}  N^{R_3^t}_{\tilde Q Q_3}
(-1)^{\ell(R_3)+\ell(\tilde Q)} 
(-1)^{(\ell(R)-r(R))/2} \frac{W_{Q_1 R } W_{Q_3 R}}{W_R}\\
&=
 q^{\frac{\kappa_{R_1}+\kappa_{R_3}}4}\sum_{ R=R^t}C^{{\rm real}}_{R_1 R}
(-1)^{(\ell(R)-r(R))/2} \ee^{-t\ell(R)/2}
\tilde C^{{\rm real}}_{R_3^t R}\,.
\end{split}
\end{equation}
This result is compatible with the real vertex formalism as described in section \ref{formalism},
in particular the sign rules discussed in subsection \ref{elusive}.

\begin{figure}
\psfrag{a}[cc][][0.7]{$O^+_V(U_2,V_2)$}
\psfrag{b}[cc][][0.7]{$O^+_V(U_1, V_1)$}
\psfrag{c}[cc][][0.7]{$O^+_V(V_1, V_2)$}
\psfrag{l1}[cc][][0.7]{$L_1$}
\psfrag{l2}[cc][][0.7]{$L_2$}
\begin{center}
\includegraphics[scale=0.8]{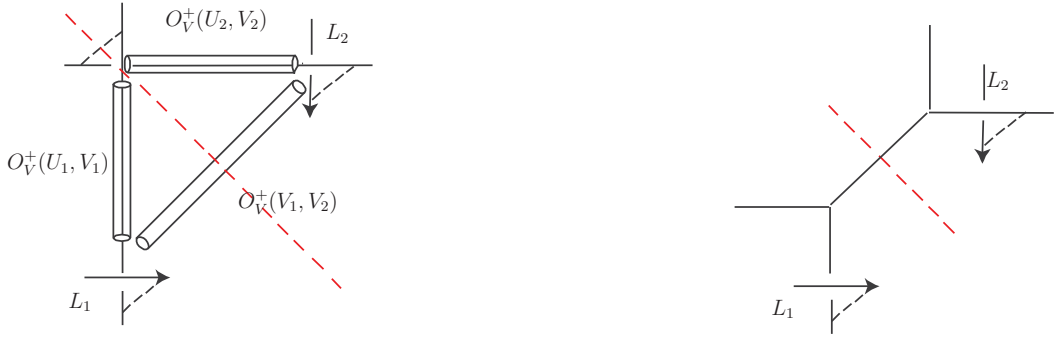}
\caption{Involution 2 with external branes.}
\label{inv2}
\end{center}
\end{figure}

Finally, we comment on involution 2 with branes on the external legs of the conifold, see figure \ref{inv2}.
Using the Ho\v rava operator \eqref{H2}, it is straightforward to derive the following amplitude
for this configuration
\begin{equation}
\label{con2}
Z^2_{\rm con.}(R_1) = \sum_{R} W_{R_1 R} \ee^{-t\ell(R)/2} q^{-\frac{\kappa_R+\kappa_{R_1}}4} 
(-1)^{(\ell(R)-c(R))/2} \,.
\end{equation}
Again, this is compatible with the formalism of section \ref{formalism}. The main interest
of the formula is the factor $q^{-\kappa_{R_1}/4}$, which again comes from the fractional framing 
of the Wilson loop insertion in Chern-Simons theory.

\section{Melting crystal}
\label{melting}

\begin{figure}
\begin{center}
\psfrag{x}[cc][][1]{$x_1$}
\psfrag{y}[cc][][1]{$x_2$}
\psfrag{z}[cc][][1]{$x_3$}
\epsfig{scale=0.3,file=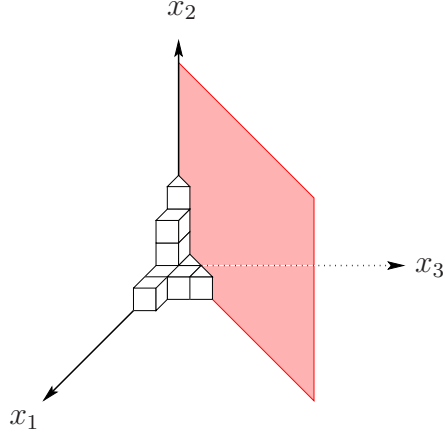}
\end{center}
\caption{Three-dimensional partition with fixed asymptotics along the three axes,
and symmetry plane at $x_1=x_3$.}
\label{closedmelting}
\end{figure}

Soon after its discovery, the topological vertex was related to a statistical mechanics
model of a melting crystal corner \cite{crystal}. The relation can be expressed in the formula
\begin{equation}
\eqlabel{expressed}
C_{R_1R_2R_3}(1/q) = q^{\frac{||R_1^t||^2+||R_2^t||^2+||R_3^t||^2}2}P_{R_1R_2R_3}(q){M(q)}^{-1}\,,
\end{equation}
where $C_{R_1R_2R_3}$ is the topological vertex, which in the context of the melting crystal is
conveniently written in terms of skew Schur functions
\begin{equation}\label{vertexschur}
C_{R_1R_2R_3}(q) = q^{\frac{\kappa_{R_3}+\kappa_{R_2}}2}
s_{R_2^t}(q^\rho) 
 \sum_{Q} s_{R_1/Q}(q^{\rho+R_2^t}) s_{R_3^t/Q}(q^{\rho+R_2})\,.
\end{equation}
On the right hand side of \eqref{expressed}, $P_{R_1R_2R_3}$ is the generating function 
counting three-dimensional partitions with fixed asymptotics along the three axes given
by the two-dimensional partitions $R_1, R_2, R_3$. The $q$-weight of each box in the
3d partition is 1 minus the number of 2d partitions containing that box. See figure \ref{closedmelting}.
Furthermore, $M(q)=\prod_{n=1}^\infty (1-q^n)^{-n}$ is the MacMahon function counting
3d partitions with trivial asymptotics, and the prefactor containing
\begin{equation}
||R_i||^2 = \sum_i {\lambda_i^2} \,,
\end{equation}
accounts for the adjustement of the framing and the gluing algorithm. 

In this section, we explain the interpretation of the real topological vertex, which we have 
developed in this paper, in the melting crystal picture. We will find that both straight
and twisted real vertex admit a melting crystal interpretation.

\subsection{Melting crystal interpretation of the real vertex}

The positive axes of figure \ref{closedmelting} can be essentially identified with the toric 
diagram of $\complex^3$. Referring back to the action of the anti-holomorphic involution
on this toric diagram (see figure \ref{c3fig} on page \pageref{c3fig}), it is then clear what
we have to do: We should count 3d partitions, with fixed asymptotics, which are invariant under 
the symmetry exchanging two of the axes. The 2d partitions constraining the asymptotics must then
satisfy the constraint $R_3=R_1^t$, $R_2=R_2^t$.

To prepare for the answer to expect, we record here the partition function of symmetric plane
partitions with empty asymptotics. This partition function was originally conjectured by 
MacMahon \cite{macmahon}; a proof was given by Macdonald \cite{macdonald} and Andrews \cite{andrews}, 
The formula is
\begin{equation}
\eqlabel{conjecture}
M^{\rm sym} (q) = \prod_{n=1}^\infty (1-q^{2n-1})^{-1} \prod_{n=1}^\infty (1-q^{2n})^{-[n/2]}\,.
\end{equation}
In this formula, symmetric partitions are weighted by the {\it total} number of boxes. From the 
orientifold point of view, it is more natural to count only the boxes on one side of the symmetry 
plane. This can be implemented by the transformation $q\to q^{1/2}$ in \eqref{conjecture}, and we
define the real MacMahon function by
\begin{equation}
\eqlabel{realmacmahon}
M^{\rm real} (q) = M^{\rm sym}(q^{1/2})\,.
\end{equation}

\begin{figure}
\begin{center}
\psfrag{A}[cc][][0.75]{$R$}
\psfrag{N}[cc][][1]{$N$}
\psfrag{t}[cc][][1]{$t=0$}
\psfrag{t2}[cc][][1]{$t=-(N-1)$}
\epsfig{scale=0.3,file=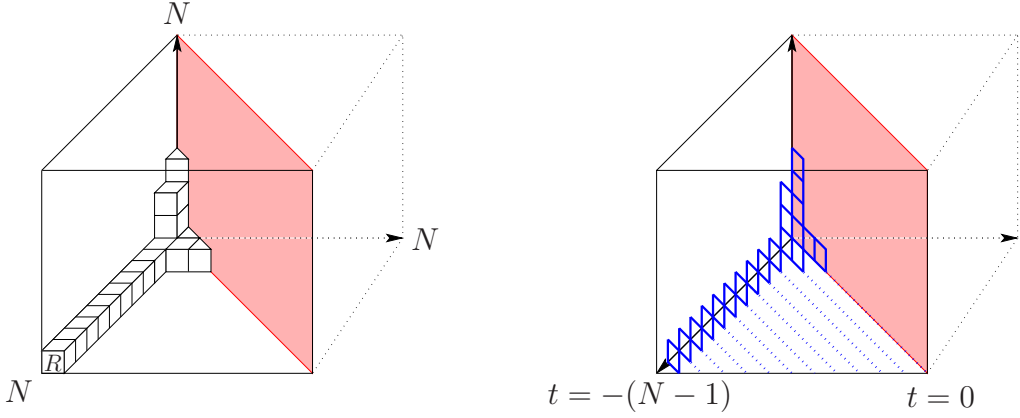}
\end{center}
\caption{Left: (Half of) a symmetrically melting crystal in a finite box of volume $N^3$ with fixed
boundary condition given by a 2d partition $R$. Right: Diagonal slicing of the setup.}
\label{symmetricmelting}
\end{figure}
Now let us study this in more detail. Following \cite{crystal}, we confine our symmetric 3d 
partitions into a box of finite size $N^3$, with boundary condition in the $(x_2,x_3)$-plane
at $x_1=N$ given by the 2d partition $R$. We assume trivial boundary condition in the $x_2$ direction.

We want to use the transfer matrix approach of \cite{crystal}, so we should slice the partition 
diagonally. This yields a stack of partitions $\nu(t)$, indexed by the coordinate $t$ of the 
corresponding slice. The $\nu(t)$ satisfy the interlacing condition
\begin{equation}
\eqlabel{intercond}
R=\nu(-N) \prec \nu(-N+1)\prec \dots \prec \nu(-1)  \prec\nu(0)\,,
\end{equation}
where two partitions $\mu$ and $\nu$ are said to interlace, $\mu\prec\nu$, if
\begin{equation}
\eqlabel{melteq3}
\mu_l\leq \nu_l\leq \cdots \leq \mu_2\leq\nu_2\leq\mu_1\leq\nu_1\,,
\end{equation}
holds, with $\mu_i$ and $\nu_i$ the $i$-th part of the partition, and $l$ is the number 
of parts. In distinction to arbitrary 3d partitions, the condition \eqref{intercond} is not
followed by a similar condition at $t>0$. Rather, the partitions at $t>0$ are determined by
those at $-t$, and the partition at $t=0$ is essentially arbitrary.

We continue to follow \cite{crystal}. To each partition $\mu$, we associate the state $\ket\mu$
in the Hilbert space of a complex fermion. Moreover, we introduce, via bosonization, the operator
$\Gamma_-(z)$ with the property
\begin{equation}
\Gamma_-(1) \ket\mu=\sum_{\nu\succ\mu} \ket\nu\,,
\end{equation}
where $\nu\succ\mu$ is equivalent to $\mu\prec\nu$.

It is then clear that we can obtain the partition function $P_{R\cdot}^{(N)}$ of the symmetric crystal in 
the cubic box of volume $N^3$ with fixed boundary $R$ at $x=N$ (or $t=-N$ in diagonal coordinates) 
via successive application of $\Gamma_-(1)$ operators on the state $\ket{R}$ and summing over all 
possible partitions $\lambda$ at $t=0$,
\begin{equation}
\eqlabel{product}
P_{R\cdot}^{(N)}(q)=q^{-\left(\!\topa{R}{2}\right) }
\sum_\lambda\biggl\langle\lambda \biggl|q^{L_0/2}\prod^{-1}_{t=-N+1}
\Gamma_-(1)q^{L_0}\biggr | R \biggr\rangle\,,
\end{equation}
where the factor of $q^{-\left(\topa{R}{2}\right)}$ with $\left(\topa{R}{2}\right):=\sum_i 
\left(\topa{R_i}{2}\right)$ accounts for the increase of boxes due to the diagonal slicing \cite{crystal}.
The extra factor $q^{L_0/2}$ counts half the number of boxes at $t=0$.

Using the commutation relation $\Gamma_-(z)=z^{-L_0}\Gamma_-(1)z^{L_0}$, we obtain
\begin{equation}
P_{R\cdot}^{(N)}(q)= q^{-\left(\topa{R}{2}\right)+(N-1/2)\ell(R) }\sum_\lambda
\biggl\langle\lambda\biggl|\prod^{N-2}_{i=0}\Ga_-(q^{-1/2-i})\biggr|R\biggr\rangle\,.
\end{equation}
Applying the fundamental identity (see for instance \cite{kac})
\begin{equation}
\prod_i\Gamma_-(x_i)\ket{\lambda}=\sum_\mu s_{\mu/\lambda}(x^{-1})\ket{\mu},
\end{equation}
we infer
\begin{equation}
P_{R\cdot}^{(N)}(q)=q^{-\left(\topa{R}{2}\right)+(N-1/2)\ell(R)}\sum_\mu s_{\mu/R}(q^{-\rho})   \,,
\end{equation}
with $q^\rho :=(q^{-1/2},q^{-3/2},\dots)$. Finally, invoking the Schur function identity (see 
for instance \cite{macdonald})
\begin{equation}
\sum_\mu s_{\mu/\lambda}(x)=\prod_i\frac{1}{1-x_i}\prod_{i<j}\frac{1}{1-x_ix_j}
\sum_\nu s_{\lambda/\nu}(x)\,,
\end{equation}
we obtain
\begin{equation}
P_{R\cdot}^{(N)}(q)= M_N^{\rm real} (q)\;
q^{-\left(\topa{R}{2}\right)+(N-1/2)\ell(R)}  \sum_\lambda s_{R/\lambda}(q^{-\rho})\,,
\end{equation}
where we have written the prefactor as
\begin{equation}
M_N^{\rm real}(q) = \prod_{i=1}^{N-1}\frac{1}{1-q^{-\rho_i}}\prod_{1\leq i<j}^{N-1}
\frac{1}{1-q^{-\rho_i-\rho_j}}=\prod_{n=1}^{N-1}\frac{1}{1-q^{n-1/2}}
\prod_{n=1}^{N-1}\frac{1}{(1-q^n)^{\lfloor n/2\rfloor}}\,.
\end{equation}
In the limit $N\to\infty$, we recover the real MacMahon function \eqref{realmacmahon} as the partition
function of the symmetrically melting crystal (up to $q\to q^{1/2}$) with empty boundary condition,
$R=\cdot~$.

For non-trivial representation $R$, we use the relation
\begin{equation}
\eqlabel{melteq7}
\frac{||R||^2}{2}:=\sum_i\frac{R_i^2}{2}=\left(\begin{matrix}R\\2\end{matrix}\right)+\ell(R)\,,
\end{equation}
to deduce (the rescaling in the limit is discussed and interpreted in detail in \cite{crystal})
\begin{equation}
\eqlabel{melteq8}
P_{R\cdot}(q):=\lim_{N\rightarrow\infty} q^{-N \ell(R)}P_{R\cdot}^{(N)}(q)=
M^{\rm real}(q) \; q^{-\frac{||R||^2}{2}}
\sum_\lambda s_{R/\lambda}(q^{-\rho})\,.
\end{equation}
This expression should be compared with the expression \eqref{rvertexschur} for the real topological 
vertex in terms of Schur functions. In the one-leg case, this simplifies to
\begin{equation}
\eqlabel{melteq9}
C^{{\rm real}}_{R \cdot}(q)=q^{\frac{||R||^2-||R^t||^2}{4}}\sum_\lambda s_{R^t/\lambda}(q^\rho)\,,
\end{equation}
where we used the relation $\kappa_R=||R||^2-||R^t||^2$.

The relation between \req{melteq8} and \req{melteq9} is explicitly
\begin{equation}
\eqlabel{melteq10}
C^{\rm real}_{R\cdot}(1/q)=q^{\frac{||R||^2+||R^t||^2}{4}} P_{R^t \cdot}(q) M^{\rm real}(q)^{-1}\,.
\end{equation}
Including the framing factors, this relation between the real vertex and symmetric crystal
melting fits beautifully into the usual relation between topological vertex and crystal melting 
expressed in \eqref{expressed}. (The seeming replacement of $R$ with $R^t$ on the right hand side
has to do with our choice of labelling the melting crystal representation with the asymptotics
along the $x_1$-axis, $R=R_1$. A choice that matches better the clockwise conventions of subsection 
\ref{elusive} is to use $R_3=R_1^t=R^t$.)

To give a similar interpretation of the twisted real vertex \eqref{twisted}, we weight each
representation in \eqref{product} by an additional minus sign $(-1)^{\ell(\lambda)}$. In other
words, we replace $q^{L_0/2}$ with $(-q)^{L_0/2}$. Repeating the steps above, we obtain
\beq
\tilde P_{R \cdot}(q)= (-1)^{\ell(R)} M^{\rm real}(1/q) \; q^{-\frac{||R||^2}{2}}
\sum_\lambda s_{R^t/\lambda}(q^{\rho})\,.
\eq
This has to be compared with the twisted real topological vertex \req{Trvertexschur}
\beq
\tilde C^{{\rm real}}_{R \cdot}(q)=(-1)^{\ell(R)}q^{\frac{||R||^2-||R^t||^2}{4}}\sum_\lambda 
s_{R/\lambda}(q^{-\rho})\,.
\eq
We deduce that
\beq
\tilde C^{\rm real}_{R\cdot}(1/q)= q^{\frac{||R||^2+||R^t||^2}{4}} \tilde P_{R^t,\cdot}(q)
M^{\rm real}(1/q)^{-1}\,.
\eq
Again, this fits with \eqref{expressed}. Note that the only difference with \eqref{melteq10} lies
in the slightly different normalization by the real MacMahon function \eqref{realmacmahon}.

\subsection{Constant map contribution in orientifolds}
\label{cmap}

It was observed in \cite{gova2} that the constant map contribution to the free energy of
the topological string on the Calabi-Yau manifold $X$ is encoded in the asymptotic expansion 
of the MacMahon function. Namely, it is known that the constant maps of genus $g$ Riemann 
surface into $X$ contribute \cite{bcov}
\begin{equation}
\eqlabel{asin}
\frac\chi 2 n^g_0 = \frac{\chi}2\int_{\calm_g} c_{g-1}^3(\cale)\,,
\end{equation}
where $\chi$ is Euler characteristic of $X$, $\calm_g$ is the moduli space of Riemann surfaces,
$\cale$ is the Hodge bundle over $\calm_g$, and $c_{g-1}$ is its $(g-1)$-st Chern class.
This Hodge integral is well-known \cite{fapa}, and one obtains 
\begin{equation}
\eqlabel{coincidence}
n^g_0 = \frac{|B_{2g} B_{2g-2}| }{2g (2g-2)(2g-2)!}\,,
\end{equation}
where $B_{2g}$ are the Bernoulli numbers. These are precisely the coefficients of the asymptotic
expansion of the MacMahon function at $g_s=-\log q\to 0$ (see appendix E of \cite{greg}):
\begin{equation}
\eqlabel{greg}
\begin{split}
\log M(q) \equiv\calf^{\rm even}(g_s)=& \sum_n \frac{q^n}{n(1-q^n)^2} \\ 
& \sim \frac{1}{g_s^2} \zeta(3) -\frac 1{12}\log g_s +  
\sum_{g\ge 2} g_s^{2g-2} n^g_0 + {\it const.}\,.
\end{split}
\end{equation}
The first two terms can be interpreted as the $g=0$ and $g=1$ contribution,
respectively. The constant has no obvious interpretation.
Moreover given the factorial  growth of the coefficients 
$n^g_0\sim (2g-1) (2g-3)!$, the above expansion  has zero convergence radius
and there can be  non-perturbative corrections
which can be worked out by means of the  Borel analysis.
After  writing the coefficients as:
\begin{equation}
n^g_0=\frac{B_{2g}}{2g(2g-2)! (2g-2)}\biggl(2(2g-2)!  \sum_{m=1} \frac{1}{(2\pi i m)^{2g-2}}
 \biggr)\,,
\end{equation}
we define the Borel transform by dividing each coefficient by its factorially divergent part 
to construct a series with finite convergence radius:
\begin{eqnarray}
B [\calf^{\rm even}] (\xi) &=& \sum_{g=2}^{+\infty} \frac{n^g_0}{(2g-3)!}\, 
\xi^{2g-2} = \sum_{g=2}^{+\infty} \frac{B_{2g}}{2g \left( 2g-2 \right)!} 
\sum_{m \in Z} \frac{\xi^{2g-2}}{\left(  2\pi i m \right)^{2g-2}}  \nonumber \\
&=& \sum_{m \in Z} \left( - \frac{1}{12} + \frac{ (2\pi i m)^2}{\xi^2} - 
\frac{1}{4}\, \frac{1}{\sinh^2 \left( \frac{\xi}{4 \pi i m } \right)} \right)\,.
\end{eqnarray} 
The inverse  of this transform is given by:
\begin{equation}
\eqlabel{bc}
\widetilde{\calf}^{\rm even} (g_s) = \frac{1}{4} \sum_{m \in Z} \int_0^{+\infty} \frac{d s}{s} 
\left(
 \frac{1}{\sin^2 \left( \frac{g_s}{4\pi  m }\, s \right)} - 
 \left( \frac{4\pi  m }{g_s} \right)^2 \frac{1}{s^2} -\frac{1}{3} \right) e^{-s}\,.
\end{equation}
Notice that for a  convergent series $F(x)$ with sum $f(x)$, the inverse Borel transform 
$\tilde{F}(x)$  is such that $\tilde{F}(x)=f(x)$. In the asymptotic case, the inverse Borel 
transform can be used, when the integral is well defined, to assign a value to the divergent sum.
However,  when  the  integral of the inverse transform is ill defined---which is the case 
for non Borel-summable series---one needs to modify the integration contour. This procedure 
is a priory not unique and it affects the reconstruction of the original series by 
introducing the so called non-perturbative ambiguity. Indeed the integral \eqref{bc} is ill 
defined since there are poles on the real axis. However, in the present case, there is a 
natural prescription to deform the contour and compute un-ambiguously the full non-perturbative 
correction since the inverse Borel transform \eqref{bc} coincides, up to a trivial change of 
variables, with the integral representation proposed by Gopakumar-Vafa for the constant map 
\cite{gova2}. The Gopakumar-Vafa  integral formula has in turn a Schwinger-like interpretation
as the four-dimensional one-loop effective action obtained by integrating out
$D0$-branes degrees of freedom in a constant self-dual gravi-photon field-strength.
From this viewpoint, it is clear that  the imaginary part of the integral \eqref{bc} 
gives the absorptive part of the action that is the non-perturbative production rate of 
$D0$-branes bound states. The imaginary part of the integral can be computed 
with the $+i \epsilon$ prescription (rotating the contour by $+\pi/2$) and 
closing the contour to pick all the residues, the result reads
\begin{eqnarray}
{\rm Im}\, \widetilde{\calf}^{\rm even}(g_s) &=& \frac{\pi}{4}\, \sum_{n=1}^{+\infty}
\sum_{m \in Z} \
 \oint_{n\pi} \frac{d \sigma}{2\pi i}\, \frac{1}{\sigma} \left( \frac{1}{\sin^2 \sigma} 
- \frac{1}{\sigma^2} - \frac{1}{3} \right) e^{-\frac{ 4\pi  m }{g_s}\, \sigma}  \nonumber \\
&=& - \frac{1}{4\pi g_s} \sum_{n,m=1}^{+\infty} \left( \frac{ 4\pi^2  m }{n} + \frac{g_s}{n^2} 
\right) e^{- \frac{4\pi^2  m n}{g_s}}\,.
\label{csim}
\end{eqnarray}
This provides  the full non-perturbative contribution to the McMahon function.\footnote{See 
\cite{ps} for more details.} The instanton action $A=4 \pi^2 m$ governs the production rate 
of bounds states of mass $2 \pi m$, the integer $m$ counts
the winding along the M-theory circle \cite{gova2}, while the integer $n$ is the instanton number.

In the previous subsection, we have seen that in the context of the real topological vertex, the
MacMahon function is replaced with its real version, $M^{\rm real}(q)$. It is natural to expect that 
this relation is more general, and that the real MacMahon function will capture the constant map 
contribution to the real topological string on a general Calabi-Yau manifold.\footnote{Note that at
this time of writing, we are not aware of a published physical derivation, nor a mathematical
theory, of this contribution. A rough estimate based along the lines of \cite{bcov} is consistent
with our present results---the perturbative contributions vanish for positive worldsheet Euler
characteristic.}
More precisely, we propose that in the large volume limit on a Calabi-Yau $X$,
the real topological string partition function behaves as
\begin{equation}\label{Gconstmap}
\calg \sim_{t\to\infty} \frac{\chi_a}{ 2} \log M^{\rm real}(q)+\frac{\chi_b}{ 2} \log{M(q)}\,,
\end{equation}
with $\chi_a+2\chi_b=\chi$ and $\chi$ as in \eqref{asin}. To explore the consequences of this conjecture, let us expand $M^{\rm real}(q)$ around
$g_s=0$. First we write
\begin{equation}
\eqlabel{split}
\begin{split}
\log M^{\rm real} (q) &= \sum_n \frac{q^n}{2n(1-q^n)^2}
+ \sum_{n \;{\rm odd}} \frac{q^{n/2}}{n(1-q^{n})} \\
& = \frac 12 \log M(q) + \calf^{\rm odd}(g_s)\,,
\end{split}
\end{equation}
which is the natural split in even and odd under $g_s\to -g_s$ ($q\to q^{-1}$). The even part 
is $1/2$ times the closed string result \eqref{greg}, which is the expected result from the 
point of view of the real topological string. For the odd part, we use the expansion
\begin{equation}
\frac{1}{2\sinh \frac x2} = \frac 1x + \sum_{k=1}^\infty (-1)^k \frac{2 x}{x^2+(2\pi k)^2}\,,
\end{equation}
to obtain
\begin{equation}
\begin{split}
\sum_{n=1}^\infty \frac{q^{n/2}}{n(1-q^n)}  &=
\sum_{n=1}^\infty \frac{1}{n} \left( \frac{1}{ng_s}
+\sum_{k=1}^\infty (-1)^k \frac{2 ng_s}{(n g_s)^2 + (2\pi k)^2} \right)\\
&= \frac{1}{g_s} \zeta(2) +
\sum_{k=1}^\infty \frac{(-1)^k}{k}\left( -\frac{g_s}{4\pi^2 k}
+\frac 12\coth\frac{2\pi^2 k}{g_s}\right) \\
&= \frac{1}{g_s}\zeta(2)  - \frac 12\log 2 +  \frac{g_s}{16\pi^2}\zeta(2)
+ \sum_{k,n=1}^\infty \frac{(-1)^k}{k} \ee^{-4\pi^2 k n/g_s}\,.
\end{split}
\end{equation}
Putting things together, this yields:
\begin{equation}
\eqlabel{putting}
\calf^{\rm odd}(g_s) = \frac {3\zeta(2)}{4g_s} - \frac 14 \log 2
+ \sum_{k,n=1}^\infty \frac{(-1)^k}{k} \bigl(\ee^{-4\pi^2 kn /g_s}
- \frac 12 \ee^{-2\pi^2 k n/g_s}\bigr)\,.
\end{equation}
We emphasize that this is a convergent expansion, with perturbative contributions vanishing
beyond one-loop. Quite interestingly, the constant appears to be violating the
behaviour under $g_s\to -g_s$ that we imposed on $\calf^{\rm odd}(g_s)$. One may check that
this is corrected by the non-perturbative contributions. 

To see this, we introduce the variable $p=\exp(-\frac{4\pi^2}{g_s})$ and perform the sum 
over $n$ in \eqref{putting} to write:
\begin{equation}
\calf^{\rm odd}(g_s) = \frac {3\zeta(2)}{4g_s} - \frac 14 \log 2
+ \sum_{k=1}^\infty \frac{(-1)^k}{k} \left(
\frac{p^k}{1-p^k}- \frac 12 \frac{p^{k/2}}{1-p^{k/2}}\right)\,.
\end{equation}
Under $g_s\to -g_s$ we have
\begin{equation}
\frac{p}{1-p}\to -\frac{1}{1-p}=-\frac{p}{1-p}-1\,,
\end{equation}
from which follows \footnote{Recall that $\sum_{k=1}^\infty \frac{(-1)^k}{k}=-\log(2)$ .}
\begin{equation}
\begin{split}
\calf^{\rm odd}(-g_s) =&-\frac {3\zeta(2)}{4g_s} - \frac 14 \log 2
- \sum_{k=1}^\infty \frac{(-1)^k}{k} \left(
\frac{p^k}{1-p^k}
- \frac 12 \frac{p^{k/2}}{1-p^{k/2}} +\frac{1}{2}   \right)\\
=
-\frac {3\zeta(2)}{4g_s} & +\frac{1}{4} \log 2
- \sum_{k=1}^\infty \frac{(-1)^k}{k} \left(
\frac{p^k}{1-p^k}  
- \frac 12 \frac{p^{k/2}}{1-p^{k/2}}    \right)=-\calf^{\rm odd}(g_s)\,.
\end{split}
\end{equation}
So in the  case of the real McMahon function it is essential to include the non-perturbative 
contribution to obtain a definite-parity asymptotic expansion. 

Finally, we shall give a Schwinger-like interpretation to the non-perturbative terms in 
$\calf^{\rm odd}(g_s)$ similar to that for the even part. To this end, we rewrite 
$\calf^{\rm odd}(g_s)$ in terms of the integral representation
\begin{equation}
\calf^{\rm odd}(g_s)=\sum_{n=-\infty}^\infty (-1)^{n} 
\int_{\epsilon}^\infty \frac{ds}{s}\frac{e^{-2\pi i s (n/2)}}{2\sinh(g_s s/2)}~,
\end{equation}
which furthermore can be split into 
\begin{equation}
\label{rsc}
\calf^{\rm odd}(g_s)=\sum_{n=-\infty}^\infty  
\int_{\epsilon}^\infty \frac{ds}{s}\frac{e^{-2\pi i s n}}{2\sinh(g_s s/2)}-
\sum_{n\,{\rm odd}} \int_{\epsilon}^\infty \frac{ds}{s}\frac{e^{-2\pi i s (n/2)}}
{2\sinh(g_s  s/2)}\,.
\end{equation}
This expression is the result of a two-dimensional Schwinger computation of integrating 
out a scalar field coupled to a constant $U(1)$ field-strength, where the scalar field has either 
integral or fractional (half-integer) charge. In order to see that one can indeed derive 
$\calf^{\rm odd}(g_s)$ from such a computation, note first that under the orientifold 
projection only the two components of the gravi-photon field $A^\mu(x)$ away from the $O4$-plane 
survive, since $A^\mu(x)$ needs to be odd under the projection. Secondly, the occurrance 
of states with fractional charge, or better momenta, can be inferred from the expectation 
that one can lift every IIA orientifold to M-theory on a $G_2$ manifold with some 
$\Z_2$ action on the M-theory circle (in our case with fixed-points). The non-trivial 
action on the M-theory circle results in fractional momentum states. Thus, 
the Schwinger interpretation of $\calf^{\rm odd}(g_s)$ is in terms of (fractional) bound states 
of $D0$-branes living away from the orientifold plane such that they feel the $\N=2$ of the bulk.

With the Schwinger representation for $\calf^{\rm odd}(g_s)$ at hand one can interpret the 
non-perturbative contribution in \eqref{putting} as we did for the McMahon function.
We evaluate the absorptive part of the integral 
\eqref{rsc} with the $+i\epsilon$ prescription 
by closing the contour to pick  the residues at the simple poles
and obtain
\begin{equation}
\sum_{k=1}\frac{(-1)^k}{k}
\Bigl(\sum_n e^{-\frac{4 \pi^2 n k}{g_s}}-\sum_{ n ~\rm{odd}}
 e^{-\frac{2 \pi^2 n k}{\lambda}}\Bigr)=
\sum_{k=1}\frac{(-1)^k}{k}
\Bigl(2 \sum_n e^{-\frac{4 \pi^2 n k}{g_s}}-\sum_{n}
 e^{-\frac{2 \pi^2 n k}{\lambda}}\Bigr)\,,
  \end{equation}
which reproduces indeed the  non-perturbative terms in the real McMahon function in 
\eqref{putting}. Notice that this time we have two families of non-perturbative contributions 
with instanton actions $4 \pi^2 n$ and $2 \pi^2 n$, which can be interpreted as controlling the 
production rates of bound states, respectively fractional bound states, of $D0$-branes.

\section{Examples}
\label{examples}

We now want to illustrate the real topological vertex formalism with its sign subtleties
explained in section \ref{formalism} at hand of a couple of instructive examples. We will
check that we obtain a consistent enumerative interpretation in terms of real Gopakumar-Vafa
invariants, as well as compatibility of the partition function under flop transitions, and 
reduction to known invariants at certain points in parameter space.

To write the expansion of the reduced free energy \eqref{realfenergy}, we denote the subset
of K\"ahler parameters that are mapped to themselves by the involution $\sigma$ defining the 
orientifold of $X$ by $Q_A$, and those that are identified pairwise by $Q'_B$. (This is an 
invariant distinction if we use an integral basis of the second cohomology of $X$.) The point 
is that the former set appears with half-integer exponents  in $\calg'^\sigma_X$, and the latter 
only with integer exponents. The real Gopakumar-Vafa invariants are denoted by 
${^X N^{(\chi)}_{d_Q,d_{Q'}}}$ with integer vectors $d_Q$, $d_{Q'}$ labelling the degree and 
$\chi\ge -1$ being related to the 5-dimensional spin in the M-theory interpretation as counting
of BPS states (see \cite{oova}). (If the BPS state is represented by a smooth real curve, 
$\chi$ is the negative of its Euler characteristic.) The expansion is
\begin{equation}
\eqlabel{Gexpansion}
\calg'^{\sigma}_X= \sum_{\substack{\chi, d_Q, d_{Q'}\geq 0 \\k\;{\rm odd}}} 
{^X N^{(\chi)}_{d_{Q},d_{Q'}}}\frac{1}{k}\biggl(2 \ii \sin \frac{k g_s}{2} 
\biggr)^{\chi} \prod_A Q_A^{k d_{Q_A}/2}\prod_B {Q'_B}^{k d_{Q'_B}}\,.
\end{equation}
A typical, though not universal, feature of this expansion is a certain correlation between $\chi$ 
and the degrees $d_{Q_A}$. Namely, $^X N^{(\chi)}_{d_Q,d_{Q'}}$ vanish unless 
$\chi\equiv \sum_A d_{Q_A} \bmod 2$. This rule can be explained from the point of view of Gromov-Witten
theory along the lines of the arguments in \cite{krwa} by the local cancellation between boundaries 
and crosscaps on fixed $\P^1$'s covered by even degree maps. However, exceptions to this 
$d_Q$-$\chi$-correlation rule are possible when there are even degree maps passing through fixed 
vertices. A rule that always holds is that $^X N^{(\chi)}_{d_Q,d_{Q'}}$ is bounded above by, and 
equal modulo 2 to, the corresponding complex invariant $^X n^{(g=\chi+1)}_{d_Q,d_{Q'},d_{Q'}}$.

\subsection{Butterfly}

\begin{figure}
\begin{center}
\psfrag{flop}[cc][][1]{flop}
\psfrag{H}[cc][][0.75]{$y$}
\psfrag{I}[cc][][0.75]{$z$}
\psfrag{F}[cc][][0.75]{$x$}
\psfrag{J}[cc][][0.75]{$x+y$}
\psfrag{K}[cc][][0.75]{$x+z$}

\includegraphics[scale=0.4]{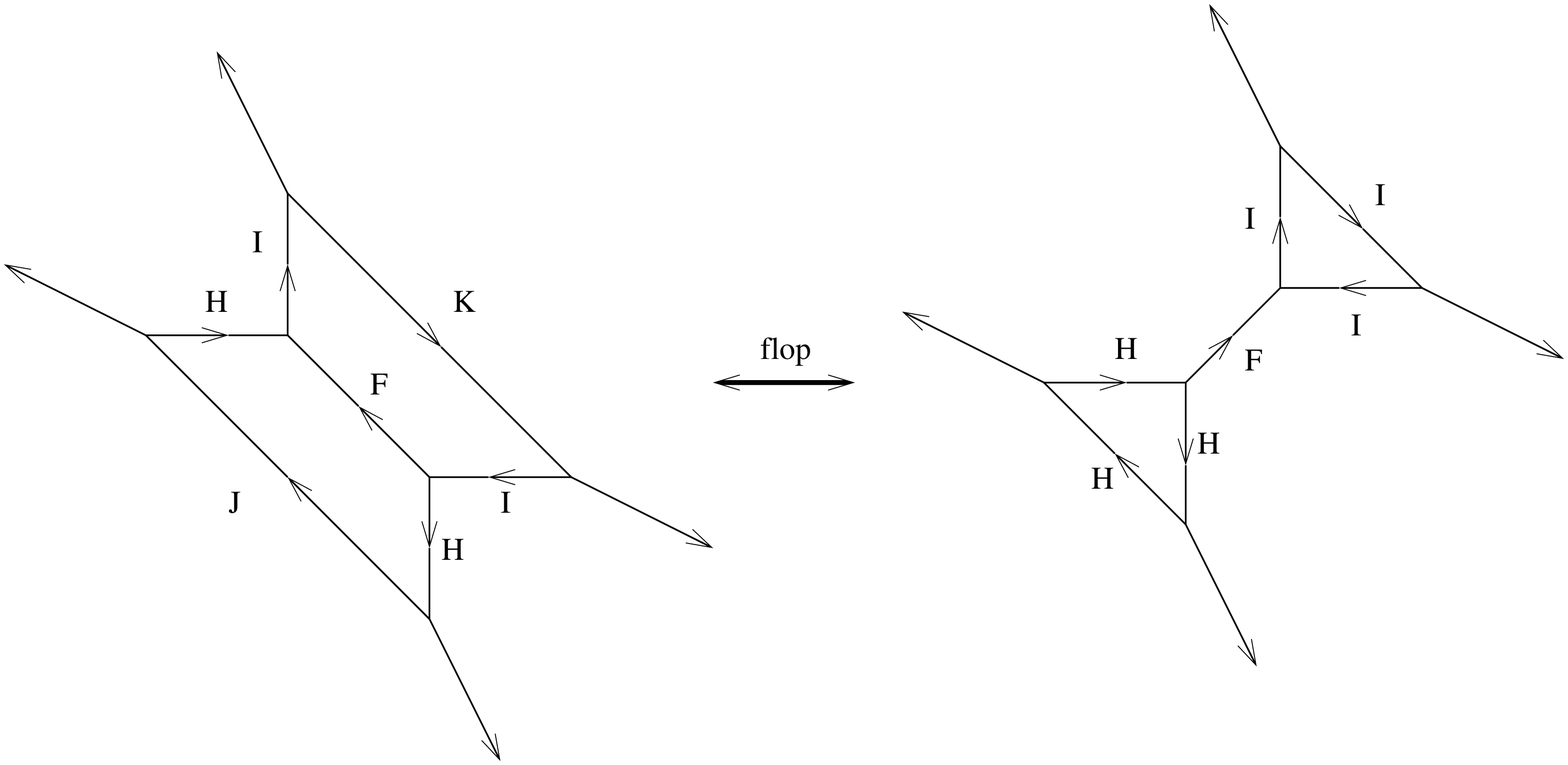}
\end{center}
\caption{The $(p,q)$-webs of the butterfly (left) and $[\P^2]^2$ (right) geometry with K\"ahler 
moduli associated to the edges. The two geometries are related via a flop of the central $\P^1$.}
\label{flyfig1}
\end{figure}

Consider the two geometries shown in terms of $(p,q)$-webs in figure \ref{flyfig1}. We will refer 
to the left geometry as butterfly, and to the right one as $[\P^2]^2$, since it consists of
two local $\P^2$'s connected via a $\P^1$ with $\O(-1)\oplus\O(-1)$ normal bundle (conifold). The 
geometries are related via a flop transition of the central $\P^1$.  These geometries have been 
used to study local mirror symmetry (at tree-level) in \cite{Chiang}. We will also refer to the 
butterfly as phase $A$ and to $[\P^2]^2$ as phase $B$.

\paragraph{Closed string}

\begin{figure}
\begin{center}
\psfrag{flop}[cc][][1]{flop}
\psfrag{I}[cc][][0.75]{1}
\psfrag{II}[cc][][0.75]{2}
\psfrag{III}[cc][][0.75]{3}
\psfrag{A}[cc][][0.75]{$R_3$}
\psfrag{B}[cc][][0.75]{$R_0$}
\psfrag{C}[cc][][0.75]{$R_6$}
\psfrag{D}[cc][][0.75]{$R_5$}
\psfrag{E}[cc][][0.75]{$R_1$}
\psfrag{G}[cc][][0.75]{$R_4$}
\psfrag{F}[cc][][0.75]{$R_2$}
\includegraphics[scale=0.4]{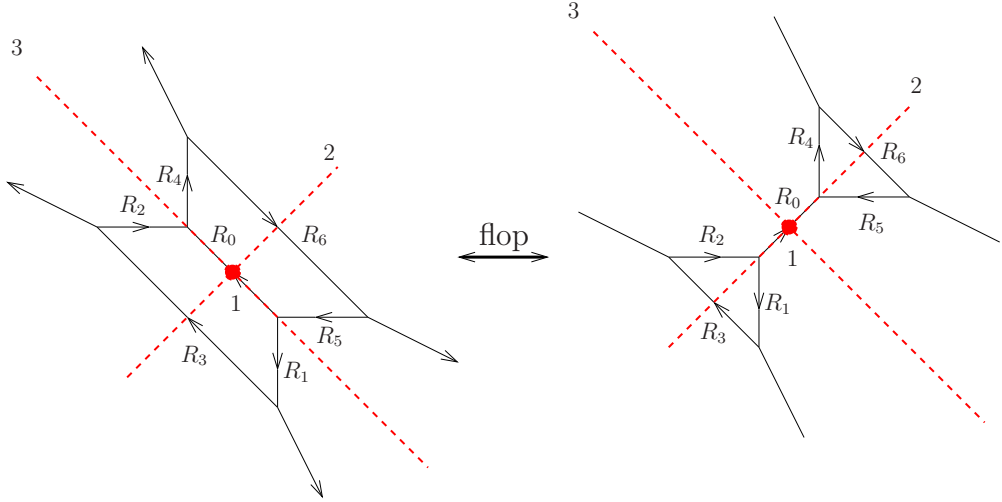}
\end{center}
\caption{The involutive $\Z_2$ symmetries of the butterfly and $[\P^2]^2$.}
\label{flyfig2}
\end{figure}

An expression for the partition function $Z_A$ of the (closed) topological string on the butterfly 
can be obtained in terms of the topological vertex  (\cf. \cite{akmv}) and reads (for the 
association of representations $R_i$ to edges see figure \ref{flyfig2})
\beq
\begin{split}
Z_A=  \sum_{R}&(-1)^{\ell(R_0)-\ell(R_3)-\ell(R_6)} q^{\sum_{i\neq 0} \k_{R_i}/2} \, 
x^{d(x)} y^{d(y)} z^{d(z)}\\
&\times C_{R_2^t R_4 R_0^t} C_{R_2 R_3^t \cdot} C_{R_3 R_1^t \cdot} C_{R_5^t R_1 R_0} 
C_{R_5 R_6^t \cdot} C_{R_6 R_4^t \cdot}  \,,
\end{split}
\eq
where $R=\{R_0,\dots, R_6\}$, 
\beq
\begin{split}
d(x)&=\ell(R_0)+\ell(R_3)+\ell(R_6)\,,\\
d(y)&=\ell(R_1)+\ell(R_2)+\ell(R_3)\,,\\
d(z)&=\ell(R_4)+\ell(R_5)+\ell(R_6)\,,\\
\end{split}
\eq
$x=e^{-t_0}$, $y=e^{-t_1}$ and $z=e^{-t_2}$, where $t_i$ are the three K\"ahler moduli of the geometry.

Similarly, the partition function $Z_B$ of $[\P^2]^2$ (cf. \cite{bouchard1}) is given by
\beq
\begin{split}
Z_B=\sum_R &(-1)^{d(x)-d(y)-d(z)} q^{\sum_{i\neq 0} \k_{R_i}}   \, x^{d(x)} y^{d(y)} z^{d(z)} \\
&\times C_{R_1 R_2^t R_0} C_{R_3 R_1^t \cdot} C_{R_2 R_3^t \cdot} C_{R_4 R_5^t R_0^t} 
C_{R_6 R_4^t \cdot} C_{R_5 R_6^t \cdot} \,,
\end{split}
\eq
with
\beq
\begin{split}
d(x)&=\ell(R_0)\,,\\
d(y)&=\ell(R_1)+\ell(R_2)+\ell(R_3)\,,\\
d(z)&=\ell(R_4)+\ell(R_5)+\ell(R_6)\,.\\
\end{split}
\eq
In \cite{Chiang} it was observed that the tree-level instanton pieces of the topological amplitudes 
of the two phases, expressed in terms of Gopakumar-Vafa invariants ${^{A/B} n}^{(g)}_{d_x,d_y,d_z}$, 
are related via \footnote{Except for ${^A n^{(0)}_{1,0,0}}$, which counts just the flopped $\P^1$.}
\beq\label{closedeq3}
{^A n^{(0)}_{d_x,d_y,d_z}}= {^B n^{(0)}_{d_y+d_z-d_x,d_y,d_z}}\,.
\eq
As it must be, and shown rigorously in \cite{Konishi}, this relation persists to all genera.

The $(p,q)$-webs possess three different $\Z_2$ symmetries, as shown in figure \ref{flyfig2}. Following the formalism outlined in section \ref{formalism} 
we can write down partition functions capturing the respective orientifolds of the theory. 

\paragraph{Involution 1}

The point-reflection involution acting on the butterfly identifies the K\"ahler moduli $y=z$ and 
the orientifold partition function is given by
\beq
\begin{split}
Z_{A}^{1}=&\sum_{R,R_0=R_0^t}(-1)^{(\ell(R_0) \pm r(R_0))/2 - \ell(R_3)} 
q^{\k_{R_1}/2+\k_{R_3}/2+\k_{R_5}/2} \, x^{d(x)} y^{d(y)}\\
&\times C_{R_5^tR_1R_0}C_{R_5R_3^t\cdot} C_{ R_3 R_1^t \cdot }\,,
\end{split}
\eq
with $R=\{R_1,R_3,R_5\}$ and
\beq
\begin{split}
d(x)&=\ell(R_0)/2+\ell(R_3)\,,\\
d(y)&=\ell(R_1)+\ell(R_3)+\ell(R_5)\,.
\end{split}
\eq
Similarly, the point reflection of $[\P^2]^2$ (which has been already discussed in \cite{bouchard1}) 
yields
\beq
Z^1_B=\sum_{R,R_0=R_0^t} (-1)^{(\ell(R_0)\mp r(R_0))/2-d(y)}q^{\k_{R_1}+\k_{R_2}+\k_{R_3}} 
C_{R_1R^t_2 R_0 }C_{R_2 R_3^t\cdot} C_{R_3 R_1^t \cdot }\, x^{d(x)} y^{d(y)}\,,
\eq
with $R=\{R_1,R_2,R_3\}$,
\beq
\begin{split}
d(x)&=\ell(R_0)/2\,,\\
d(y)&=\ell(R_1)+\ell(R_2)+\ell(R_3)\,.
\end{split}
\eq
We expand the resulting real free energies defined as in \req{realfenergy} into real Gopakumar-Vafa 
invariants ${^{A/B}N^{(\chi)}_{d_x,d_y}}$, following \req{Gexpansion}. We obtain the invariants 
listed in table \ref{flyinv1tab} of appendix \ref{appendix}. Note that the
$d_Q$-$\chi$-correlation holds in this example. For $\chi$ odd, we reproduce the results for $c=1$, and
for $\chi$ even, those for $c=2$, of \cite{bouchard1}.

We observe that the real BPS numbers of the two phases are related via
\beq\label{flypointfloprelation}
{^A N_{d_x,d_y}^{(\chi)}}={^B N_{2d_y-d_x,d_y}^{(\chi)}}\,,
\eq
which is just \req{closedeq3} under the identification $d_y=d_z$. Note that in order that 
\req{flypointfloprelation} hold exactly, \ie, not only up to a sign, the pre-sign of the $r$-type 
sign needs to switch under the flop.

\paragraph{Involution 2}

This involution acts as the identity on the moduli. The butterfly has three type 2 fixed edges 
(corresponding to three invariant $S^1$'s as fixed-point locus), such that we have to insert three 
$c$-type signs. According to the rules in section \ref{elusive}, there are two consistent global 
choices. We prefer the following expresssion:
\beq
\begin{split}
Z^{2}_A=\sum_R&(-1)^{(\ell(R_0)\pm c(R_0)+\ell(R_3)\mp c(R_3)+\ell(R_6)\mp c(R_6^t))/2}
q^{-\k_{R_0}/4+ \k_{R_1}/2+\k_{R_3}/4+\k_{R_5}/2+3\k_{R_6}/4}\\
&\times C_{R_5^t R_1 R_0} C_{R_5 R_6^t\cdot} C_{R_3 R_1^t \cdot} \, x^{d(x)} y^{d(y)} z^{d(z)}\,,
\end{split}
\eq
with $R=\{R_0,R_1,R_3,R_5,R_6\}$ and
\beq
\begin{split}
d(x)&=\ell(R_0)/2+\ell(R_3)/2+\ell(R_6)/2\,,\\
d(y)&=\ell(R_1)+\ell(R_3)/2\,,\\
d(z)&=\ell(R_5)+\ell(R_6)/2\,.\\
\end{split}
\eq

Expansion of the partition function into real Gopakumar-Vafa invariants as in \req{Gexpansion}
delivers the invariants listed in table \ref{flyinv2GV} of the appendix.
 
Turning to phase B, we have two type 2 fixed legs and two real vertices connected via an (isolated) 
type 3 fixed leg (see figure \ref{flyfig2}). Thus, we need to insert two $c$-type signs and make
the right choice between straight and twisted real vertex. The rules of section \ref{elusive} 
allow
\beq
\begin{split}
Z^{2}_B& =\sum_{R,R_0=R_0^t}
(-1)^{\ell(R_1)+\ell(R_4)+(\ell(R_0)\mp r(R_0)+\ell(R_3)\pm c(R_3)+\ell(R_6)\pm c(R_6^t))/2} \\
&\times
q^{5 \k_{R_1}/4+\k_{R_3}/4+5\k_{R_4}/4+\k_{R_6}/4} \times   
C_{R_3 R_1^t \cdot} C_{R_6 R_4^t \cdot} C^{\rm real}_{R_1 R_0} 
\tilde C^{\rm real}_{R_4 R_0}  \, x^{d(x)} y^{d(y)} z^{d(z)}\,,
\end{split}
\eq
with $R=\{R_1,R_3,R_4,R_6\}$ and
\beq
\begin{split}
d(x)&=\ell(R_0)/2\,,\\
d(y)&=\ell(R_1)+\ell(R_3)/2\,,\\
d(z)&=\ell(R_4)+\ell(R_6)/2\,.\\
\end{split}
\eq
But we wish to emphasize that exchanging $c(R_i)$ with $c(R_i^t)$ and straight with twisted
real vertex gives the same result. Expansion gives the real Gopakumar-Vafa invariants listed
in table \ref{P22inv2GV} in the appendix.
Since all moduli are mapped to themselves, we expect that the relation \req{closedeq3} persists, \ie,
\beq\label{inv2floprel}
{^A N^{(\chi)}_{d_x,d_y,d_z}}= {^B N^{(\chi)}_{d_y+d_z-d_x,d_y,d_z}}\,.
\eq
As we infer by comparing tables \ref{flyinv2GV} and \ref{P22inv2GV}, this is indeed the case. 
Another check on the consistency of the obtained invariants is given by the relations
\beq
\begin{split}
{^A N^{(\chi)}_{d,0,d}}={^A N^{(\chi)}_{d,d,0}}= {^{\P^2} N^{(\chi)}_{d}}\,,\\
{^B N^{(\chi)}_{0,d,0}}={^B N^{(\chi)}_{0,0,d}}= {^{\P^2} N^{(\chi)}_{d}}\,,\\
\end{split}
\eq
counting the real invariants of the individual $\P^2$ in the geometry.

\paragraph{Involution 3}

Similarly to involution 1, this involution projects the K\"ahler parameters to $y=z$. In fact, from 
a quotient space perspective of the action on $[\P^2]^2$, the only difference to involution 1 acting 
on $[\P^2]^2$ lies in the action on the central $\P^1$. Whereas involution 1 acts without fixed-point 
on the $\P^1$, involution 3 has an $S^1$ fixed locus. Since locally around the fixed-point locus the 
geometry corresponds to the conifold, and following the discussion of section \ref{formalism}, we 
expect that involution 3 yields the same real Gopakumar-Vafa invariants as involution 1.

Indeed, the orientifold partition function, where we inserted a single $c$-type sign for the type 2 
fixed-leg, is given by
\beq
Z_B^{3}=\sum_{R,R_0} (-1)^{d(y)+(\ell(R_0)\mp c(R_0))/2}q^{-\k_{R_0}/4 +\k_{R_1}+\k_{R_2}+\k_{R_3}} 
C_{ R_1 R_2^t R_0} C_{R_2 R_3^t \cdot} C_{R_3 R_1^t \cdot}  \, x^{d(x)} y^{d(y)}\,,
\eq
with $R=\{R_1,R_2,R_3\}$ and
\beq
\begin{split}
d(x)&=\ell(R_0)/2\,,\\
d(y)&=\ell(R_1)+\ell(R_2)+\ell(R_3)\,.
\end{split}
\eq
The resulting free energy is equal to that of involution 1, as expected.
For involution 3 acting on the butterfly we obtain
\beq
\begin{split}
Z^{3}_A=\sum_{R_0=R_0^t,R} & (-1)^{(\ell(R_0)\pm r(R_0))/2-\ell(R_3)} 
q^{3\k_{R_1}/4+\k_{R_2}/4+\k_{R_3}/2}  \, x^{d(x)} y^{d(y)} \\
&\times C_{R_2 R_3^t \cdot} C_{R_3 R_1^t \cdot } C^{\rm real}_{R_2^t R_0} \tilde
C^{\rm real}_{R_1^t R_0} \,, 
\end{split}
\eq
with $R=\{R_1,R_2,R_3\}$ as before and
\beq
\begin{split}
d(x)&=\ell(R_0)/2+\ell(R_3)\,,\\
d(y)&=\ell(R_1)+\ell(R_2)+\ell(R_3)\,.
\end{split}
\eq
Again, $Z^3_A=Z^1_A$. Note that in order that the invariants of the two geometries 
agree exactly (including sign), the pre-signs of the $r$-type and $c$-type sign need to be 
opposite.

\subsection{Hybridfly}

\begin{figure}
\begin{center}
\psfrag{f}[cc][][1]{flop}
\psfrag{H}[cc][][0.75]{$x+y$}
\psfrag{F}[cc][][0.75]{$y$}
\psfrag{A}[cc][][0.75]{$x$}
\psfrag{D}[cc][][0.75]{$u$}
\psfrag{E}[cc][][0.75]{$u$}
\psfrag{B}[cc][][0.75]{$x+u$}
\psfrag{J}[cc][][0.75]{$x+z$}
\psfrag{G}[cc][][0.75]{$z$}

\includegraphics[scale=0.4]{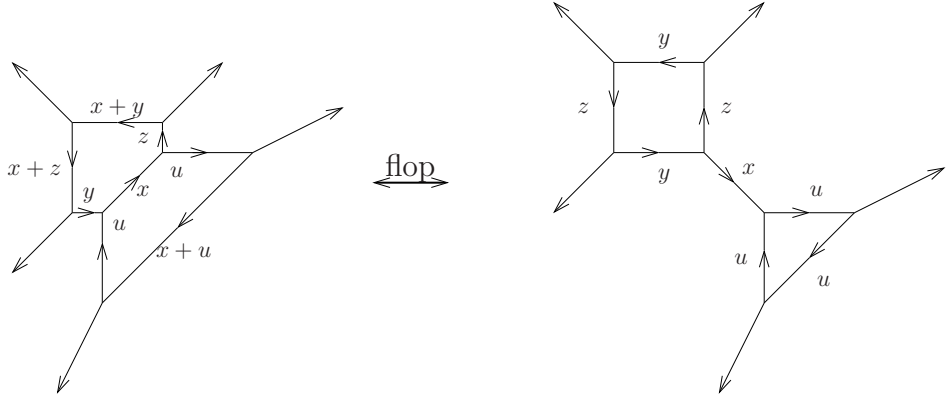}
\end{center}
\caption{The $(p,q)$-webs of the hybridfly (left) and $[\P^2][\mathbb F_0]$ (right) geometry with 
K\"ahler moduli associated to the edges. The two geometries are related via a flop of the central 
$\P^1$.}
\label{hybridflyfig1}
\end{figure}

The next more complicated example is the 4-parameter geometry shown in the left of figure 
\ref{hybridflyfig1}. We will refer to it as hybridfly. The flopped geometry is a $\P^2$ connected 
to a $\mathbb F_0$ via a conifold. We therefore denote the flopped geometry as $[\mathbb F_0][\P^2]$. 
We will also refer to the hybridfly as phase $A$ and to $[\mathbb F_0][\P^2]$ as phase $B$. 

\paragraph{Closed string geometry}

\begin{figure}
\begin{center}
\psfrag{f}[cc][][1]{flop}
\psfrag{I}[cc][][0.75]{1}
\psfrag{A}[cc][][0.75]{$R_0$}
\psfrag{B}[cc][][0.75]{$R_6$}
\psfrag{C}[cc][][0.75]{$R_6$}
\psfrag{D}[cc][][0.75]{$R_5$}
\psfrag{E}[cc][][0.75]{$R_7$}
\psfrag{G}[cc][][0.75]{$R_1$}
\psfrag{F}[cc][][0.75]{$R_4$}
\psfrag{H}[cc][][0.75]{$R_2$}
\psfrag{J}[cc][][0.75]{$R_3$}
\includegraphics[scale=0.4]{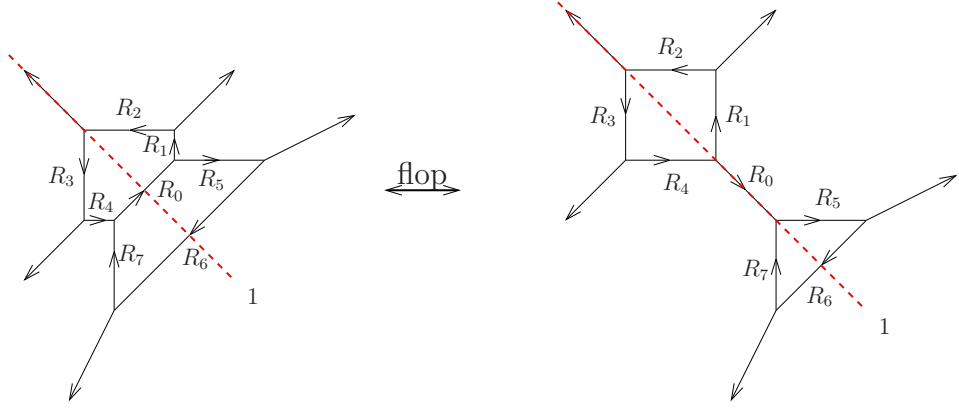}
\end{center}
\caption{The single involutive $\Z_2$ symmetry of the hybridfly and $[\mathbb F_0][\P^2]$.}
\label{hybridflyfig2}
\end{figure}

The corresponding (closed) topological string partition 
functions can be easily obtained to be given by (see figure \ref{hybridflyfig2} for the association 
of representations to edges)
\beq
\begin{split}
Z_A=&\sum_R (-1)^{\ell(R_0)+\ell(R_1)+\ell(R_4)-\ell(R_6)} 
q^{(-\kappa_{R_2}-\kappa_{R_3}+\kappa_{R_5}+2\kappa_{R_6}+\kappa_{R_7})/2} \, 
x^{d(x)}y^{d(y)}z^{d(z)}u^{d(u)}\\
&\times C_{R_1 R_5 R_0^t} C_{R_1^t R_2\cdot} C_{R_2^t R_3\cdot} C_{R_3^t R_4\cdot}
C_{R_7^t R_4^t R_0} C_{R_7 R_6^t\cdot}C_{R_6R_5^t\cdot}\,,
\end{split}
\eq
with $R=\{R_0,\dots,R_7\}$ and
\beq
\begin{split}
d(x)&=\ell(R_0)+\ell(R_2)+\ell(R_3)+\ell(R_6)\,,\\
d(y)&=\ell(R_2)+\ell(R_4)\,,\\
d(z)&=\ell(R_1)+\ell(R_3)\,,\\
d(u)&=\ell(R_5)+\ell(R_6)+\ell(R_7)\,.
\end{split}
\eq
Similarly, we obtain for the flopped geometry
\beq
\begin{split}
Z_B&=\sum_R (-1)^{l(R_0)-d(u)} q^{-(\kappa_{R_1}+\kappa_{R_2}+\kappa_{R_3}+\kappa_{R_4})/2}
q^{\kappa_{R_5}+\kappa_{R_6}+\kappa_{R_7}}\, x^{d(x)}y^{d(y)}z^{d(z)}u^{d(u)}\\
&\times C_{R_4^t R_1 R_0} C_{R_1^t R_2\cdot} C_{R_2^t R_3 \cdot} C_{R_3^t R_4\cdot} 
C_{R_5 R_7^t R_0^t} C_{R_6 R_5^t\cdot} C_{R_7 R_6^t\cdot}\,,
\end{split}
\eq
with
\beq
\begin{split}
d(x)&=\ell(R_0)\,,\\
d(y)&=\ell(R_2)+\ell(R_4)\,,\\
d(z)&=\ell(R_1)+\ell(R_3)\,,\\
d(u)&=\ell(R_5)+\ell(R_6)+\ell(R_7)\,.
\end{split}
\eq
The relation between the Gopakumar-Vafa invariants ${^{A/B} n^{(g)}_{d_x,d_y,d_z,d_u}}$ of the
two geometries reads
\beq
\eqlabel{hybridfloprel}
{^{A} n^{(g)}_{d_x,d_y,d_z,d_u}}={^{B} n^{(g)}_{d_y+d_z+d_u-d_x,d_y,d_z,d_u}}\,.
\eq
These two geometries have only one involutive symmetry, as indicated in figure \ref{hybridflyfig2}.

\paragraph{Involution 1}

Let us start with the hybridfly. The involution projects the moduli $y=z$ and maps the remaining
two moduli to themselves. We need to insert two $c$-type signs and one real vertex. 
\beq
\begin{split}
Z^1_A=\sum_R & (-1)^{(\ell(R_0)\pm c(R_0^t)+ \ell(R_6)\pm c(R_6)/2+\ell(R_1)} 
q^{(\kappa_{R_0}-3\kappa_{R_2}+2\kappa_{R_5}+\kappa_{R_6})/4}  \\
&\times 
C_{R_1 R_5 R_0^t} C_{R_1^t R_2\cdot} C_{R_6 R_5^t \cdot} C^{\rm real}_{R_2^t\cdot}\,
\, x^{d(x)} z^{d(z)} u^{d(u)} \,, 
\end{split}
\eq
with $R=\{R_0,R_1,R_2,R_5,R_6\}$ and
\beq
\begin{split}
d(x)&=\ell(R_0)/2+\ell(R_2)+\ell(R_6)/2\,,\\
d(z)&=\ell(R_1)+\ell(R_2)\,,\\
d(u)&=\ell(R_5)+\ell(R_6)/2\,.
\end{split}
\eq
The real Gopakumar-Vafa invariants extracted from the corresponding free energy 
are listed in table \ref{hybridflyGV} of the appendix.

In the flopped phase, we obtain
\beq\label{Z1flophexa}
\begin{split}
Z^1_B=\sum_{R,R_0=R_0^t} & (-1)^{(\ell(R_0\mp r(R_0))/2 +\ell(R_5) +(\ell(R_6)\pm c(R_6))/2} 
q^{(-\kappa_{R_1}-3\kappa_{R_2}+5\kappa_{R_5}-\kappa_{R_6})/4} \\ 
&\times C_{R_1^t R_2\cdot} C_{R_6R_5^t\cdot} C^{\rm real}_{R_2^t\cdot} \tilde 
C^{\rm real}_{R_1^t R_0} C^{\rm real}_{R_5 R_0}\,  x^{d(x)} z^{d(z)} u^{d(u)}
\end{split}
\eq
with $R=\{R_1,R_2,R_5,R_6\}$ and
\beq
\begin{split}
d(x)&=\ell(R_0)/2\,,\\
d(z)&=\ell(R_1)+\ell(R_2)\,,\\
d(u)&=\ell(R_5)+\ell(R_6)/2\,.
\end{split}
\eq
The Gopakumar-Vafa invariants are listed in table \ref{F0P2GV}. The relation to the numbers in table
\ref{hybridflyGV} is
\beq
{^{A} N^{(\chi)}_{d_x,d_z,d_u}}={^{B} N^{(\chi)}_{2d_z+d_u-d_x,d_z,d_u}}\,.
\eq
As a consistency check on the obtained invariants, we recover 
\beq
\begin{split}
{^{A} N^{(\chi)}_{0,0,d}}&={^{B} N^{(\chi)}_{d,0,d}}={^{\P^2}N^{(\chi)}_{d}}\,,\\
{^{A} N^{(\chi)}_{0,d,0}}&={^{B} N^{(\chi)}_{2d,d,0}}={^{\mathbb F_0}N^{(\chi)}_{d}}\,,
\end{split}
\eq 
\ie, the real invariants of the $\P^2$ and $\mathbb F_0$ in the geometry. Note that $Z^1$ of 
$\mathbb F_0$ can be easily obtained by setting $R_5=R_6=R_0=\cdot$ in \req{Z1flophexa}. The 
resulting real invariants are listed in table \ref{F0GV}. 

We note that the last non-vanishing invariants for fixed $d$ in table \ref{F0GV} show a very 
simple structure:
\begin{equation}
^{\mathbb F_0} N^{(d-1)^2-1}_{d} = 1
\,, \quad \text{for $d$ even;}
\qquad
^{\mathbb F_0} N^{(d-1)^2-2}_{d} = 2\,, \quad \text{for $d$ odd}\,.
\end{equation}
These results can be verified in the computational scheme for GV invariants developed in 
\cite{kkv}. In the notation of section 8.4 of that reference, we are interested
in the real version of $n_d^r$ with the maximal possible value of $r$ for fixed $d$. (Our
$d$ is $a=b$ of \cite{kkv}, and $\chi=r-1$.) For $a$ even, we have $r=(a-1)^2$, and the relevant 
moduli space is $\P^{(a+1)^2-1}$, with $e(\RP^{{\rm even}})=1$. For $a$ odd, on the other hand, 
since $e(\RP^{\rm odd})=0$, we need to look at $\delta=1$, \ie, $r=(a-1)^2-1$. Then by the first 
line of equation (5.4), we need $e(\calc)$, where $\calc$ is the universal curve. $\calc\to
\P^1\times \P^1$ with fiber $\P^{(a+1)^2-2}$. Because the involution exchanges the two $\P^1$'s,
we have $e(\calc^{\rm real})=e(\RP^{(a+1)^2-2}) \cdot e(\CP^1)=2$.

\subsection{Pentafly}

\begin{figure}
\begin{center}
\psfrag{A}[cc][][0.75]{$x$}
\psfrag{B}[cc][][0.75]{$u+x$}
\psfrag{C}[cc][][0.75]{$w+x$}
\psfrag{D}[cc][][0.75]{$u$}
\psfrag{E}[cc][][0.75]{$w$}
\psfrag{F}[cc][][0.75]{$z$}
\psfrag{H}[cc][][0.75]{$z+x$}
\psfrag{J}[cc][][0.75]{$y+x$}
\psfrag{G}[cc][][0.75]{$y$}
\psfrag{f}[cc][][1]{flop}
\includegraphics[scale=0.5]{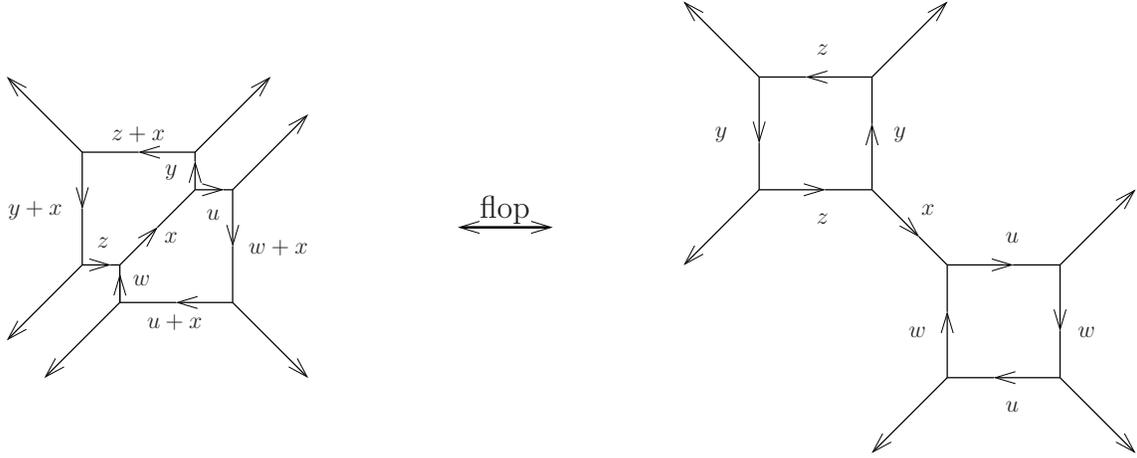}
\end{center}
\caption{The $(p,q)$-webs of the pentafly (left) and $[\mathbb F_0]^2$ (right) with K\"ahler moduli 
associated to the edges. The geometries are related via a flop of the central $\P^1$.}
\label{pentaflyfig1}
\end{figure}

As a final example, we consider the $5$-parameter geometries shown in figure \ref{pentaflyfig1},
which are again related via a flop of the central $\P^1$. 

\paragraph{Closed string}

The geometry in the left of the figure will be denoted as pentafly, and the flopped geometry as 
$[\mathbb F_0]^2$, since this geometry consists of two $\mathbb F_0$ connected via a conifold. We 
will refer to the pentafly also as phase $A$ and to $[\mathbb F_0]^2$ as phase $B$. Note that phase 
$A$ consists of two $\mathcal B_2$. Here, $\mathcal B_2$ denotes the $2$-point blowup of $\P^2$. 
Thus, we expect to recover at specific values in parameter space the known results of $\P^2$.

\begin{figure}
\begin{center}
\psfrag{f}[cc][][1]{flop}
\psfrag{I}[cc][][0.75]{1}
\psfrag{II}[cc][][0.75]{2}
\psfrag{III}[cc][][0.75]{3}
\psfrag{A}[cc][][0.75]{$R_0$}
\psfrag{B}[cc][][0.75]{$R_1$}
\psfrag{C}[cc][][0.75]{$R_2$}
\psfrag{D}[cc][][0.75]{$R_3$}
\psfrag{E}[cc][][0.75]{$R_4$}
\psfrag{F}[cc][][0.75]{$R_8$}
\psfrag{G}[cc][][0.75]{$R_7$}
\psfrag{H}[cc][][0.75]{$R_6$}
\psfrag{J}[cc][][0.75]{$R_5$}
\includegraphics[scale=0.5]{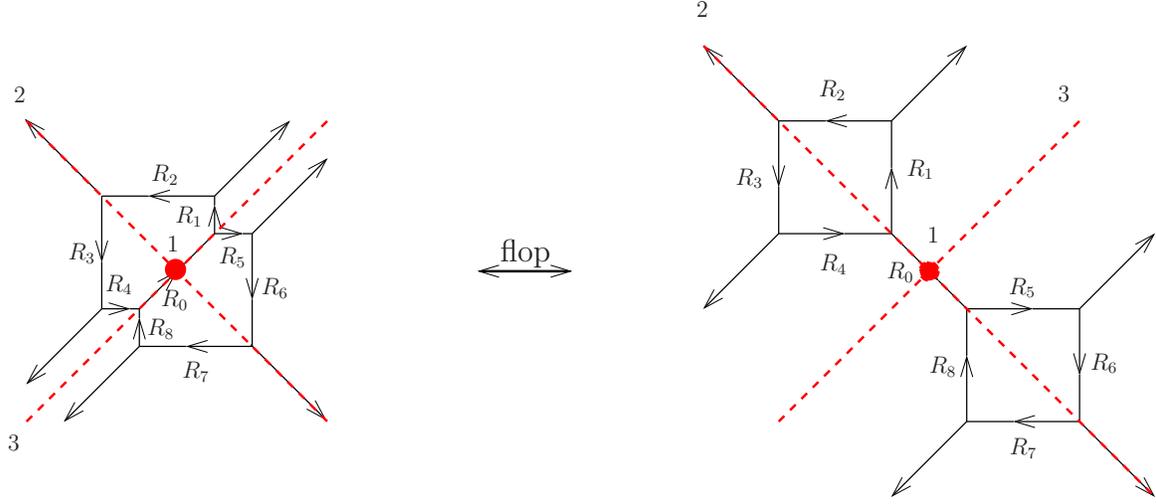}
\end{center}
\caption{The involutive $\Z_2$ symmetries of the pentafly and $[\mathbb F_0]^2$.}
\label{pentaflyfig2}
\end{figure}

The corresponding (closed) topological string partition functions can be easily obtained (see 
figure \ref{pentaflyfig2} for the association of representations to edges). We infer for the 
pentafly,
\beq
\begin{split}
Z_A=\sum_R & (-1)^{\ell(R_0)+\ell(R_1)+\ell(R_4)+\ell(R_5)+\ell(R_8)} 
q^{(-\kappa_{R_2}-\kappa_{R_3}+\kappa_{R_6}+\kappa_{R_7})/2} \\
&\times  \, x^{d(x)}y^{d(y)}z^{d(z)}u^{d(u)}w^{d(w)}\\
&\qquad \times C_{R_1 R_5 R_0^t}C_{R_1^t R_2 \cdot}C_{R_2^t R_3 \cdot }
C_{R_3^t R_4 \cdot } C_{R_8^t R_4^t R_0 } C_{R_5^t \cdot R_6}C_{R_6^t \cdot R_7}
C_{R_7^t \cdot R_8}\,,\\
\end{split}
\eq
with $R=\{R_0,\dots,R_8\}$ and
\beq
\begin{split}
d(x)&=\ell(R_0)+\ell(R_2)+\ell(R_3)+\ell(R_6)+\ell(R_7)\,,\\
d(y)&=\ell(R_1)+\ell(R_3)\,,\\
d(z)&=\ell(R_2)+\ell(R_4)\,,\\
d(u)&=\ell(R_5)+\ell(R_7)\,,\\
d(w)&=\ell(R_6)+\ell(R_8)\,.\\
\end{split}
\eq
Meanwhile, the partition function of $[\mathbb F_0]^2$ reads
\beq
\begin{split}
Z_B=\sum_R& (-1)^{d(x)}q^{-(\kappa(R_1)+\k(R_2)+\k(R_3)+\k(R_4))/2 }
q^{(\kappa_{R_5}+\k_{R_6}+\k_{R_7}+\k_{R_8})/2 } \\
&\times x^{d(x)} y^{d(y)} z^{d(z)}u^{d(u)} w^{d(w)} \\
&\qquad \times C_{R_4^t R_1 R_0 }  C_{R_1^t R_2\cdot} C_{R_2^t R_3\cdot} 
C_{R_3^t R_4\cdot} C_{R_6 R_5^t\cdot} C_{R_5 R_8^t R_0^t} C_{R_7 R_6^t\cdot} 
C_{R_8 R_7^t \cdot}\,,
\end{split}
\eq
with 
\beq
\begin{split}
d(x)&=\ell(R_0)\,,\\
d(y)&=\ell(R_1)+\ell(R_3) \,,\\
d(z)&=\ell(R_2)+\ell(R_4)\,,\\
d(u)&=\ell(R_5)+\ell(R_7)\,,\\
d(w)&=\ell(R_6)+\ell(R_8)\,.\\
\end{split}
\eq
The relation between the Gopakumar-Vafa invariants ${^{A/B} n^{(g)}_{d_x,d_y,d_z,d_u,d_w}}$
under the flop transition relating phase $A$ and $B$ is
\beq
{^A n^{(g)}_{d_x,d_y,d_z,d_u,d_w}}={^B n^{(g)}_{d_y+d_z+d_u+d_w-d_x,d_y,d_z,d_u,d_w}}\,.
\eq
The two geometries possess three involutive $\Z_2$ symmetries which we illustrated in figure 
\ref{pentaflyfig2}. For each involution, we discuss in the following the respective orientifold
topological partition function.

\paragraph{Involution 1}

The point-reflection acts via identifying $y=w$ and $z=u$, while $x$ is mapped to itself. We deduce 
that the real partition function of the pentafly reads
\beq
\begin{split}
Z^{1}_A=\sum_{R, R_0=R_0^t} & (-1)^{(\ell(R_0)\pm r(R_0))/2+\ell(R_1)+\ell(R_4)}
q^{-(\kappa_{R_2}+\kappa_{R_3})/2}\, x^{d(x)} y^{d(y)} z^{d(z)}\\
&\times C_{R_1 R_4^t R_0^t} C_{R_1^t R_2 \cdot} C_{R_2^t R_3 \cdot} C_{R_3^t R_4 \cdot}\,,
\end{split}
\eq
with $R=\{R_1,\dots,R_4\}$ and
\beq
\begin{split}
d(x)&=\ell(R_0)/2+\ell(R_2)+\ell(R_3)\,,\\
d(y)&=\ell(R_1)+\ell(R_3)\,,\\
d(z)&=\ell(R_2)+\ell(R_4)\,.\\
\end{split}
\eq
Similarly, we obtain for $[\mathbb F_0]^2$
\beq
\begin{split}
Z^1_B=\sum_{R,R_0=R_0^t} & (-1)^{(\ell(R_0)\mp r(R_0))/2} 
q^{-(\kappa_{R_1}+\k_{R_2}+\k_{R_3}+\k_{R_4})/2 }\, x^{d(x)} y^{d(y)} z^{d(z)}\\
&\times C_{R_4^t R_1 R_0 }  C_{R_1^t R_2\cdot} C_{R_2^t R_3\cdot} C_{R_3^t R_4\cdot} \,,
\end{split}
\eq
with $R$ as for the pentafly and
\beq
\begin{split}
d(x)&=\ell(R_0)/2\,,\\
d(y)&=\ell(R_1)+\ell(R_3) \,,\\
d(z)&=\ell(R_2)+\ell(R_4)\,.\\
\end{split}
\eq
As usual for this type of involution, we inserted a single $r$-type sign into the projected 
partition functions.

We expand both partition functions into real Gopakumar-Vafa invariants and list the results in 
tables \ref{pentaflypointGV} and \ref{F02pointGV}. The invariants of 
both phases are related via (in order to match the signs, we invert the $r$-type sign under 
the flop)
\beq
\eqlabel{floprelation}
{^A N^{(\chi)}_{d_x,d_y,d_z}}={^B N^{(\chi)}_{2d_y+2d_z-d_x,d_y,d_z}}\,.
\eq

\paragraph{Involution 2}

From figure \ref{pentaflyfig2} we infer that this involution projects the K\"ahler parameters to
$x=y$ and $u=w$. 
The real topological string partition function is
\beq\label{pentainv2Z}
\begin{split}
Z^{2}_A=\sum_R& (-1)^{(\ell(R_0)\mp c(R_0))/2+\ell(R_1)+\ell(R_5)}
q^{(\kappa_{R_0}-3\kappa_{R_2}+\kappa_{R_6})/4} \, x^{d(x)}y^{d(y)}u^{d(u)}\\
&\times C_{R_1 R_5 R_0^t} C_{R_1^t R_2 \cdot} C_{R_6 R_5^t \cdot} C^{\rm real}_{R_2^t\cdot}
\tilde C^{\rm real}_{R_6\cdot} \,,
\end{split}
\eq
with $R=\{R_0,R_1,R_2,R_5,R_6\}$ and
\beq\label{pentainv2d}
\begin{split}
d(x)&=\ell(R_0)/2+\ell(R_2)+\ell(R_6)\,,\\
d(y)&=\ell(R_1)+\ell(R_2)\,,\\
d(u)&=\ell(R_5)+\ell(R_6)\,.
\end{split}
\eq
Expansion of the resulting free energies yields the invariants listed in table \ref{pentainv2GV}. 
Note that since the pentafly consists of two $\mathcal B_2$'s, we 
expect to recover the real Gopakumar-Vafa invariants ${^{\mathcal B_2} N^{(\chi)}_{d_x,d_y}}$ of 
$\mathcal B_2$. Setting $R_5=R_6=\cdot$ in \req{pentainv2Z} and \req{pentainv2d}, the partition 
function reduces to the partition function of real $\mathcal B_2$ and the resulting invariants 
indeed satisfy
\beq
{^B N^{(\chi)}_{d_x,d_y,0}}={^B N^{(\chi)}_{d_x,0,d_y}}={^{\mathcal B_2} N^{(\chi)}_{d_x,d_y}}\,.
\eq
A non-trivial check on the consistency of the obtained invariants lies in the fact that we recover 
the known invariants of local $\P^2$, \ie,
\beq
{^{\mathcal B_2} N^{(\chi)}_{d_x,d_x}}={^{\P^2} N^{(\chi)}_{d_x}}\,.
\eq

Let us turn to the flopped geometry. 
\beq
\begin{split}
Z^2_B(\sigma,\tau)=\sum_{R,R_0=R_0^t}& (-1)^{(\ell(R_0)\mp r(R_0))/2}  
 q^{(-\k_{R_1}-3\k_{R_2}+3\k_{R_5}+\k_{R_6})/4}\, x^{d(x)}y^{d(y)}u^{d(u)}\\
& \times C_{R_1^t R_2\cdot } C_{R_6 R_5^t\cdot} C^{\rm real}_{R_2^t\cdot}
\tilde C^{\rm real}_{R_1^t R_0} C^{\rm real}_{R_5 R_0} \tilde C^{\rm real}_{R_6\cdot}\,,
\end{split}
\eq
with $R=\{R_1,R_2,R_5,R_6\}$ and
\beq
\begin{split}
d(x)&=\ell(R_0)/2\,,\\
d(y)&=\ell(R_1)+\ell(R_2)\,,\\
d(u)&=\ell(R_5)+\ell(R_6)\, .
\end{split}
\eq
The invariants are listed in table \ref{F02inv2GV}.

Similarly as for the pentafly, setting $R_5=R_6=\cdot$, we calculate the partition function 
of a flop of real $\mathcal B_2$, which is $\mathbb F_0$ with a conifold attached. 
We recover the invariants of the two $\mathbb F_0$ in the geometry as follows
\beq
{^B N^{(\chi)}_{0,d,0}}={^B N^{(\chi)}_{0,0,d}}={^{\mathbb F_0}N^{(\chi)}_d}\,.
\eq
Also, the invariants of the two phases are related via
\beq
{^A N^{(\chi)}_{d_x,d_y,d_u}}={^B N^{(\chi)}_{2d_y+2d_u-d_x,d_y,d_u}}\,,
\eq
as expected.

\paragraph{Involution 3}

The projected partition function of the pentafly under involution 3 is given by
\beq
\begin{split}
Z^3_A=\sum_{R,R_0=R_0^t} & (-1)^{(\ell(R_0)\pm r(R_0))/2+\ell(R_1)+\ell(R_4)} 
q^{-(\kappa_{R_2}+\kappa_{R_3})/2+(\kappa_{R_1}-\kappa_{R_4})/4} \\
&\times C_{R_1^t R_2 \cdot} C_{R_2^t R_3\cdot} C_{R_3^t R_4 \cdot} C^{\rm real}_{R_1 R_0}
\tilde C^{\rm real}_{R_4 R_0} \, x^{d(x)} y^{d(y)} z^{d(z)}  \,,
\end{split}
\eq
with $R=\{R_1,\dots,R_4\}$,
\beq
\begin{split}
d(x)&=\ell(R_0)/2+\ell(R_2)+\ell(R_3)\,,\\
d(y)&=\ell(R_1)+\ell(R_3)\,,\\
d(z)&=\ell(R_2)+\ell(R_4)\,.\\
\end{split}
\eq
When expanding the partition function according to \eqref{Gexpansion}, we notice that while the
$^XN^{(\chi)}_{d_x d_y d_z}$ are all integer, they {\it do not} satisfy the $d_Q$-$\chi$ correlation
prominent in all previous examples (here, $d_Q=d_x$). This is somewhat unexpected, but in fact
not in violation of any fundamental principle. The simplest geometry with this feature is the
closed topological vertex of \cite{bryankarp}. 

Let us compare to the flopped $[\mathbb F_0]^2$ geometry. The projected partition function is given by
\beq
\begin{split}
Z^3_B=\sum_R& (-1)^{(\ell(R_0)+c(R_0))/2} q^{-\k_{R_0}/4-(\kappa_{R_1}+
\k_{R_2}+\k_{R_3}+\k_{R_4})/2 }\, x^{d(x)} y^{d(y)} z^{d(z)}\\
&\times C_{R_4^t R_1 R_0 }  C_{R_1^t R_2\cdot} C_{R_2^t R_3\cdot} C_{R_3^t R_4\cdot} \,,
\end{split}
\eq
Again, the expansion shows a violation of $d_Q$-$\chi$ correlation. The terms that do satisfy 
the correlation coincide to those of involution 1, while all terms are related to those of the pentafly 
via the flop \eqref{floprelation}.

We note that the terms that violate $d_Q$-$\chi$ correlation change sign under the replacement of 
$C^{\rm real}$ with $\tilde C^{\rm real}$ and $c(R)$ with $c(R^t)$, while those that satisfy 
$d_Q$-$\chi$ correlation are invariant under this replacement. We can write this in terms of the 
formula
\begin{equation}
\eqlabel{finalrelation}
Z(q^{1/2},Q^{1/2},Q) = \tilde Z(-q^{1/2},-Q^{1/2},Q')\,,
\end{equation}
where $Z$ and $\tilde Z$ are the partition functions evaluated with the two consistent real vertex
prescriptions discussed in section \ref{elusive}.

\section{Conclusions}
\label{conclusions}

In this paper, we have developed the real topological vertex formalism proposed in \cite{krwa}
to a complete computational prescription for the evaluation of real topological string amplitudes
on local toric Calabi-Yau threefolds. The real vertex is essentially a squareroot of the ordinary
vertex amplitude of \cite{akmv}. We have uncovered several new sign subtleties that arise in the
gluing of the real vertex to the global expression. In particular, we have seen that we actually 
require two different real vertices with slightly different sign insertions. 

We have also discussed the interpretation of the real vertex in Chern-Simons theory on the orbifold
$S^3/\zet_2$ with fixed point locus in codimension two, giving a complete realization of the 
existence of the two types of real vertex. We have also given an interpretation via the melting 
crystal picture of \cite{crystal}, which consists in considering a symmetrically melting crystal.
Via this connection, we have made and studied a proposal for the constant map contribution to real 
topological string on general Calabi-Yau orientifolds.

Among the few open questions, let us mention just two: It would be interesting to study also the 
melting crystal representation of the real vertex with
non-trivial representation on the fixed leg (non-trivial asymptotics along the $x_2$-axis).
It is rather straightforward to write down the vertex operator formula, but it is not immediately 
clear how to relate the resulting expressions with the squareroots in \eqref{rvertexschur} and 
\eqref{Trvertexschur}.

We have seen in the final example in section
\ref{examples} that there are cases in which the real topological partition function does not 
satisfy ``$d_Q$-$\chi$ correlation'' (see beginning of section \ref{examples} for the definition
of this notion). This correlation might have been expected based on the experience with real 
localization techniques \cite{tadpole,krwa}. In particular, local tadpole cancellation was 
implemented in this context by cancelling even degree real maps between holes and crosscaps on
the worldsheet, and by neglecting any possible contribution from a non-trivial intersection theory 
on the moduli space of real curves. (The only Hodge integrals required for the computations in 
\cite{krwa} are those on moduli spaces of complex curves.) The violation of $d_Q$-$\chi$ correlation 
from the real topological vertex seems to indicate that this assumption might not be correct in general. 
It would be interesting to pursue this further.

\begin{acknowledgments}
We would like to thank Marcos Mari\~no for Chern-Simons consultations at initial stages of this 
work, and in particular for reminding us of Ho\v rava's work. We thank Chris Beasley, Jim Bryan, 
and Cumrun Vafa for valuable discussions. The work of D.K. was supported in part by an EU 
Marie-Curie EST fellowship, the Max-Planck society and the WPIRC initiative by  MEXT of 
Japan. D.K. thanks the ASC at LMU and CERN for hospitality during part of this project. 
S.P.\ thanks ASC at LMU for hospitality. J.W.\ is grateful to McGill University and to SUNY 
Stonybrook during the 2009 Simons Workshop for hospitality while some of the signs of this 
work were being fixed.
\end{acknowledgments}

\newpage
\appendix

\section{Real Gopakumar-Vafa invariants}
\label{appendix}

\begin{table}[!hp]
\begin{center}
\tiny{
\begin{tabular}{|c|c|c|c|c|c|c|c|c|c|}
\hline
$d\slash \chi$&$0$&$1$&$2$&$3$&$4$&$5$&$6$&$7$&$8$\\
\hline
$0$&&&$1$&$8$&$69$&$608$&$5475$&$50136$&$465173$\\
$2$&&&&$2$&$76$&$1544$&$24696$&$382934$&$5324640$\\
$4$&&&&&$39$&$2020$&$65892 $&$1651424$&$35161960 $\\
$6$&&&&&$10$&$1586$&$111660$&$4914874$&$164667808$\\
$8$&&&&&$1$&$756$&$132105 $&$10723220$&$582872279$\\
$10$&&&&&&$212$&$111774$&$17629822$&$1607349528$\\
$12$&&&&&&$32$&$68342 $&$22182896$&$3518490422$\\
$14$&&&&&&$2$&$30194$&$21562774$&$6195809584 $\\
$16$&&&&&&&$9530$&$16278148$&$8866082190$\\
$18$&&&&&&&$2092$&$9561340$&$10391531036$\\
$20$&&&&&&&$303$&$4361964$&$10036474120$\\
$22$&&&&&&&$26$&$1536200$&$8023729064$\\
$24$&&&&&&&$1$&$412728$&$5325069376$\\
$26$&&&&&&&&$82898$&$2937580602$\\
$28$&&&&&&&&$12036$&$1346409352$\\
$30$&&&&&&&&$1192$&$511497566$\\
$32$&&&&&&&&$72$&$160302439$\\
$34$&&&&&&&&$2$&$41130990$\\
$36$&&&&&&&&&$8542684$\\
$38$&&&&&&&&&$1412510$\\
$40$&&&&&&&&&$181428$\\
$42$&&&&&&&&&$17436$\\
$44$&&&&&&&&&$1179$\\
$46$&&&&&&&&&$50$\\
$48$&&&&&&&&&$1$\\
$50$&&&&&&&&&\\
\hline
\end{tabular}
}
\caption{${^{\mathbb F_0} N^{(\chi)}_{d}}$ of the diagonal involution of $\mathbb F_0$ (two real vertices).}
\label{F0GV}
\end{center}
\end{table}

The vectors in the following tables are to be understood as a list of real Gopakumar-Vafa invariants 
for higher $\chi$. When the invariants satisfy the $d_Q$-$\chi$ correlation, we only list the values
for $\chi\equiv \sum_A d_{Q_A}\bmod 2$, since the remaining invariants are identically zero. The first 
entry in each list corresponds to $\chi=-1$, or $\chi=0$. When the last entry of the list is zero,
all invariants of higher $\chi$ also vanish. (Otherwise, the list might be truncated to fit into the
table.)

\begin{table}[!hp]
\begin{center}
\tiny{
\begin{tabular}{|c|ccccc|}
\hline
$d_x\! \setminus\! d_y$&0&1&2&3&4\\
\hline
0&$0$&$0$&$0$&$0$&$0$\\
1&$-1$&$-2$&$-3$&$-4$&$-5$\\
2&$0$&$0$&$0$&$-(0,1,0)$&$-(0,2,0)$\\
3&$0$&$0$&$5$&$(30,7,0)$&$(112,59,9,0)$\\
4&$0$&$0$&$0$&$4$&$(0,11,6,1,0)$\\
5&$0$&$0$&$0$&$-(32,9,0)$&$-(369,315,103,12)$\\
6&$0$&$0$&$0$&$0$&$0$\\
7&$0$&$0$&$0$&$0$&$(286,288,108,14)$\\
8&$0$&$0$&$0$&$0$&$0$\\
\hline
\end{tabular}
\hspace{0.1cm}
\begin{tabular}{|c|ccccc|}
\hline
$d_x\!\setminus\! d_y$&0&1&2&3&4\\
\hline
0&$0$&$0$&$0$&$0$&$0$\\
1&$1$&$-2$&$5$&$-(32,9,0)$&$(286,288,108,14)$\\
2&$0$&$0$&$0$&$0$&$0$\\
3&$0$&$0$&$-3$&$(30,7,0)$&$-(369,315,103,12)$\\
4&$0$&$0$&$0$&$-(0,1,0)$&$(0,11,6,1,0)$\\
5&$0$&$0$&$0$&$-4$&$(112,59,9,0)$\\
6&$0$&$0$&$0$&$0$&$-(0,2,0)$\\
7&$0$&$0$&$0$&$0$&$-5$\\
8&$0$&$0$&$0$&$0$&$0$\\
\hline
\end{tabular}
}
\caption{Left table: ${^A N_{d_x,d_y}^{(\chi)}}$ for involution 1 \& 3 of the butterfly.  
Right table: ${^B N_{d_x,d_y}^{(\chi)}}$ of the corresponding $[\P^2]^2$ involutions.}
\label{flyinv1tab}
\end{center}
\end{table}
 
\begin{sidewaystable}[!hp]

\tiny{

\begin{tabular}{|c|ccccccc|}
\hline
$\substack{d_x=0\\d_y\slash d_z}$&0&1&2&3&4&5&6\\
\hline
0&$0$&$0$&$0$&$0$&$0$&$0$&$0$\\
1&$0$&$0$&$0$&$0$&$0$&$0$&$0$\\
2&$0$&$0$&$0$&$0$&$0$&$0$&$0$\\
3&$0$&$0$&$0$&$0$&$0$&$0$&$0$\\
4&$0$&$0$&$0$&$0$&$0$&$0$&$0$\\
5&$0$&$0$&$0$&$0$&$0$&$0$&$0$\\
6&$0$&$0$&$0$&$0$&$0$&$0$&$0$\\
\hline
\end{tabular}
\hspace{0.25cm}
\begin{tabular}{|c|ccccccc|}
\hline
$\substack{d_x=1\\d_y\slash d_z}$&0&1&2&3&4&5&6\\
\hline
0&$-1$&$-1$&$-1$&$-1$&$-1$&$-1$&$-1$\\
1&$-1$&$0$&$0$&$0$&$0$&$0$&$0$\\
2&$-1$&$0$&$-1$&$-1$&$-1$&$-1$&$-1$\\
3&$-1$&$0$&$-1$&$0$&$0$&$0$&$0$\\
4&$-1$&$0$&$-1$&$0$&$-1$&$-1$&$-1$\\
5&$-1$&$0$&$-1$&$0$&$-1$&$0$&$0$\\
6&$-1$&$0$&$-1$&$0$&$-1$&$0$&$-1$\\
\hline
\end{tabular}
\hspace{0.25cm}
\begin{tabular}{|c|ccccccc|}
\hline
$\substack{d_x=2\\d_y\slash d_z}$&0&1&2&3&4&5&6\\
\hline
0&$0$&$0$&$0$&$(0,1,0)$&$0$&$(0,4,1,0)$&$0$\\
1&$0$&$0$&$0$&$0$&$0$&$0$&$0$\\
2&$0$&$0$&$0$&$(0,1,0)$&$0$&$(0,5,1,0)$&$0$\\
3&$(0,1,0)$&$0$&$(0,1,0)$&$(0,1,0)$&$(0,1,0)$&$(0,1,0)$&$(0,1,0)$\\
4&$0$&$0$&$0$&$(0,1,0)$&$0$&$(0,5,1,0)$&$0$\\
5&$(0,4,1,0)$&$0$&$(0,5,1,0)$&$(0,1,0)$&$(0,5,1)$&$(0,5,1,0)$&$(0,5,1)$\\
6&$0$&$0$&$0$&$(0,1,0)$&$0$&$(0,5,1,0)$&$0$\\
\hline
\end{tabular}
\vspace{0.25cm}\\
\begin{tabular}{|c|ccccccc|}
\hline
$\substack{d_x=3\\d_y\slash d_z}$&0&1&2&3&4&5&6\\
\hline
0&$0$&$0$&$0$&$1$&$2$&$(5,1,0)$&$(9,2,0)$\\
1&$0$&$0$&$0$&$0$&$0$&$0$&$0$\\
2&$0$&$0$&$1$&$2$&$(6,1,0)$&$(11,2,0)$&$(22,8,1,0)$\\
3&$1$&$0$&$2$&$-(0,1,0)$&$4$&$-(0,5,1,0)$&$6$\\
4&$2$&$0$&$(6,1,0)$&$4$&$(18,7,1,0)$&$(18,3,0)$&$(48,30,9,1,0)$\\
5&$(5,1,0)$&$0$&$(11,2,0)$&$-(0,5,1,0)$&$(18,3,0)$&$-(0,22,9,1,0)$&$(30,5,0)$\\
6&$(9,2,0)$&$0$&$(22,8,1,0)$&$6$&$(48,30,9,1,0)$&$(30,5,0)$&$(103,94,46,11,1,0)$\\
\hline
\end{tabular}
\vspace{0.25cm}\\
\begin{tabular}{|c|ccccccc|}
\hline
$\substack{d_x=4\\d_y\slash d_z}$&0&1&2&3&4&5&6\\
\hline
0&$0$&$0$&$0$&$0$&$(0,3,1,0)$&$-(0,4,1,0)$&$(0,35,57,36,10,1,0)$\\
1&$0$&$0$&$0$&$0$&$0$&$0$&$0$\\
2&$0$&$0$&$0$&$-(0,1,0)$&$0$&$-(0,18,8,1,0)$&$(0,34,57,36,10,1,0)$\\
3&$0$&$0$&$-(0,1,0)$&$-(0,2,0)$&$-(0,6,1,0)$&$-(0,11,2,0)$&$-(0,22,8,1,0)$\\
4&$(0,3,1,0)$&$0$&$0$&$-(0,6,1,0)$&$-(0,9,6,1,0)$&$-(0,60,38,10,1,0)$&$-(0,27,26,9,1,0)$\\
5&$-(0,4,1,0)$&$0$&$-(0,18,8,1,0)$&$-(0,11,2,0)$&$-(0,60,38,10,1,0)$&$-(0,60,21,2,0)$&$-(0,165,138,57,12,1,0)$\\
6&$(0,35,57,36,10,1,0)$&$0$&$(0,34,57,36,10,1,0)$&$-(0,22,8,1,0)$&$-(0,27,26,9,1,0)$&$-(0,165,138,57,12,1,0)$&$-(0,153,252,182,68,13,1,0)$\\
\hline
\end{tabular}
\vspace{0.25cm}\\
\begin{tabular}{|c|ccccccc|}
\hline
$\substack{d_x=5\\d_y\slash d_z}$&0&1&2&3&4&5&6\\
\hline
0&$0$&$0$&$0$&$0$&$0$&$-(5,10,6,1,0)$&$-(14,40,57,36,10,1,0)$\\
1&$0$&$0$&$0$&$0$&$0$&$0$&$0$\\
2&$0$&$0$&$0$&$0$&$-2$&$-(10,2,0)$&$-(45,54,59,36,10,1,0)$\\
3&$0$&$0$&$0$&$(0,1,0)$&$-5$&$(0,18,8,1,0)$&$-(35,42,57,36,10,1,0)$\\
4&$0$&$0$&$-2$&$-5$&$-(27,9,1,0)$&$(-42,19,27,9,1,0)$&$-(192,155,88,39,10,1,0)$\\
5&$-(5,10,6,1,0)$&$0$&$-(10,2,0)$&$(0,18,8,1,0)$&$(-42,19,27,9,1,0)$&$(0,186,169,67,13,1,0)$&$(-198,-24,40,17,2,0)$\\
6&$-(14,40,57,36,10,1,0)$&$0$&$-(45,54,59,36,10,1,0)$&$-(35,42,57,36,10,1,0)$&$-(192,155,88,39,10,1,0)$&$(-198,-24,40,17,2,0)$&$-(818,852,498,179,36,3,0)$\\
\hline
\end{tabular}
\vspace{0.25cm}\\
\begin{tabular}{|c|ccccccc|}
\hline
$\substack{d_x=6\\d_y\slash d_z}$&0&1&2&3&4&5&6\\
\hline
0&$0$&$0$&$0$&$0$&$0$&$0$&$-(0,44,63,37,10,1,0)$\\
1&$0$&$0$&$0$&$0$&$0$&$0$&$0$\\
2&$0$&$0$&$0$&$0$&$0$&$(0,4,1,0)$&$-(0,70,114,72,20,2,0)$\\
3&$0$&$0$&$0$&$0$&$(0,2,0)$&$(0,10,2,0)$&$(0,45,54,59,36,10,1,0)$\\
4&$0$&$0$&$0$&$(0,2,0)$&$0$&$(0,72,45,11,1,0)$&$-(0,82,174,126,38,4,0)$\\
5&$0$&$0$&$(0,4,1,0)$&$(0,10,2,0)$&$(0,72,45,11,1,0)$&$-(0,-105,41,104,54,12,1,0)$&$(0,570,685,521,276,89,15,1)$\\
6&$-(0,44,63,37,10,1,0)$&$0$&$-(0,70,114,72,20,2,0)$&$(0,45,54,59,36,10,1,0)$&$-(0,82,174,126,38,4,0)$&$(0,570,685,521,276,89,15,1)$&$(0,36,-96,-88,-16,4,1,0)$\\
\hline
\end{tabular}

}
\caption{${^A N_{d_x,d_y,d_z}^{(\chi)}}$ for involution 2 of the butterfly.}
\label{flyinv2GV}
\end{sidewaystable}

\begin{sidewaystable}
\tiny{
\begin{tabular}{|c|ccccccc|}
\hline
$\substack{d_x=0\\d_y\slash d_z}$&0&1&2&3&4&5&6\\
\hline
0&$0$&$-1$&$0$&$1$&$(0,3,1)$&$-(5,10,6)$&$-(0,44,63)$\\
1&$-1$&$0$&$0$&$0$&$0$&$0$&$0$\\
2&$0$&$0$&$0$&$0$&$0$&$0$&$0$\\
3&$1$&$0$&$0$&$0$&$0$&$0$&$0$\\
4&$(0,3,1)$&$0$&$0$&$0$&$0$&$0$&$0$\\
5&$-(5,10,6)$&$0$&$0$&$0$&$0$&$0$&$0$\\
6&$-(0,44,63)$&$0$&$0$&$0$&$0$&$0$&$0$\\
\hline
\end{tabular}
\hspace{0.25cm}
\begin{tabular}{|c|ccccccc|}
\hline
$\substack{d_x=1\\d_y\slash d_z}$&0&1&2&3&4&5&6\\
\hline
0&$1$&$0$&$-1$&$(0,1,0)$&$2$&$-(0,4,1)$&$-(14,40,57)$\\
1&$0$&$0$&$0$&$0$&$0$&$0$&$0$\\
2&$-1$&$0$&$1$&$-(0,1,0)$&$2$&$(0,4,1)$&$(14,40,57)$\\
3&$(0,1,0)$&$0$&$-(0,1,0)$&$(0,1,0)$&$(0,2,0)$&$-(0,4,1)$&$-(0,14,40)$\\
4&$2$&$0$&$-2$&$(0,2,0)$&$4$&$-(0,8,2)$&$-(28,80,114)$\\
5&$-(0,4,1,0)$&$0$&$(0,4,1,0)$&$-(0,4,1,0)$&$-(0,8,2)$&$(16,8,1)$&$(0,56,174)$\\
6&$-(14,40,57)$&$0$&$(14,40,57)$&$-(0,14,40)$&$-(28,80,114)$&$(0,56)$&$(196,1120)$\\
\hline
\end{tabular}
\vspace{0.25cm}\\
\begin{tabular}{|c|ccccccc|}
\hline
$\substack{d_x=2\\d_y\slash d_z}$&0&1&2&3&4&5&6\\
\hline
0&$0$&$0$&$0$&$-1$&$0$&$(5,1,0)$&$(0,35,57,36,10,1)$\\
1&$0$&$0$&$0$&$0$&$0$&$0$&$0$\\
2&$0$&$0$&$0$&$2$&$0$&$-(10,2,0)$&$-(0,70,114,72,20,2)$\\
3&$-1$&$0$&$2$&$-(0,2,0)$&$-5$&$(0,10,2,0)$&$(32,79,114,72,20,2,0)$\\
4&$0$&$0$&$0$&$-5$&$0$&$(25,5,0)$&$(0,175,285,180,50,5,0)$\\
5&$(5,1,0)$&$0$&$-(10,2,0)$&$(0,10,2,0)$&$(25,5,0)$&$-(0,50,20,2,0)$&$-(160,427,649,474,172,30,2,0)$\\
6&$(0,35,57,36,10,1)$&$0$&$-(0,70,114,72,20,2)$&$(32,79,114,72,20,2)$&$(0,175,285,180,50,5,0)$&$-(160,427,649,474,172,30,2)$&$-(0,2240,6728,11310,12826)$\\
\hline
\end{tabular}
\vspace{0.25cm}\\
\begin{tabular}{|c|ccccccc|}
\hline
$\substack{d_x=3\\d_y\slash d_z}$&0&1&2&3&4&5&6\\
\hline
0&$0$&$0$&$0$&$0$&$1$&$(0,4,1,0)$&$(9,2,0)$\\
1&$0$&$0$&$0$&$0$&$0$&$0$&$0$\\
2&$0$&$0$&$-1$&$(0,1,0)$&$(6,1,0)$&$-(0,18,8,1)$&$-(45,54,59,36,10,1)$\\
3&$0$&$0$&$(0,1,0)$&$-(0,1,0)$&$-(0,6,1,0)$&$(0,18,8,1,0)$&$(0,45,54,59,36,10,1,0)$\\
4&$-1$&$0$&$(6,1,0)$&$-(0,6,1,0)$&$-(27,9,1,0)$&$(0,72,45,11,1,0)$&$(190,324,402,275,96,16,1,0)$\\
5&$(0,4,1,0)$&$0$&$-(0,18,8,1,0)$&$(0,18,8,1,0)$&$(0,72,45,11,1,0)$&$-(0,180,168,67,13,1,0)$&
$-(0,490,1072,1470,1168,527,134,18,1,0)$\\
6&$(9,2,0)$&$0$&$-(45,54,59,36,10,1)$&$(0,45,54,59,36,10)$&$(190,324,402,275,96,16)$&$-(0,490,1072,1470,1168,527)$&$-(1314,4297,8082,9600,8269)$\\
\hline
\end{tabular}
\vspace{0.25cm}\\
\begin{tabular}{|c|ccccccc|}
\hline
$\substack{d_x=4\\d_y\slash d_z}$&0&1&2&3&4&5&6\\
\hline
0&$0$&$0$&$0$&$0$&$0$&$-1$&$0$\\
1&$0$&$0$&$0$&$0$&$0$&$0$&$0$\\
2&$0$&$0$&$0$&$-1$&$0$&$(11,2,0)$&$(0,34,57,36,10,1,0)$\\
3&$0$&$0$&$-1$&$(0,1,0)$&$4$&$-(0,11,2,0)$&$-(35,42,57,36,10,1,0)$\\
4&$0$&$0$&$0$&$4$&$-(9,6,1,0)$&$(-42,19,27,9,1,0)$&$-(0,82,174,126,38,4,0)$\\
5&$-1$&$0$&$(11,2,0)$&$-(0,11,2,0)$&$(-42,19,27,9,1,0)$&$-(-105,41,104,54,12,1,0)$&$(329,356,495,402,158,29,2,0)$\\
6&$0$&$0$&$(0,34,57,36,10,1)$&$-(35,42,57,36,10,1)$&$-(0,82,174,126,38,4)$&$(329,356,495,402,158,29,2)$&$(0,2056,5258,7092,6925,5010)$\\
\hline
\end{tabular}
\vspace{0.25cm}\\
\begin{tabular}{|c|ccccccc|}
\hline
$\substack{d_x=5\\d_y\slash d_z}$&0&1&2&3&4&5&6\\
\hline
0&$0$&$0$&$0$&$0$&$0$&$0$&$-1$\\
1&$0$&$0$&$0$&$0$&$0$&$0$&$0$\\
2&$0$&$0$&$0$&$0$&$-1$&$(0,5,1)$&$(22,8,1)$\\
3&$0$&$0$&$0$&$0$&$(0,1,0)$&$-(0,5,1,0)$&$-(0,22,8,1)$\\
4&$0$&$0$&$1$&$(0,1,0)$&$(18,7,1,0)$&$-(0,60,38)$&$-(192,155,88)$\\
5&$0$&$0$&$(0,5,1)$&$-(0,5,1,0)$&$-(0,60,38)$&$(0,186,169)$&$(0,570,685)$\\
6&$1$&$0$&$(22,8,1)$&$-(0,22,8,1)$&$-(192,155,88)$&$(0,570,685)$&$(1786,3163)$\\
\hline
\end{tabular}
\hspace{0.25cm}
\begin{tabular}{|c|ccccccc|}
\hline
$\substack{d_x=6\\d_y\slash d_z}$&0&1&2&3&4&5&6\\
\hline
0&$0$&$0$&$0$&$0$&$0$&$0$&$0$\\
1&$0$&$0$&$0$&$0$&$0$&$0$&$0$\\
2&$0$&$0$&$0$&$0$&$0$&$-1$&$0$\\
3&$0$&$0$&$0$&$0$&$0$&$(0,1)$&$6$\\
4&$0$&$0$&$0$&$0$&$0$&$(18,3,0)$&$-(0,27,26)$\\
5&$0$&$0$&$-1$&$(0,1)$&$(18,3,0)$&$-(0,60,21,2)$&$-(198,24,-40)$\\
6&$0$&$0$&$0$&$6$&$-(0,27,26)$&$-(198,24,-40)$&$(0,36,-96,-88)$\\
\hline
\end{tabular}
}
\caption{${^B N_{d_x,d_y,d_z}^{(\chi)}}$ for involution 2 of $[\P^2]^2$.}
\label{P22inv2GV}
\end{sidewaystable}

\begin{sidewaystable}
\tiny{
\begin{tabular}{|c|cccc|}
\hline
$\substack{d_x=0\\d_z\slash d_u}$&0&1&2&3\\
\hline
0&$0$&$0$&$0$&$0$\\
1&$0$&$0$&$0$&$0$\\
2&$0$&$0$&$0$&$0$\\
3&$0$&$0$&$0$&$0$\\
4&$0$&$0$&$0$&$0$\\
5&$0$&$0$&$0$&$0$\\
6&$0$&$0$&$0$&$0$\\
\hline
\end{tabular}
\hspace{0.25cm}
\begin{tabular}{|c|cccc|}
\hline
$\substack{d_x=1\\d_z\slash d_u}$&0&1&2&3\\
\hline
0&$-1$&$1$&$0$&$0$\\
1&$-1$&$0$&$0$&$0$\\
2&$-1$&$1$&$0$&$0$\\
3&$-1$&$1$&$0$&$0$\\
4&$-1$&$1$&$0$&$0$\\
5&$-1$&$1$&$0$&$0$\\
6&$-1$&$1$&$0$&$0$\\
\hline
\end{tabular}
\hspace{0.25cm}
\begin{tabular}{|c|cccc|}
\hline
$\substack{d_x=2\\d_z\slash d_u}$&0&1&2&3\\
\hline
0&$0$&$0$&$0$&$0$\\
1&$0$&$0$&$0$&$0$\\
2&$0$&$0$&$0$&$0$\\
3&$(0,1,0)$&$-(0,1,0)$&$0$&$0$\\
4&$0$&$0$&$0$&$0$\\
5&$(0,4,1,0)$&$-(0,5,1,0)$&$(0,1,0)$&$0$\\
6&$0$&$0$&$0$&$0$\\
\hline
\end{tabular}
\hspace{0.25cm}
\begin{tabular}{|c|cccc|}
\hline
$\substack{d_x=3\\d_z\slash d_u}$&0&1&2&3\\
\hline
0&$0$&$0$&$1$&$-1$\\
1&$0$&$0$&$0$&$0$\\
2&$0$&$-1$&$2$&$-1$\\
3&$1$&$-2$&$2$&$0$\\
4&$2$&$-(6,1,0)$&$(6,1,0)$&$-2$\\
5&$(5,1,0)$&$-(11,2,0)$&$(8,1,0)$&$-2$\\
6&$(9,2,0)$&$-(22,8,1,0)$&$(18,7,1,0)$&$-(5,1,0)$\\
\hline
\end{tabular}
\vspace{0.25cm}\\
\begin{tabular}{|c|cccc|}
\hline
$\substack{d_x=4\\d_z\slash d_u}$&0&1&2&3\\
\hline
0&$0$&$0$&$(0,1,0)$&$-(0,4,1,0)$\\
1&$0$&$0$&$0$&$0$\\
2&$0$&$0$&$0$&$-(0,4,1,0)$\\
3&$0$&$(0,1,0)$&$-(0,2,0)$&$(0,1,0)$\\
4&$(0,3,1,0)$&$0$&$-(0,3,1,0)$&$0$\\
5&$-(0,4,1,0)$&$(0,18,8,1,0)$&$-(0,24,9,1,0)$&$(0,11,2,0)$\\
6&$(0,35,57,36,10,1,0)$&$-(0,34,57,36,10,1,0)$&$-(0,9,6,1,0)$&$(0,8,6,1,0)$\\
\hline
\end{tabular}
\vspace{0.25cm}\\
\begin{tabular}{|c|cccc|}
\hline
$\substack{d_x=5\\d_z\slash d_u}$&0&1&2&3\\
\hline
0&$0$&$0$&$0$&$(4,4,1,0)$\\
1&$0$&$0$&$0$&$0$\\
2&$0$&$0$&$-1$&$(9,5,1,0)$\\
3&$0$&$0$&$-3$&$(8,4,1,0)$\\
4&$0$&$2$&$-(12,2,0)$&$(30,13,2,0)$\\
5&$-(5,10,6,1,0)$&$(10,2,0)$&$(-24,5,6,1,0)$&$(39,10,1,0)$\\
6&$-(14,40,57,36,10,1,0)$&$(45,54,59,36,10,1,0)$&$-(87,37,5,0)$&$(115,63,14,1,0)$\\
\hline
\end{tabular}
\vspace{0.25cm}\\
\begin{tabular}{|c|cccc|}
\hline
$\substack{d_x=6\\d_z\slash d_u}$&0&1&2&3\\
\hline
0&$0$&$0$&$0$&$(0,8,2,0)$\\
1&$0$&$0$&$0$&$0$\\
2&$0$&$0$&$0$&$(0,8,2,0)$\\
3&$0$&$0$&$(0,1,0)$&$-(0,9,5,1,0)$\\
4&$0$&$0$&$0$&$-(0,-4,3,1,0)$\\
5&$0$&$-(0,4,1,0)$&$(0,36,16,2,0)$&$-(108,76,21,2,0)$\\
6&$-(44,63,37,10,1,0)$&$(0,70,114,72,20,2,0)$&$-(0,50,102,70,20,2,0)$&$-(0,30,38,12,1,0)$\\
\hline
\end{tabular}

}
\caption{${^A N_{d_x,d_z,d_u}^{(\chi)}}$ for the involution of the hybridfly.}
\label{hybridflyGV}

\end{sidewaystable}

\begin{sidewaystable}
\tiny{
\begin{tabular}{|c|cccc|}
\hline
$\substack{d_x=0\\d_z\slash d_u}$&0&1&2&3\\
\hline
0&$0$&$0$&$(0,1,0)$&$(0,8,2,0)$\\
1&$-1$&$0$&$0$&$0$\\
2&$0$&$0$&$0$&$0$\\
3&$1$&$0$&$0$&$0$\\
4&$(0,3,1,0)$&$0$&$0$&$0$\\
5&$-(5,10,6,1,0)$&$0$&$0$&$0$\\
6&$-(0,44,63,37,10,1,0)$&$0$&$0$&$0$\\
\hline
\end{tabular}
\hspace{0.25cm}
\begin{tabular}{|c|cccc|}
\hline
$\substack{d_x=1\\d_z\slash d_u}$&0&1&2&3\\
\hline
0&$1$&$1$&$1$&$(4,4,1,0)$\\
1&$0$&$0$&$0$&$0$\\
2&$-1$&$-1$&$-1$&$-(4,4,1,0)$\\
3&$(0,1,0)$&$(0,1,0)$&$(0,1,0)$&$(0,4,4,1,0)$\\
4&$2$&$2$&$2$&$(8,8,2,0)$\\
5&$-(0,4,1,0)$&$-(0,4,1,0)$&$-(0,4,1,0)$&$-(0,16,20,8,1,0)$\\
6&$-(14,40,57,36,10,1,0)$&$-(14,40,57,36,10,1,0)$&$-(14,40,57,36,10,1,0)$&$-(56,216,402,412,241,80)$\\
\hline
\end{tabular}
\vspace{0.25cm}\\
\begin{tabular}{|c|cccc|}
\hline
$\substack{d_x=2\\d_z\slash d_u}$&0&1&2&3\\
\hline
0&$0$&$0$&$0$&$-(0,4,1,0)$\\
1&$0$&$0$&$0$&$0$\\
2&$0$&$0$&$0$&$(0,8,2,0)$\\
3&$-1$&$-2$&$-3$&$-(10,8,2,0)$\\
4&$0$&$0$&$0$&$-(0,20,5,0)$\\
5&$(5,1,0)$&$(10,2,0)$&$(15,3,0)$&$(50,50,18,2,0)$\\
6&$(0,35,57,36)$&$(0,70,114,72,20)$&$(105,171,108,30)$&$(0,478,918,895,502)$\\
\hline
\end{tabular}
\hspace{0.25cm}
\begin{tabular}{|c|cccc|}
\hline
$\substack{d_x=3\\d_z\slash d_u}$&0&1&2&3\\
\hline
0&$0$&$0$&$0$&$-1$\\
1&$0$&$0$&$0$&$0$\\
2&$0$&$1$&$2$&$(9,5,1,0)$\\
3&$0$&$-(0,1,0)$&$-(0,2,0)$&$-(0,9,5,1,0)$\\
4&$-1$&$-(6,1,0)$&$-(12,2,0)$&$-(46,36,11,1,0)$\\
5&$(0,4,1,0)$&$(0,18,8,1)$&$(0,36,16,2)$&$(0,130,142,64,13,1)$\\
6&$(9,2,0)$&$(45,54,59,36,10)$&$(90,108,118,72,20)$&$(334,668,840,673,329)$\\
\hline
\end{tabular}
\vspace{0.25cm}\\
\begin{tabular}{|c|cccc|}
\hline
$\substack{d_x=4\\d_z\slash d_u}$&0&1&2&3\\
\hline
0&$0$&$0$&$0$&$0$\\
1&$0$&$0$&$0$&$0$\\
2&$0$&$0$&$0$&$-(0,4,1,0)$\\
3&$0$&$1$&$2$&$(8,4,1,0)$\\
4&$0$&$0$&$-(0,3,1,0)$&$-(0,-4,3,1,0)$\\
5&$-1$&$-(11,2,0)$&$(-24,5,6,1,0)$&$(-88,-24,14,8,1,0)$\\
6&$0$&$-(34,57,36,10,1,0)$&$-(0,50,102,70,20,2,0)$&$-(0,340,593,538,279,84,14,1)$\\
\hline
\end{tabular}
\vspace{0.25cm}\\
\begin{tabular}{|c|cccc|}
\hline
$\substack{d_x=5\\d_z\slash d_u}$&0&1&2&3\\
\hline
0&$0$&$0$&$0$&$0$\\
1&$0$&$0$&$0$&$0$\\
2&$0$&$0$&$0$&$-1$\\
3&$0$&$0$&$0$&$(0,1,0)$\\
4&$0$&$1$&$(6,1,0)$&$(30,13,2,0)$\\
5&$0$&$-(0,5,1,0)$&$-(0,24,9,1,0)$&$-(0,108,76,21,2,0)$\\
6&$-1$&$-(22,8,1,0)$&$-(87,37,5,0)$&$-(366,317,152,51,11,1,0)$\\
\hline
\end{tabular}
\vspace{0.25cm}\\
\begin{tabular}{|c|cccc|}
\hline
$\substack{d_x=6\\d_z\slash d_u}$&0&1&2&3\\
\hline
0&$0$&$0$&$0$&$0$\\
1&$0$&$0$&$0$&$0$\\
2&$0$&$0$&$0$&$0$\\
3&$0$&$0$&$0$&$0$\\
4&$0$&$0$&$0$&$0$\\
5&$0$&$1$&$(8,1,0)$&$(39,10,1,0)$\\
6&$0$&$0$&$-(9,6,1,0)$&$-(0,30,38,12,1,0)$\\
\hline
\end{tabular}

}
\caption{${^B N_{d_x,d_z,d_u}^{(\chi)}}$ for the involution of $[\mathbb F_0][\P^2]$.}
\label{F0P2GV}
\end{sidewaystable}

\begin{table}
\tiny{
\begin{tabular}{|c|c|c|c|c|}
\hline
$\substack{d_x=0\\d_y\slash d_z }$&$0$&$1$&$2$&$3$\\
\hline
\hline
$0$&$0$&$0$&$0$&$0$\\
$1$&$0$&$0$&$0$&$0$\\
$2$&$0$&$0$&$0$&$0$\\
$3$&$0$&$0$&$0$&$0$\\
\hline
\end{tabular}
\hspace{0.25cm}
\begin{tabular}{|c|c|c|c|c|}
\hline
$\substack{d_x=1\\d_y\slash d_z }$&$0$&$1$&$2$&$3$\\
\hline
\hline
$0$&$-1$&$1$&$0$&$0$\\
$1$&$1$&$-1$&$0$&$0$\\
$2$&$0$&$0$&$0$&$0$\\
$3$&$0$&$0$&$0$&$0$\\
\hline
\end{tabular}
\hspace{0.25cm}
\begin{tabular}{|c|c|c|c|c|}
\hline
$\substack{d_x=2\\d_y\slash d_z }$&$0$&$1$&$2$&$3$\\
\hline
\hline
$0$&$0$&$0$&$0$&$0$\\
$1$&$0$&$0$&$0$&$0$\\
$2$&$0$&$0$&$0$&$0$\\
$3$&$0$&$0$&$0$&$0$\\
\hline
\end{tabular}
\vspace{0.25cm}\\
\begin{tabular}{|c|c|c|c|c|}
\hline
$\substack{d_x=3\\d_y\slash d_z }$&$0$&$1$&$2$&$3$\\
\hline
\hline
$0$&$0$&$0$&$0$&$0$\\
$1$&$0$&$3$&$-3$&$0$\\
$2$&$0$&$-3$&$3$&$0$\\
$3$&$0$&$0$&$0$&$0$\\
\hline
\end{tabular}
\hspace{0.25cm}
\begin{tabular}{|c|c|c|c|c|}
\hline
$\substack{d_x=4\\d_y\slash d_z }$&$0$&$1$&$2$&$3$\\
\hline
\hline
$0$&$0$&$0$&$0$&$0$\\
$1$&$0$&$0$&$0$&$0$\\
$2$&$0$&$0$&$(0,1,0)$&$0$\\
$3$&$0$&$0$&$0$&$0$\\
\hline
\end{tabular}
\hspace{0.25cm}
\begin{tabular}{|c|c|c|c|c|}
\hline
$\substack{d_x=5\\d_y\slash d_z }$&$0$&$1$&$2$&$3$\\
\hline
\hline
$0$&$0$&$0$&$0$&$0$\\
$1$&$0$&$0$&$5$&$-5$\\
$2$&$0$&$5$&$-(30,6,0)$&$(30,6,0)$\\
$3$&$0$&$-5$&$(30,6,0)$&$-(30,6,0)$\\
\hline
\end{tabular}
\vspace{0.25cm}\\
\begin{tabular}{|c|c|c|c|c|}
\hline
$\substack{d_x=6\\d_y\slash d_z }$&$0$&$1$&$2$&$3$\\
\hline
\hline
$0$&$0$&$0$&$0$&$0$\\
$1$&$0$&$0$&$0$&$0$\\
$2$&$0$&$0$&$0$&$(0,4,1,0)$\\
$3$&$0$&$0$&$(0,4,1,0)$&$-(0,8,2,0)$\\
\hline
\end{tabular}

}

\caption{${^{A} N^{(\chi)}_{d_x,d_y,d_z}}$ for involution 1 \& 3  of the pentafly.
Note that involution 3 has additional non-vanishing invariants that do not satisfy the
$d_Q$-$\chi$-correlation implicit in these tables. Namely, the additional invariants are 
${^{A} N^{(-1)}_{0,1,0}}=-\,{^{A} N^{(-1)}_{0,0,1}}=1$.}
\label{pentaflypointGV}
\end{table}

\begin{table}
\tiny{
\begin{tabular}{|c|c|c|c|c|}
\hline
$\substack{d_x=0\\d_y\slash d_z }$&$0$&$1$&$2$&$3$\\
\hline
\hline
$0$&$0$&$0$&$0$&$0$\\
$1$&$0$&$0$&$0$&$0$\\
$2$&$0$&$0$&$0$&$0$\\
$3$&$0$&$0$&$0$&$0$\\
\hline
\end{tabular}
\hspace{0.25cm}
\begin{tabular}{|c|c|c|c|c|}
\hline
$\substack{d_x=1\\d_y\slash d_z }$&$0$&$1$&$2$&$3$\\
\hline
\hline
$0$&$1$&$1$&$0$&$0$\\
$1$&$1$&$3$&$5$&$7$\\
$2$&$0$&$5$&$(35,8,0)$&$(135,72,11,0)$\\
$3$&$0$&$7$&$(135,72,11,0)$&$(1100,1304,662,160,15,0)$\\
\hline
\end{tabular}
\hspace{0.25cm}
\begin{tabular}{|c|c|c|c|c|}
\hline
$\substack{d_x=2\\d_y\slash d_z }$&$0$&$1$&$2$&$3$\\
\hline
\hline
$0$&$0$&$0$&$0$&$0$\\
$1$&$0$&$0$&$0$&$0$\\
$2$&$0$&$0$&$0$&$0$\\
$3$&$0$&$0$&$0$&$0$\\
\hline
\end{tabular}
\vspace{0.25cm}\\
\begin{tabular}{|c|c|c|c|c|}
\hline
$\substack{d_x=3\\d_y\slash d_z }$&$0$&$1$&$2$&$3$\\
\hline
\hline
$0$&$0$&$0$&$0$&$0$\\
$1$&$0$&$-1$&$-3$&$-5$\\
$2$&$0$&$-3$&$-(30,6,0)$&$-(147,66,9,0)$\\
$3$&$0$&$-5$&$-(174,66,9,0)$&$-(1494,1509,681,150,13,0)$\\
\hline
\end{tabular}
\hspace{0.25cm}
\begin{tabular}{|c|c|c|c|c|}
\hline
$\substack{d_x=4\\d_y\slash d_z }$&$0$&$1$&$2$&$3$\\
\hline
\hline
$0$&$0$&$0$&$0$&$0$\\
$1$&$0$&$0$&$0$&$0$\\
$2$&$0$&$0$&$(0,1,0)$&$(0,4,1,0)$\\
$3$&$0$&$0$&$(0,4,1,0)$&$(0,40,30,9,1,0)$\\
\hline
\end{tabular}
\vspace{0.25cm}\\
\begin{tabular}{|c|c|c|c|c|}
\hline
$\substack{d_x=5\\d_y\slash d_z }$&$0$&$1$&$2$&$3$\\
\hline
\hline
$0$&$0$&$0$&$0$&$0$\\
$1$&$0$&$0$&$0$&$0$\\
$2$&$0$&$0$&$3$&$(30,6,0)$\\
$3$&$0$&$0$&$(30,6,0)$&$(504,341,95,10,0)$\\
\hline
\end{tabular}
\hspace{0.25cm}
\begin{tabular}{|c|c|c|c|c|}
\hline
$\substack{d_x=2\\d_y\slash d_z }$&$0$&$1$&$2$&$3$\\
\hline
\hline
$0$&$0$&$0$&$0$&$0$\\
$1$&$0$&$0$&$0$&$0$\\
$2$&$0$&$0$&$0$&$0$\\
$3$&$0$&$0$&$0$&$-(8,2,0)$\\
\hline
\end{tabular}

}

\caption{${^{B} N^{(\chi)}_{d_x,d_y,d_z}}$ for involution 1 \& 3 of $[\mathbb F_0]^2$.
Note that involution 3 has additional non-vanishing invariants that do not satisfy the
$d_Q$-$\chi$-correlation implicit in these tables. Namely, the additional invariants are 
${^{B}N^{(-1)}_{2,1,0}}=- \, {^{B}N^{(-1)}_{2,0,1}}=1$. }
\label{F02pointGV}
\end{table}

\begin{table}
\tiny{
\begin{tabular}{|c|c|c|c|c|}
\hline
$\substack{d_x=0\\d_y\slash d_z }$&$0$&$1$&$2$&$3$\\
\hline
\hline
$0$&$0$&$0$&$0$&$0$\\
$1$&$0$&$0$&$0$&$0$\\
$2$&$0$&$0$&$0$&$0$\\
$3$&$0$&$0$&$0$&$0$\\
\hline
\end{tabular}
\hspace{0.25cm}
\begin{tabular}{|c|c|c|c|c|}
\hline
$\substack{d_x=1\\d_y\slash d_z }$&$0$&$1$&$2$&$3$\\
\hline
\hline
$0$&$-1$&$1$&$0$&$0$\\
$1$&$1$&$-1$&$0$&$0$\\
$2$&$0$&$0$&$0$&$0$\\
$3$&$0$&$0$&$0$&$0$\\
\hline
\end{tabular}
\hspace{0.25cm}
\begin{tabular}{|c|c|c|c|c|}
\hline
$\substack{d_x=2\\d_y\slash d_z }$&$0$&$1$&$2$&$3$\\
\hline
\hline
$0$&$0$&$0$&$0$&$0$\\
$1$&$0$&$0$&$0$&$0$\\
$2$&$0$&$0$&$0$&$0$\\
$3$&$0$&$0$&$0$&$0$\\
\hline
\end{tabular}
\vspace{0.25cm}\\
\begin{tabular}{|c|c|c|c|c|}
\hline
$\substack{d_x=3\\d_y\slash d_z }$&$0$&$1$&$2$&$3$\\
\hline
\hline
$0$&$0$&$0$&$1$&$-1$\\
$1$&$0$&$1$&$-2$&$1$\\
$2$&$1$&$-2$&$1$&$0$\\
$3$&$-1$&$1$&$0$&$0$\\
\hline
\end{tabular}
\hspace{0.25cm}
\begin{tabular}{|c|c|c|c|c|}
\hline
$\substack{d_x=4\\d_y\slash d_z }$&$0$&$1$&$2$&$3$\\
\hline
\hline
$0$&$0$&$0$&$(0,1,0)$&$-(0,4,1,0)$\\
$1$&$0$&$0$&$0$&$(0,4,1,0)$\\
$2$&$(0,1,0)$&$0$&$-(0,1,0)$&$0$\\
$3$&$-(0,4,1,0)$&$(0,4,1,0)$&$0$&$0$\\
\hline
\end{tabular}
\vspace{0.25cm}\\
\begin{tabular}{|c|c|c|c|c|}
\hline
$\substack{d_x=5\\d_y\slash d_z }$&$0$&$1$&$2$&$3$\\
\hline
\hline
$0$&$0$&$0$&$0$&$(4,4,1,0)$\\
$1$&$0$&$0$&$1$&$-(9,5,1,0)$\\
$2$&$0$&$1$&$-4$&$(8,1,0)$\\
$3$&$(4,4,1,0)$&$-(9,5,1,0)$&$(8,1,0)$&$-4$\\
\hline
\end{tabular}
\hspace{0.25cm}
\begin{tabular}{|c|c|c|c|c|}
\hline
$\substack{d_x=6\\d_y\slash d_z }$&$0$&$1$&$2$&$3$\\
\hline
\hline
$0$&$0$&$0$&$0$&$(0,8,2,0)$\\
$1$&$0$&$0$&$0$&$-(0,8,2,0)$\\
$2$&$0$&$0$&$0$&$(0,4,1,0)$\\
$3$&$(0,8,2,0)$&$-(0,8,2,0)$&$(0,4,1,0)$&$0$\\
\hline
\end{tabular}

}
\caption{${^{A} N^{(\chi)}_{d_x,d_z,d_u}}$ for involution 2 of the pentafly.}
\label{pentainv2GV}
\end{table}

\begin{table}
\tiny{
\begin{tabular}{|c|c|c|c|c|}
\hline
$\substack{d_x=0\\d_y\slash d_z }$&$0$&$1$&$2$&$3$\\
\hline
\hline
$0$&$0$&$0$&$(0,1,0)$&$(0,8,2,0)$\\
$1$&$0$&$0$&$0$&$0$\\
$2$&$(0,1,0)$&$0$&$0$&$0$\\
$3$&$(0,8,2,0)$&$0$&$0$&$0$\\
\hline
\end{tabular}
\hspace{0.25cm}
\begin{tabular}{|c|c|c|c|c|}
\hline
$\substack{d_x=1\\d_y\slash d_z }$&$0$&$1$&$2$&$3$\\
\hline
\hline
$0$&$1$&$1$&$1$&$(4,4,1,0)$\\
$1$&$1$&$1$&$1$&$(4,4,1,0)$\\
$2$&$1$&$1$&$1$&$(4,4,1,0)$\\
$3$&$(4,4,1,0)$&$(4,4,1,0)$&$(4,4,1,0)$&$(16,32,24,8,1,0)$\\
\hline
\end{tabular}
\vspace{0.25cm}\\
\begin{tabular}{|c|c|c|c|c|}
\hline
$\substack{d_x=2\\d_y\slash d_z }$&$0$&$1$&$2$&$3$\\
\hline
\hline
$0$&$0$&$0$&$0$&$-(0,4,1,0)$\\
$1$&$0$&$0$&$0$&$-(0,8,2,0)$\\
$2$&$0$&$0$&$0$&$-(0,12,3,0)$\\
$3$&$-(0,4,1,0)$&$-(0,8,2,0)$&$-(0,12,3,0)$&$-(0,80,52,16,2,0)$\\
\hline
\end{tabular}
\vspace{0.25cm}\\
\begin{tabular}{|c|c|c|c|c|}
\hline
$\substack{d_x=3\\d_y\slash d_z }$&$0$&$1$&$2$&$3$\\
\hline
\hline
$0$&$0$&$0$&$0$&$-1$\\
$1$&$0$&$-1$&$-2$&$-(9,5,1,0)$\\
$2$&$0$&$-2$&$-4$&$-(18,10,2,0)$\\
$3$&$-1$&$-(9,5,1,0)$&$-(18,10,2,0)$&$-(74,87,43,10,1,0)$\\
\hline
\end{tabular}
\hspace{0.25cm}
\begin{tabular}{|c|c|c|c|c|}
\hline
$\substack{d_x=4\\d_y\slash d_z }$&$0$&$1$&$2$&$3$\\
\hline
\hline
$0$&$0$&$0$&$0$&$0$\\
$1$&$0$&$0$&$0$&$(0,4,1,0)$\\
$2$&$0$&$0$&$-1$&$(0,4,1,0)$\\
$3$&$0$&$(0,4,1,0)$&$(0,4,1,0)$&$(0,48,24,7,1,0)$\\
\hline
\end{tabular}
\vspace{0.25cm}\\

\begin{tabular}{|c|c|c|c|c|}
\hline
$\substack{d_x=5\\d_y\slash d_z }$&$0$&$1$&$2$&$3$\\
\hline
\hline
$0$&$0$&$0$&$0$&$0$\\
$1$&$0$&$0$&$0$&$1$\\
$2$&$0$&$0$&$1$&$(8,1,0)$\\
$3$&$0$&$1$&$(8,1,0)$&$(46,19,3,0)$\\
\hline
\end{tabular}
\hspace{0.25cm}
\begin{tabular}{|c|c|c|c|c|}
\hline
$\substack{d_x=6\\d_y\slash d_z }$&$0$&$1$&$2$&$3$\\
\hline
\hline
$0$&$0$&$0$&$0$&$0$\\
$1$&$0$&$0$&$0$&$0$\\
$2$&$0$&$0$&$0$&$0$\\
$3$&$0$&$0$&$0$&$0$\\
\hline
\end{tabular}
}
\caption{${^{B} N^{(\chi)}_{d_x,d_y,d_z}}$ for involution 2 of $[\mathbb F_0]^2$.}
\label{F02inv2GV}

\end{table}

\end{document}